\documentclass[manuscript,screen]{acmart}

\usepackage[utf8]{inputenc}
\usepackage{geometry}
\usepackage{newunicodechar}
\newunicodechar{↔}{\leftrightarrow}      
\usepackage{graphicx}
\usepackage{textcomp}
\usepackage{subcaption}
\usepackage{xspace}
\usepackage{multirow}
\usepackage{xcolor}

\definecolor{olive}{RGB}{128, 128, 80}
\definecolor{darkgreen}{rgb}{0.0, 0.5, 0.0}

\definecolor{tabGreen}{RGB}{34, 139, 34}
\definecolor{tabRed}{RGB}{220, 20, 60}
\definecolor{tabBlue}{RGB}{65, 105, 225}

\usepackage{fancyvrb}
\usepackage{rotating}
\usepackage[many]{tcolorbox}    	
\usepackage{alltt}
\tcbuselibrary{listings}
\usepackage{tikz}
\usepackage{soul}
\usepackage[normalem]{ulem}
\usepackage{tabularx}
\usepackage{makecell}
\usepackage{url}
\usepackage{booktabs}
\usepackage{pifont}
\usepackage{paralist}
\usepackage{ifthen}
\usepackage{enumitem}
\usepackage[table]{xcolor}
\usepackage[export]{adjustbox}
\usepackage{amsmath,amsfonts,amssymb}
\usepackage{algorithmic}
\usepackage{comment}
\usepackage{ifxetex}
\usepackage{pgfplots}
\pgfplotsset{compat=1.18}
\usepackage{pgfplotstable}
\usetikzlibrary{patterns}
\usepackage{stfloats}
\usepackage{ragged2e}

\tcbset{
  enhanced,
  breakable,
  colback=yellow!8,
  colframe=orange!70!black,
  coltitle=black,
  boxrule=0.6pt,
  arc=3mm,
  left=6pt,right=6pt,top=6pt,bottom=6pt,
}

\newtcolorbox{highlightbox}[1][]{
  title=#1,
  fonttitle=\bfseries,
}

\ifxetex
  \usepackage{fontspec}
  \usepackage{xeCJK}
\fi

\def\BibTeX{{\rm B\kern-.05em{\sc i\kern-.025em b}\kern-.08em
    T\kern-.1667em\lower.7ex\hbox{E}\kern-.125emX}}

\newcommand{\ie}{\emph{i.e.,}\xspace}
\newcommand{\eg}{\emph{e.g.,}\xspace}
\newcommand{\etc}{etc.\xspace}
\newcommand{\etal}{\emph{et~al.}\xspace}
\newcommand{\secref}[1]{Section~\ref{#1}\xspace}

\newcommand{\figref}[1]{Fig.~\ref{#1}\xspace}

\newcommand{\tabref}[1]{Table~\ref{#1}\xspace}

\tcbset{
  on line,
  boxsep=1pt,
  left=2pt,
  right=2pt,
  top=1pt,
  bottom=1pt,
  colframe=blue!70!black, 
  colback=blue!70!black,  
  coltext=white,          
  rounded corners,
  boxrule=0pt             
}

\newcommand{\MTfft}{\textit{MT-FFT}\xspace}
\newcommand{\MTqlora}{\textit{MT-QLoRA}\xspace}

\newcommand{\STqlora}{\textit{ST-QLoRA}\xspace}
\newcommand{\MT}{\textit{Multi-task}\xspace}
\newcommand{\ST}{Single-task\xspace}
\newcommand{\CS}{\textsc{CS}$_{\text{\textit{Java,Python}}}$}
\newcommand{\CG}{\textsc{CG}$_{\text{\textit{Java,Python}}}$}
\newcommand{\CT}{\textsc{CT}$_{\text{\textit{Java,C\#}}}$}

\renewcommand{\rq}[1]{\textit{RQ\textsubscript{#1}}}

\newcommand{\lcms}{\textit{LCMs}\xspace}

\newcommand{\passatone}{\emph{pass@1}\xspace}
\newcommand{\java}{\textit{Java}\xspace}
\newcommand{\python}{\textit{Python}\xspace}
\newcommand{\vs}{\emph{vs.}\xspace}

\newcommand{\revlong}[1]{\begingroup\color{black}#1\endgroup}
\newcommand{\rev}[1]{\textcolor{black}{#1}}
\newcommand{\revtwo}[1]{\textcolor{black}{#1}}

\definecolor{promptgreen}{RGB}{128, 128, 80}
\definecolor{promptbg}{RGB}{255, 255, 248}

\newtcolorbox{evalprompt}[1][]{
    enhanced,
    colback=promptbg,
    colframe=promptgreen,
    colbacktitle=promptgreen!30!white,
    coltitle=black,
    fonttitle=\bfseries\sffamily,
    title=#1,
    boxrule=1pt,
    arc=0pt,
    left=10pt,
    right=10pt,
    top=8pt,
    bottom=8pt,
}

\newtcolorbox{summarybox}[1][]{
    colback=gray!10, 
    colframe=black, 
    arc=5mm, 
    colupper=black, 
    boxrule=1pt,
    title={\textbf{#1}}, 
    coltitle=white,
    colbacktitle=black,
    toptitle=1mm,
    bottomtitle=1mm,
    left=6pt, 
    right=6pt, 
    top=6pt, 
    bottom=6pt,
    before upper={\justifying}
}

\begin{document}


\title{Parameter-Efficient Multi-Task Fine-Tuning in Code-Related Tasks}

\author{Md Zahidul Haque}
\affiliation{%
  \institution{William \& Mary}
  \department{Computer Science}
  \city{Williamsburg}
  \state{VA}
  \country{USA}
}
\email{mhaque@wm.edu}

\author{Saima Afrin}
\affiliation{%
  \institution{William \& Mary}
  \department{Computer Science}
  \city{Williamsburg}
  \state{VA}
  \country{USA}
}
\email{safrin@wm.edu}

\author{Antonio Mastropaolo}
\affiliation{%
  \institution{William \& Mary}
  \department{Computer Science}
  \city{Williamsburg}
  \state{VA}
  \country{USA}
}
\email{amastropaolo@wm.edu}

\renewcommand{\shortauthors}{Haque, Afrin and Mastropaolo}

\newcommand{\ANTONIO}[1]{{\color{red}\textbf{[ANTONIO: #1]}}}
\newcommand{\ALVI}[1]{{\color{purple}\textbf{[ALVI: #1]}}}
\newcommand{\saima}[1]{{\color{blue}\textbf{[SAIMA: #1]}}}

\begin{abstract}

Large Language Models (LLMs) have proven highly effective in automating software engineering tasks, bridging natural language and code semantics to achieve notable results in code generation and summarization. However, their scale incurs substantial computational costs, making full fine-tuning impractical. Parameter-Efficient Fine-Tuning (PEFT) methods like QLoRA enable efficient specialization with lower resource demands. Recent studies show QLoRA-optimized Large Code Models (LCMs) perform strongly across diverse tasks, yet it remains unclear whether this effectiveness persists when a single model is QLoRA fine-tuned for multiple code-related tasks. The interaction between Multi-task fine-tuning and QLoRA optimization, and how transfer learning affects correctness and quality of generated artifacts, remains largely unexplored.
We investigate Multi-task QLoRA fine-tuning \revtwo{--- a representative parameter-efficient method from the LoRA family ---} across three representative tasks: code generation, translation, and summarization. We evaluate functional correctness through execution-based and similarity-based metrics, complemented by comprehensive code quality analysis--an aspect largely overlooked in prior work. Our findings show that Multi-task QLoRA effectively leverages transfer learning, achieving competitive or superior performance \rev{at the 1.5B, 3B, and 7B configurations} relative to both Single-task QLoRA and Multi-task full fine-tuning. Larger models demonstrate more consistent balance between correctness and quality, whereas smaller models preserve functionality but exhibit a higher incidence of quality-related issues.

\end{abstract}



\maketitle


\section{Introduction}

The emergence of Large Code Models (\lcms)--specialized architectures designed for code understanding and generation--has fundamentally expanded the automation capabilities within software engineering (SE). As adaptations of Large Language Models (LLMs) for code-intensive applications, \lcms have redefined the boundaries of what practitioners and developers can accomplish when supported by intelligent coding assistants capable of automating a variety of activities, including the generation of various software artifacts, such as code and code documentation \cite{ciniselli2021empirical, du2024evaluating, ahmed2022few, leclair2020improved}. 
By leveraging large-scale pre-training on source code and natural language corpora, \lcms capture latent aspects of program semantics that enable them to automate key software engineering tasks and streamline workflows across the development lifecycle \cite{wang2025software}. These capabilities have transformed \lcms into powerful tools for practitioners, facilitating effective knowledge transfer and fundamentally reshaping how developers understand, write, and reason about code.

Facilitated by the introduction and widespread adoption of \textit{``on-the-fly''} adaptation techniques--such as prompt engineering methods that enable task-specific calibration without retraining or modifying the model's internal parameters--\lcms have quickly established themselves as flexible and accessible tools for a broad range of software engineering applications. These methods allow practitioners to guide model behavior through natural language instructions, examples, or structured information (\ie prompts), reducing the need for costly fine-tuning cycles while maintaining strong performance across diverse tasks.

However, despite the impressive adaptability offered by prompt-based interaction, recent studies~\cite{weyssow2023exploring,liu2022few,ahmed2024automatic,ahmed2022few,afrin2025resource} have shown that fine-tuning remains essential for achieving consistent, high-quality performance--particularly in specialized downstream code-related tasks such as code summarization and code generation. In this regard, fine-tuned \lcms not only exhibit greater robustness and contextual precision but, when optimized efficiently, can attain these improvements with substantially lower resource demands through techniques such as Parameter-Efficient Fine-Tuning (PEFT)~\cite{weyssow2025exploring, ayupov2022parameter, lu2023llama, su2024distilled, shi2023towards}.

PEFT methods achieve targeted model adaptation by updating only a small subset of parameters, thus removing the need for complete model retraining. This strategy offers a highly favorable balance between efficiency and performance, achieving substantial reductions in computational cost while maintaining--or even improving--accuracy on downstream code-related tasks \cite{ayupov2022parameter, chen2023pass, choi2023codeprompt, wang2025beyond, zlotchevski2022exploring, goel2022cross, saberi2024utilization, wang2023prompt, wang2023one, liu2023empirical}. 

If, on the one hand, this behavior represents a clear advantage in terms of efficiency and scalability, on the other hand, it raises important questions about the extent to which these improvements hold across heterogeneous tasks. Most existing evaluations of PEFT methods--including QLoRA\cite{dettmers2024qlora}, which has been consistently identified as one of the most effective fine-tuning approaches for software engineering automation \cite{afrin2025systematic}--have been conducted in \ST settings, typically focusing on code generation or summarization. In such cases, task boundaries are clearly defined, and model objectives remain uniform. However, real-world software engineering workflows often involve multiple, interdependent tasks, such as generating, translating, and summarizing code, which demand that models generalize effectively across different modalities and reasoning processes.

On top of that, introducing additional parameters through QLoRA optimization may disrupt the delicate balance between parameter efficiency and representational stability, particularly during joint fine-tuning across diverse tasks. This concern aligns with the findings of Afrin \etal~\cite{afrin2025quantization}, who demonstrated that quantization can reshape a model's parameter space by altering its topological structure, potentially affecting output quality. Since QLoRA inherently couples learnable parameter injection with quantization, it becomes crucial to examine whether these combined processes compromise output quality, consistency, and reliability--an aspect particularly relevant to modern software engineering practices, where models must handle heterogeneous and interdependent tasks within a unified framework. Understanding this interaction is essential to determine whether QLoRA maintains both efficiency and stability across diverse code-related applications. In practice, this requires models to perform effectively on tasks such as natural language to code generation (NL-to-Code), code summarization (Code-to-NL), and code transformations (Code-to-Code) (\eg code translation). Ensuring consistent performance across these modalities demands a shared representational space that captures the semantic and syntactic relationships between programming and natural languages. Evaluating QLoRA's capacity to preserve this coherence under \MT fine-tuning is therefore key to developing scalable, resource-efficient \lcms capable of supporting real-world software engineering workflows. 

Motivated by these considerations, this paper investigates the effect of \MT QLoRA fine-tuning across \rev{a representative set of} code-related tasks, evaluating \textit{code generation}, \textit{code summarization}, and \textit{code translation} under a unified framework \rev{spanning the three canonical input-output modalities for \lcms (NL$\rightarrow$Code, Code$\rightarrow$NL, and Code$\rightarrow$Code)}. Specifically, we adopt Qwen2.5-Coder--a model family widely recognized for its performance in code-oriented applications--as the base \textit{CLM}. To ensure a controlled and balanced evaluation, we employ two task-aligned datasets (one for \python and one for \java) for the Code-to-NL (code summarization) and NL-to-Code tasks (code generation), thereby strengthening the generalizability of our findings. For Code-to-Code applications, we focus on code translation, following prior research~\cite{ayupov2022parameter, choi2023codeprompt, liu2023empirical, wang2025beyond, wang2022no, wang2023prompt} by selecting \java and C\# as the source and target languages.

Our experimental design comprises three distinct training configurations applied across \rev{four} model scales from the Qwen2.5-Coder family (0.5B, 1.5B, \rev{3B, and 7B} parameters). For comprehensive evaluation, we implement: (i) QLoRA fine-tuning for both \MT and \ST scenarios, (ii) full-parameter fine-tuning (FFT) for \MT scenarios. \rev{This yields 20 Qwen2.5-Coder models: 4 \MT QLoRA models (one per size, trained on all three tasks), 12 \ST QLoRA models (3 tasks $\times$ 4 sizes), and 4 \MT FFT models (one per size).} \revtwo{To assess whether our findings hold beyond a single model family, we additionally replicate the RQ$_1$ comparison on CodeLlama-7B-Instruct (1 \MT QLoRA and 3 \ST QLoRA models) under an identical protocol.} \rev{We further train four 3B QLoRA variants for our ablation studies --- one using balanced task sampling and three restricted to two-task subsets --- bringing the total to 28 specialized models.}
We systematically analyze differences in functional correctness and non-functional quality, while controlling for potential confounding factors to ensure a rigorous and fair comparison across all configurations.

Our comprehensive evaluation reveals nuanced trade-offs between correctness and quality in \MT QLoRA configurations. While \MT QLoRA demonstrates comparable functional correctness to \ST variants \rev{(particularly at the 1.5B, 3B, and 7B configurations, while the 0.5B configuration preserves functionality but shows greater variability in code quality)}, it achieves this with substantially reduced memory footprint. When contrasted with full-parameter \MT fine-tuning, QLoRA maintains competitive performance while requiring significantly fewer computational resources.

To the best of our knowledge, this constitutes \rev{one of the first systematic investigations} examining both functional correctness and non-functional quality attributes of \MT QLoRA \revtwo{--- one representative method from the LoRA family ---} across \rev{the three studied code-related tasks}. Specifically, our contributions encompass:

\rev{This work makes contributions along two complementary axes. From a \emph{methodological} perspective, we present a systematic empirical characterization of \MT QLoRA across model scales for three code-related tasks. From an \emph{evaluation} perspective, we couple standard functional-correctness metrics with a non-functional quality analysis using multiple static analyzers, providing a more holistic view than functional metrics alone.}

\begin{itemize}[topsep=0pt, itemsep=0pt, leftmargin=1.5em]
\item A systematic empirical comparison of \textbf{\MT versus \ST QLoRA} configurations across code generation, summarization, and translation, quantifying differences in functional correctness and resource utilization across the Qwen2.5-Coder model family.
\item A rigorous evaluation contrasting \textbf{\MT QLoRA with \MT full fine-tuning}, establishing the practical trade-offs between parameter efficiency and functional performance across varying model scales.
\item \rev{A systematic} \textbf{non-functional quality assessment} of \MT QLoRA outputs, employing: (i) static analyzers (Pylint, PMD, Roslyn, Lizard, SonarCloud) for generation and translation tasks, and (ii) LLM-as-judge evaluation for summarization to assess summary quality across multiple dimensions.
\item A publicly available \textbf{replication package}~\cite{replication} containing curated datasets, fine-tuned models spanning the Qwen2.5-Coder family, evaluation harnesses, and comprehensive documentation to facilitate reproducibility and extension of our findings.
\end{itemize}

\revtwo{These contributions are scoped to QLoRA as a LoRA-family method; we do not test whether the observed patterns hold for PEFT approaches beyond that family (\eg adapters, prompt or prefix tuning).}



\smallskip



\section{Background and Related Work}
\label{sec:related_work}


This section provides an overview of recent advancements that form the foundation of our investigation. We begin by examining large code models and their role in automating software engineering tasks. We then discuss parameter-efficient fine-tuning (PEFT) techniques, with particular emphasis on QLoRA, which enable efficient adaptation of large models to specialized domains. Finally, we review recent findings on the quality of LLM-generated software artifacts, demonstrating that functional correctness alone provides an insufficient evaluation framework. While our work focuses on \MT QLoRA optimization, we note that \MT learning approaches for code models have evolved substantially since the introduction of text-to-text transfer transformers (T5)~\cite{raffel2020exploring, wang2021codet5, mastropaolo2021studying, mastropaolo2022using}; readers seeking a comprehensive synthesis of this evolution are directed to the systematic literature review by Watson \etal~\cite{watson2022systematic}.

\subsection{Large Code Models}
\label{subsec:clm_finetuning}

Recent advances in large language models have substantially expanded capabilities for code automation. Specialized large code models (\lcms) including CodeGen~\cite{nijkamp2022codegen}, AlphaCode~\cite{li2022competition}, PolyCoder~\cite{xu2022systematic}, CodeLlama~\cite{roziere2023code}, DeepSeek-Coder~\cite{zhu2024deepseek}, and Qwen2.5-Coder~\cite{hui2024qwen2} demonstrate proficiency across diverse software engineering tasks~\cite{hou2024large}, including code generation, summarization, translation, bug fixing, and test generation. General-purpose language models such as GPT-5~\cite{OpenAI2025GPT5}, Claude~\cite{anthropic2024claude}, and Gemini~\cite{team2023gemini} similarly exhibit strong performance on code-related tasks, establishing their utility as benchmarks for evaluating model reasoning, cross-language generalization, and coding capabilities. \revlong{Throughout this paper, we use ``\textit{LCM}'' to denote code-specialized models such as Qwen2.5-Coder, CodeLlama, and DeepSeek-Coder, and reserve ``\textit{LLM}'' for general-purpose models such as GPT-5, Claude, and Gemini, as well as for established methodological terminology (\eg LLM-as-a-judge).}

Although \lcms are often used without fine-tuning--adapting instead through instruction-based prompting that dynamically shapes model behavior rather than adjusting weights via backpropagation--this approach typically achieves only superficial adaptation with limited specialization for complex or domain-specific tasks~\cite{qin2023chatgpt, wei2022emergent}. Fine-tuning remains the most effective strategy for optimal performance and reliability, as it enables models to internalize domain-specific representations and capture nuanced task dependencies that prompting alone cannot achieve~\cite{shin2025prompt, weyssow2025exploring}.

\subsection{Parameter-Efficient Fine-Tuning}
\label{subsec:peft_multitask}
The substantial computational demands of fine-tuning billion parameter code models have motivated research into optimization strategies that reduce resource requirements while preserving model effectiveness. In a recent literature mapping, Shi \etal \cite{shi2025efficient} identified four key areas for optimizing large code models: data reduction, model-centric approaches, system-centric methods, and program-centric techniques. Parameter-Efficient Fine-Tuning (PEFT) falls within the model-centric category, optimizing model adaptation by updating only a subset of parameters rather than the entire model. The success of QLoRA for various coding tasks--as demonstrated in the systematic literature review by Afrin \etal \cite{afrin2025systematic}--can be attributed to its ability to achieve both performance improvements and efficiency gains in almost all evaluated applications across Automated Program Repair, Code Completion, Code Generation, and Code Summarization. This reliability stems from QLoRA's integration of 4-bit quantization with low-rank adaptation, enabling effective fine-tuning of large models in resource-constrained settings.

\revtwo{Within the LoRA family --- the most widely adopted PEFT approach for code-related tasks~\cite{afrin2025systematic} --- we study QLoRA as one representative method. Two considerations motivate this choice. First, on \textbf{efficiency}, QLoRA's 4-bit quantization of the frozen base fits 7B-scale fine-tuning onto a single commodity GPU, whereas standard LoRA, which keeps the base model in 16-bit precision, requires substantially more memory and is correspondingly less practical in the resource-constrained settings PEFT is meant to serve. Second, on \textbf{quality}, prior evidence supports QLoRA as a practical alternative to LoRA: head-to-head comparisons report comparable functional correctness for QLoRA and LoRA~\cite{weyssow2025exploring}, while recent empirical work on 4-bit weight quantization~\cite{afrin2025quantization} shows that both functional correctness and non-functional code quality can be preserved up to the 33B/34B scale on \java and \python benchmarks. These findings motivate our selection of QLoRA but do not establish that quantization is neutral under multi-task fine-tuning. Accordingly, our experiments characterize the differences between \ST and \MT fine-tuning under QLoRA specifically; they do not isolate the effect of quantization or establish that the same patterns would hold for standard LoRA, other LoRA-family variants, or PEFT approaches beyond the LoRA family (\eg adapters, prompt or prefix tuning).}

Recent research has applied PEFT methods to various code automation tasks, though primarily in \ST configurations. Studies have explored Adapter tuning \cite{wang2022adamix}, LoRA, and Prompt Tuning for tasks including code search, summarization, clone detection, defect detection, and code translation~\cite{sun2024source, wang2023prompt, wang2023one, ayupov2022parameter, chen2023pass, choi2023codeprompt, afrin2025resource, liu2023empirical, liu2024delving}. These investigations demonstrate that parameter-efficient methods can achieve performance comparable to or exceeding full fine-tuning while updating only a fraction of model parameters. However, existing PEFT research for code focuses exclusively on \ST scenarios where models are optimized for one specific objective. Whether parameter-efficient methods can maintain effectiveness when learning multiple code-related tasks simultaneously remains unexplored.

\subsection{Quantized Low-Rank Adaptation (QLoRA) of Large Code Models}
\label{sec:qlora}
Dettmers \etal~\cite{dettmers2024qlora} proposed QLoRA, an approach combining LoRA with LLM quantization\footnote{Quantization is a model compression technique that reduces the precision of numerical representations (\eg from 16-bit floating point to 8-bit or 4-bit formats) to decrease memory usage and computational cost. By representing model weights with fewer bits, quantization enables the fine-tuning and deployment of large models on resource-constrained hardware while maintaining near-original performance~\cite{dettmers2022gpt3, frantar2022gptq}.}.
QLoRA introduces key innovations including (i) the 4-bit NormalFloat (NF4) data type, (ii) Double Quantization (DQ), and (iii) Paged Optimizers, enabling efficient fine-tuning that reduces memory usage while maintaining high performance~\cite{dettmers2024qlora}. The method quantizes pre-trained model weights to 4-bit precision using NF4, a data type optimized for neural network weight distributions. Double quantization further reduces memory footprint by quantizing both model weights and quantization constants, while Paged Optimizers manage memory spikes during gradient checkpointing to prevent out-of-memory errors. A detailed explanation of QLoRA and the fine-tuning process is provided in \secref{sec:design_qlora} and \secref{sec:finetune_design}.

\subsubsection{QLoRA-based Optimization of \lcms}
Limited research has investigated QLoRA's efficiency for \rev{\lcms}. Yang \etal~\cite{yang2024multi} applied QLoRA to models including CodeLlama \cite{codellama2}, StarChat-alpha \cite{Tunstall2023starchat-alpha}, and Mistral-Instruct-7B \cite{mistral} for automatic program repair (APR), demonstrating its effectiveness in defect repair tasks. Weyssow \etal~\cite{weyssow2023exploring} compared PEFT techniques to In-Context Learning (ICL) for code generation, finding PEFT methods superior, and investigated QLoRA applicability across CodeLlama 7B, 13B, and 34B \python models using 8-bit and 4-bit quantization. On the other hand, Afrin \etal~\cite{afrin2025resource} explored QLoRA for code summarization tasks, demonstrating proficient performance when models are required to process code as input, capture its intent, and project that understanding into natural language output.

Liu \etal \cite{liu2024mftcoder} introduced MFTCoder, a \MT fine-tuning framework that applies LoRA and QLoRA across five code-related tasks: code completion, text-to-code generation, code comment generation, code translation, and unit test case generation. The framework incorporates multiple loss functions to address data imbalance, task difficulty variations, and convergence speed inconsistencies across tasks. Experimental results on CodeLlama-13B-Python demonstrated that their \MT approach outperformed both \ST fine-tuning and mixed-task data training across multiple benchmarks, achieving 74.4\% pass@1 on HumanEval with their CodeFuse-CodeLlama-34B model.


Despite these advances, several limitations persist in the current state of research. First, MFTCoder relies on synthetically generated training data through Self-Instruct and GPT-based generation for specific tasks, particularly code exercises datasets, raising questions about generalizability to real-world code distributions. Second, evaluation focuses exclusively on functional correctness metrics--pass@1 for generation tasks, BLEU scores for translation and summarization, and GPT-4-based assessment for code comments--without examining the quality attributes of code produced by generation and translation tasks, or the linguistic quality of natural language summaries beyond similarity scores.


To address these gaps, our work extends the evaluation of \MT QLoRA in three key directions. First, we conduct a systematic comparison across model scales (0.5B, 1.5B, 3B\rev{, 7B}) to understand how capacity interacts with \MT learning under parameter-efficient constraints. Second, we provide the first comprehensive non-functional quality assessment of \MT QLoRA outputs, employing static analyzers (Pylint, PMD, Roslyn, Lizard, SonarCloud) for code generation and translation tasks, and LLM-as-judge evaluation for summarization quality beyond lexical similarity. Third, we contrast \MT QLoRA against \MT full fine-tuning to establish the trade-offs between parameter efficiency and model effectiveness across different code-related tasks and model scales. \tabref{tab:related_work_comparison} summarizes the scope of existing QLoRA-based parameter-efficient fine-tuning studies for code-related tasks and highlights the gaps addressed by our work.

\begin{table*}[t]
\centering
\small
\caption{Comparison of parameter-efficient fine-tuning studies for code-related tasks. CG = Code Generation, CS = Code Summarization, CT = Code Translation, APR = Automatic Program Repair, CC = Code Completion, UT = Unit Test Generation.}
\label{tab:related_work_comparison}
\begin{tabular}{lccccccccccc}
\toprule
\multirow{2}{*}{\textbf{Study}} & \multirow{2}{*}{\textbf{Method}} & \multirow{2}{*}{\textbf{\MT}} & \multicolumn{6}{c}{\textbf{Tasks}} & \multirow{2}{*}{\textbf{Functional}} & \multirow{2}{*}{\textbf{Quality}} \\
\cmidrule{4-9}
& & & CG & CS & CT & APR & CC & UT & & \\
\midrule
Weyssow \etal~\cite{weyssow2023exploring} & QLoRA & \ding{55} & \ding{51} & & & & & & \ding{51} & \ding{55} \\
Yang \etal~\cite{yang2024multi} & QLoRA & \ding{55} & & & & \ding{51} & & & \ding{51} & \ding{55} \\
Afrin \etal~\cite{afrin2025resource} & QLoRA & \ding{55} & & \ding{51} & & & & & \ding{51} & \ding{55} \\
Liu \etal~\cite{liu2024mftcoder} & LoRA/QLoRA & \ding{51} & \ding{51} & \ding{51} & \ding{51} & & \ding{51} & \ding{51} & \ding{51} & \ding{55} \\
\midrule
\textbf{Our work} & QLoRA & \ding{51} & \ding{51} & \ding{51} & \ding{51} & & & & \ding{51} & \ding{51} \\
\bottomrule
\end{tabular}
\end{table*}


\section{Study Methodology}

The primary objective of this study is to comprehensively examine the impact of \MT QLoRA fine-tuning on the functional correctness and non-functional quality attributes of \lcms outputs, compared to \ST approaches. While QLoRA has demonstrated promise in reducing computational overhead for \ST scenarios, it remains crucial to understand whether these efficiency gains persist and at what cost when models must balance multiple, potentially competing objectives within a shared parameter space. To systematically investigate these concerns, we formulate two research questions:
\smallskip

\noindent \textbf{RQ}$_1$: How do \MT QLoRA-optimized models compare to \ST QLoRA models in terms of functional correctness and overall output quality across code generation, summarization, and translation tasks? \\
\noindent In \rq{1}, we measure performance variations when specializing \lcms with QLoRA under \MT versus \ST configurations, examining whether joint training on complementary code-related tasks enhances generalization or introduces performance trade-offs compared to task-specific optimization. 

\smallskip

\noindent \textbf{RQ}$_2$: How does \MT QLoRA fine-tuning compare to \MT full fine-tuning in terms of functional correctness and non-functional quality across code generation, summarization, and translation tasks? \\ 
\noindent In \rq{2}, we evaluate whether the parameter efficiency of \MT QLoRA comes at the cost of performance degradation compared to \MT full fine-tuning, assessing the trade-offs between computational efficiency and model effectiveness across different code-related tasks.

\smallskip



\subsection{\rev{Large Code Models}}

To address the research questions outlined above, we employ the Qwen2.5-Coder-Instruct model family across \rev{four} parameter scales: 0.5B, 1.5B, \rev{3B, and 7B}. Qwen2.5-Coder represents a code-specialized adaptation of the Qwen2.5 architecture, pre-trained on over 5.5 trillion tokens encompassing source code, code-grounded text, and synthetic data~\cite{hui2024qwen2}. The model family implements a carefully balanced data mixture of 70\% code, 20\% text, and 10\% mathematical content, designed to maintain strong coding capabilities while preserving general language understanding. We select Qwen2.5-Coder over alternatives such as CodeLlama and DeepSeek-Coder based on its demonstrated effectiveness in recent software engineering automation research \cite{zhuo2024bigcodebench, quan2025codeelo}--where it has shown strong performance across diverse code understanding and generation tasks. \revtwo{We note that this choice concerns the primary model family used throughout the study; CodeLlama is additionally employed as a second family in \secref{sec:rq1}, where we replicate the RQ$_1$ comparison to test whether our findings depend on a particular architecture, tokenizer, and pretraining distribution.}
On top of that, its robust instruction-following capabilities and competitive benchmark results make the Qwen2.5-Coder-Instruct family a reliable foundation for evaluating \MT scenarios where models must differentiate between task types--such as generation, summarization, and translation--and adapt their behavior according to natural language directives. Furthermore, its open availability, comprehensive documentation, and active community support facilitate reproducibility and enable rigorous experimental validation, aligning closely with the objectives of this study.

\revtwo{\textbf{Second model family.} For the cross-family replication, we use CodeLlama-7B-Instruct~\cite{roziere2023code}, selected because it differs from Qwen2.5-Coder along the three axes relevant to external validity --- architecture, tokenizer, and pretraining corpus --- while remaining a widely adopted, instruction-tuned code model at a scale matching our largest Qwen2.5-Coder configuration. We train one \MTqlora model on the joint three-task mixture and three \STqlora models (one per task), using the same adapter configuration (\tabref{tab:qlora_config}), hyperparameters (\tabref{tab:training_config}), data, and evaluation procedure as the Qwen2.5-Coder-7B runs, so that model family is the only varying factor. Because the two families differ in tokenization and pretraining data, we compare the direction of the \STqlora--\MTqlora differences rather than absolute scores. This replication covers RQ$_1$ only; RQ$_2$ and the ablations remain Qwen2.5-Coder-only.}

\subsection{Tasks and Datasets}
We ground our analysis on three fundamental code-related tasks that span different data modalities and represent core capabilities in modern software engineering workflows: \textbf{\textit{code generation (NL-to-Code)}}, \textbf{\textit{code summarization (Code-to-NL)}}, and \textbf{\textit{code translation (Code-to-Code)}}. These tasks collectively assess a model's ability to bridge natural language and code while transforming across programming languages, providing a comprehensive evaluation framework that captures the diverse reasoning patterns required for effective software engineering automation.

\subsubsection{Code Generation:}
\label{sec:codegen}


For code generation, we evaluate performance across two programming languages--\python and \java--to provide a comprehensive and balanced assessment. This dual-language approach enhances the empirical robustness of our analysis and strengthens the generalizability of our findings by capturing model behavior across syntactically and semantically distinct language paradigms commonly employed in code generation research \cite{yu2024codereval, cassano2023multipl}.


\textbf{Training Data:} We utilize the Code-to-Text dataset from the CodeXGLUE benchmark~\cite{lu2021codexglue}, adapting it for code generation through task inversion. CodeXGLUE was originally designed for code summarization and comprises pairs of code methods with corresponding natural language descriptions. We selected CodeXGLUE based on its established use in prior research on \rev{\lcms} for code-related tasks~\cite{afrin2025resource, wang2025beyond} and its comprehensive coverage of both \java and \python implementations.


During this adaptation process, we filtered instances where: (i) the docstring field was empty or missing, (ii) the code field was empty or missing, or (iii) signature extraction failed to produce a valid result. The third condition applies when code does not conform to expected structural patterns--specifically, \python code lacking lines beginning with \texttt{def} or \texttt{async def}, or \java code missing an opening brace to delimit the method signature from the body. These filtering criteria ensure that all retained instances contain valid input-output pairs with complete and parseable signatures. As shown in Table~\ref{tab:dataset_statistics}, this process yields 164,923 \java and 251,818 \python training instances, with corresponding validation sets of 5,183 and 13,914 instances, respectively.

\textbf{Evaluation Data:} For evaluation, we employ CoderEval~\cite{yu2024codereval}, a benchmark specifically designed for assessing code generation capabilities and widely used in the current body of research \cite{tambon2025bugs, wang2025beyond, paul2024benchmarks}. Originally, CoderEval featured 460 code generation problems (230 per language), where each problem comprises: (i) a natural language description specifying functional requirements, (ii) a reference implementation demonstrating a correct solution, and (iii) a test suite for evaluating functional correctness of generated code.

Before adopting CoderEval, we conducted quality assurance following Crupi \etal~\cite{crupi2025effectiveness}. Our final evaluation set contains 184 \java and 190 \python problems with verified test suites (reported as Test instances in \tabref{tab:dataset_statistics}), ensuring that metrics capture genuine functional correctness rather than artifacts of test suite deficiencies.

\subsubsection{Code Translation:}
\label{sec:codetrans}
For code translation, we employ the benchmark dataset from CodeXGLUE~\cite{lu2021codexglue}, which provides paired code snippets for migration between \java and C\#.

\textbf{Training Data:} The CodeXGLUE code translation dataset comprises paired \java and C\# code snippets that implement equivalent functionality, enabling bidirectional translation tasks (Java$\rightarrow$C\#, C\#$\rightarrow$Java). These tasks focus on preserving semantic equivalence while adapting syntax and idioms to the target language. Each pair serves as an input-output example where source code in one language corresponds to its functionally equivalent implementation in the other. As shown in Table~\ref{tab:dataset_statistics}, the dataset includes 10{,}300 training instances per direction (20{,}600 total) and 1{,}000 validation instances (500 per direction).

\textbf{Evaluation Data:} For evaluation, we rely on the test split provided by CodeXGLUE, comprising 1{,}000 Java$\rightarrow$C\# and 1{,}000 C\#$\rightarrow$Java translation pairs (2{,}000 total instances). Each test instance consists of a source code snippet paired with its reference target implementation.

\subsubsection{Code Summarization:}
\label{sec:codesumm}

For code summarization, we employ the CodeXGLUE Code-to-Text dataset~\cite{lu2021codexglue} for training and a curated subset from CoderEval~\cite{yu2024codereval} for evaluation. The CoderEval subset was created by inverting the input-output pairs from the code generation task, following the approach adopted in recent research~\cite{vitale2025optimizing,crupi2025effectiveness}.

\textbf{Training Data:}
For code summarization, we employ the same Code-to-Text dataset from the CodeXGLUE benchmark~\cite{lu2021codexglue}, which comprises paired examples of source code methods and their corresponding natural language summaries. The distinction from the code generation configuration lies in the input-output orientation: here, the input is the code snippet and the output is its semantically aligned natural language description--the inverse of the configuration used for code generation. Consequently, the dataset composition mirrors that of the code generation task in terms of size and language distribution, encompassing both \python and Java. Training, validation, and test splits are reported in Table~\ref{tab:dataset_statistics}.

\textbf{Evaluation Data:} For evaluation, we employ the curated CoderEval subset from Crupi \etal~\cite{crupi2025effectiveness}, constructed by selecting the 100 longest functions per language based on statement count, as meaningful summaries are particularly valuable for complex code. While Crupi \etal reported 198 instances (99 per language) after deduplication, their replication package contains 199 instances: 99 \java and 100 \python functions. We utilize all 199 instances (\tabref{tab:dataset_statistics}), each comprising a function with its corresponding developer-written summary.
The original CoderEval reference summaries contain documentation styles that differ from our training distribution. \java summaries consistently include Javadoc formatting with structured annotations (\eg \texttt{@param}, \texttt{@return} tags), inline markup (\eg \texttt{{@code}}), and HTML elements, while \python summaries frequently contain multi-line docstrings with parameter sections, return specifications, or verbose explanations spanning multiple sentences. Since our training data from CodeXGLUE comprises concise, single-sentence natural language summaries without such annotations or extended descriptions, \revlong{evaluating against raw CoderEval summaries with similarity-based metrics (BLEU, ROUGE-L, chrF) would penalize all systems uniformly for the absence of structural decoration that no model trained on conventional summarization data is expected to emit, conflating summary quality with documentation-formatting compliance.} To ensure a fair and consistent evaluation, we preprocess the CoderEval reference summaries for both languages \rev{in line with the curation approach used by Crupi \etal~\cite{crupi2025effectiveness}}: extracting only the first sentence from multi-sentence docstrings (to match the single-sentence summary format), removing structured annotations, and condensing verbose explanations into concise summaries. \revlong{This preprocessing is applied identically to the references used by every evaluated configuration --- \ST QLoRA, \MT QLoRA, and \MT full fine-tuning, across all model scales --- which share the same CodeXGLUE training distribution, so the inter-configuration comparison is unaffected by the choice of preprocessing rule. Model-generated summaries are evaluated as-produced; no normalization is applied on the system-output side.} The processed evaluation dataset is included in our replication package~\cite{replication}.


\begin{table*}[h]
\centering
\caption{Dataset statistics across all three code-related tasks. Training and Validation instances are from CodeXGLUE~\cite{lu2021codexglue}.}
\label{tab:dataset_statistics}
\begin{tabular}{@{}llrrr@{}}
\toprule
\textbf{Task} & \textbf{Language} & \textbf{Train} & \textbf{Validation} & \textbf{Test} \\
\midrule
\multirow{3}{*}{\textbf{Code Summarization (CS)}} 
    & \java & 164,923 & 5,183 & 99\textsuperscript{*} \\
    & \python & 251,820 & 13,914 & 100\textsuperscript{*} \\
    \cmidrule(lr){2-5}
    & \textit{Total} & \textit{416,743} & \textit{19,097} & \textit{199} \\
\midrule
\multirow{3}{*}{\textbf{Code Generation (CG)}} 
    & \java & 164,923 & 5,183 & 184\textsuperscript{*} \\
    & \python & 251,818 & 13,914 & 190\textsuperscript{*} \\
    \cmidrule(lr){2-5}
    & \textit{Total} & \textit{416,741} & \textit{19,097} & \textit{374} \\
\midrule
\multirow{3}{*}{\textbf{Code Translation (CT)}} 
    & \java $\rightarrow$ C\# & 10,300 & 500 & 1,000 \\
    & C\# $\rightarrow$ \java & 10,300 & 500 & 1,000 \\
    \cmidrule(lr){2-5}
    & \textit{Total} & \textit{20,600} & \textit{1,000} & \textit{2,000} \\
\bottomrule
\multicolumn{5}{@{}l@{}}{\textsuperscript{*}Test instances from CoderEval~\cite{yu2024codereval} (see Sections~\ref{sec:codegen} and~\ref{sec:codesumm}).}
\end{tabular}
\end{table*}

\subsection{QLoRA Fine-Tuning in a Nutshell}
\label{sec:design_qlora}

We employ QLoRA~\cite{dettmers2024qlora}, a parameter-efficient fine-tuning approach that combines low-rank adaptation (LoRA) with quantization to enable lightweight task-specific model specialization while keeping the base parameters frozen. LoRA modules are configured with rank $r = 8$, scaling factor $\alpha = 16$, and dropout probability $p = 0.1$, as summarized in Table~\ref{tab:qlora_config}. These adapters are injected into seven projection layers within each transformer block: the attention projections (\texttt{q\_proj}, \texttt{k\_proj}, \texttt{v\_proj}, \texttt{o\_proj}) and the feed-forward projections (\texttt{gate\_proj}, \texttt{up\_proj}, \texttt{down\_proj}). The scaling factor $\alpha$ regulates the contribution of adapter updates, with an effective learning rate proportional to $\alpha / r$. During training for code generation, summarization, and translation tasks, only the adapter parameters are updated, substantially reducing the number of trainable weights compared to full fine-tuning while preserving model performance.

\begin{table}[t]
\centering
\small
\caption{QLoRA adapter configuration parameters.}
\label{tab:qlora_config}
\begin{tabular}{lll}
\toprule
\textbf{Parameter} & \textbf{Description} & \textbf{Value} \\
\midrule
\texttt{r} & Rank of adapter matrices & 8 \\
\texttt{lora\_alpha} & Adapter scaling factor & 16 \\
\texttt{lora\_dropout} & Dropout rate for adapters & 0.1 \\
\texttt{target\_modules} & Layers with adapter injection & 
\makecell[l]{\texttt{q/k/v/o\_proj}, \\ \texttt{gate/up/down\_proj}} \\
\bottomrule
\end{tabular}
\end{table}

\subsection{Training Procedure}
\label{sec:finetune_design}

We conduct experiments under two primary training paradigms: \textit{\MT fine-tuning}, where a single model learns across all three code-related tasks simultaneously, and \textit{\ST fine-tuning}, where separate models are trained independently for each task. For clarity, we adopt the following notation throughout this paper: MT denotes \MT configurations, ST denotes \ST configurations, and FFT refers to full fine-tuning where all model parameters are updated. Consequently, \MTfft represents \MT full fine-tuning, \STqlora represents \ST QLoRA fine-tuning, and \MTqlora represents \MT QLoRA fine-tuning.

\textbf{\ST Training Strategy.}
For \STqlora, we train independent models for each of the experimented tasks. Each \ST model is specialized with the same task-specific system prompt and data processing pipeline as its \MT counterpart, differing only in excluding training instances from other tasks during fine-tuning.  

\textbf{\MT Training Strategy.}
For \MT configurations (\MTfft, \MTqlora), we first create a unified training corpus that includes instances from \CS, \CG, and \CT. We concatenate the processed datasets and apply shuffling with a fixed random seed (seed=$42$) to ensure randomized task exposure during training. This approach results in naturally balanced task sampling without explicit task weighting, as the model encounters examples from all tasks with probability proportional to their representation in the combined dataset--a well-known and widely used approach to tackle code-related tasks \cite{mastropaolo2021studying,mastropaolo2022using,ciniselli2021empirical}.

\textbf{Training Configuration.}
We maintain consistent hyperparameters across all experiments (\tabref{tab:training_config}), following Afrin \etal~\cite{afrin2025resource}. Training employs a per-device batch size of 2 with gradient accumulation over 16 steps, yielding an effective batch size of 32. All models are trained for 5 epochs using the AdamW optimizer with a learning rate of $1\times 10^{-4}$, cosine learning rate scheduling, and 1{,}000 warmup steps. We set the maximum sequence length to 512 tokens with packing enabled, which concatenates short instances and truncates long ones to improve GPU utilization by minimizing padding overhead.

For code translation, which comprises approximately 21K instances compared to approximately 416K instances for other tasks, we adjust the evaluation frequency to 625 steps and reduce warmup steps to 100 to maintain proportionality with the smaller dataset size (approximately one epoch given $\sim$644 steps per epoch). We evaluate on 100 samples randomly selected from the 1{,}000 validation instances, representing 5--10\% proportional sampling across tasks.

QLoRA fine-tuning employs 8-bit AdamW to reduce memory consumption and disables gradient checkpointing, as 4-bit quantization provides sufficient memory savings. Full fine-tuning uses standard PyTorch AdamW with gradient checkpointing enabled to manage higher memory requirements.

\textbf{Early Stopping and Model Selection.}
We apply early stopping based on validation performance: SacreBLEU~\cite{post2018call} for summarization and CodeBLEU~\cite{ren2020codebleu} for generation and translation in \ST models, or averaged combined metrics for \MT models. Training stops after three consecutive evaluations without improvement. Evaluation metrics are detailed in \secref{sec:eval}.
All experiments ran on Ubuntu 22.04 LTS with 8 NVIDIA A40 GPUs (46GB each) and CUDA 12.2.

\begin{table}[h]
\centering
\footnotesize
\caption{Training hyperparameters for all experimental configurations.}
\label{tab:training_config}
\resizebox{0.5\textwidth}{!}{%
\begin{tabular}{lll}
\toprule
\textbf{Parameter} & \textbf{Description} & \textbf{Value} \\
\midrule
\texttt{batch\_size} & Per-device training batch & 2 \\
\texttt{gradient\_accum} & Gradient accumulation steps & 16 \\
\texttt{learning\_rate} & Initial learning rate & $1 \times 10^{-4}$ \\
\texttt{lr\_scheduler} & Learning rate schedule & \texttt{cosine} \\
\texttt{warmup\_steps} & Linear warmup steps & 1,000$^*$ \\
\texttt{num\_epochs} & Training epochs & 5 \\
\texttt{max\_seq\_length} & Maximum sequence length & 512 \\
\texttt{packing} & Sequence packing & \texttt{True} \\
\texttt{patience} & Early stopping patience & 3 \\
\midrule
\multicolumn{3}{l}{\textit{Method-Specific Settings}} \\
\texttt{optimizer} & QLoRA & \texttt{adamw\_bnb\_8bit} \\
\texttt{optimizer} & FFT & \texttt{adamw\_torch} \\
\texttt{gradient\_ckpt} & QLoRA & \texttt{False} \\
\texttt{gradient\_ckpt} & FFT & \texttt{True} \\
\bottomrule
\end{tabular}
}
\\[0.2cm]
\footnotesize{$^*$Translation uses 100 warmup steps and 625 evaluation steps due to smaller dataset size.}
\end{table}

\subsection{Evaluation Procedure}
\label{sec:eval}
Having established our experimental setup, we now evaluate the outputs produced by different model configurations across \rev{the three studied} code-related tasks using established metrics from the literature. We structure this section by grouping tasks based on output modality: \textit{code generation} and \textit{code translation} produce source code and are evaluated for functional correctness and code quality, while \textit{code summarization} produces natural language and is evaluated for linguistic and semantic quality. This organization allows us to apply task-appropriate metrics while maintaining coherence in evaluation methodology.

\subsubsection{Evaluating Correctness and Quality of Automatically Generated Code}  
The evaluation of AI-generated code encompasses two complementary dimensions: (i) \textit{correctness}, which measures whether the code executes as intended and produces expected outputs, and (ii) \textit{quality}, which assesses whether the code adheres to syntactic, semantic, and stylistic standards. Correctness captures functional reliability and behavioral equivalence with respect to ground-truth implementations through execution-based testing. Quality encompasses attributes such as readability, maintainability, and adherence to coding standards--factors that may not affect immediate execution but are crucial for long-term software maintenance and evolution. Together, these dimensions provide a comprehensive assessment of a model's ability to produce code that is both functionally accurate and practically valuable in real-world software engineering contexts.\\

\noindent \textbf{Measuring Correctness via pass@k and CodeBLEU.}  
We evaluate the functional correctness of automatically generated code using two complementary metrics: \textit{pass@k}~\cite{chen2021evaluating} and \textit{CodeBLEU}~\cite{ren2020codebleu}.

The \textit{pass@k} metric estimates the probability that at least one of $k$ generated solutions passes all unit tests for a given problem. We set $k = 1$, meaning only the model's top-ranked output is evaluated. This strict configuration provides a conservative measure of the model's ability to synthesize functionally valid programs that satisfy all task requirements.

\textit{CodeBLEU} extends the traditional BLEU metric~\cite{papineni2002bleu} by incorporating code-aware components that capture the structural and semantic characteristics of programming languages. Specifically, it combines: (1) $n$-gram matching for surface-level similarity, (2) weighted $n$-gram matching based on syntax tokens derived from abstract syntax trees (ASTs), (3) data-flow matching to model variable dependencies, and (4) keyword matching for programming-specific operators and constructs. This metric provides deeper insights into the syntactic and semantic fidelity of generated code relative to reference implementations.

For \textit{code generation}, we use \textit{\passatone} as the primary metric to assess whether the top-ranked output passes all functional tests in the CoderEval benchmark~\cite{yu2024codereval}. We complement this with \textit{CodeBLEU} scores to capture structural and semantic alignment with reference implementations.

For \textit{code translation}, the evaluation focuses on the model's ability to accurately transform code from a source language into a target language while preserving its semantics and structure. Since executable test suites are unavailable for the CodeXGLUE translation dataset, we rely exclusively on \textit{CodeBLEU} to assess functional correctness by measuring how well the translated code preserves the semantic intent and structural characteristics of the reference implementation across languages. \\

\noindent \textbf{Measuring Quality Attributes of Code.}  
To assess the non-functional quality of the generated code, we rely on a set of well-established static analysis tools--following the approach adopted in recent research \cite{afrin2025quantization, siddiq2024quality, liu2024refining, kharma2025security, liu2023your}. These tools quantify structural, stylistic, and maintainability attributes that influence the readability and robustness of model-generated code. For clarity, we report each indicator using its short form in parentheses (\eg \textit{Cyclomatic Complexity (CyC)}). Specifically, we employ \textsc{Lizard}, \textsc{PMD}, \textsc{Pylint}, \textsc{SonarCloud}, and \textsc{Roslyn} to evaluate different aspects of code complexity, style conformance, and maintainability across \java, \python, and C\#.

\textbf{Lizard.} \cite{yin2019lizard} is a language-agnostic code complexity analyzer that processes source code without requiring compilation infrastructure. The tool performs lexical analysis to extract structural metrics from function implementations. We apply Lizard to all three languages (\java, \python, C\#) in our study.

\textbf{PMD.} \cite{pmd2025} is a static analysis tool for \java that evaluates code quality using rule-based detection. We apply PMD to \java outputs generated in code generation and translation tasks.

\textbf{Pylint.} \cite{PylintTeam2025} performs static analysis on \python code to enforce style guidelines and identify potential defects~\cite{siddiq2024quality, liu2024refining}.

\textbf{SonarCloud.} \cite{sonarsource2025sonarcloud} performs cloud-based static analysis to detect code quality issues and technical debt across multiple programming languages. We apply SonarCloud to \java, \python, and C\# outputs to obtain cross-language quality metrics that complement the language-specific insights from other tools. Since our code generation and translation tasks produce function-level outputs, we employ compilation wrappers to enable SonarCloud's full analysis capabilities.

\textbf{Roslyn.}~\cite{MicrosoftRoslyn2025} provides Microsoft's open compiler infrastructure for C\#, exposing compilation APIs that enable deep code inspection. The platform performs syntactic, semantic, and structural analysis by leveraging the same logic used for compilation. This supports tasks including pattern detection, rule validation, and automated correction suggestions. We leverage Roslyn to categorize issues in C\# outputs.

\tabref{tab:static_analysis_tools} summarizes the metrics computed by each tool along with the supported programming languages.

\begin{table*}[h]
\centering
\footnotesize
\caption{Overview of static analysis tools and metrics used for non-functional code quality evaluation.}
\label{tab:static_analysis_tools}
\resizebox{0.9\textwidth}{!}{%
\begin{tabular}{p{2cm}p{3cm}p{8.2cm}ccc}
\toprule
\textbf{Tool} & \textbf{Metric} & \textbf{Description} & \textbf{Java} & \textbf{\python} & \textbf{C\#} \\
\midrule
\multirow{10}{*}{\textbf{Lizard}~\cite{yin2019lizard}}
& \textit{Detection Rate} (DR) & Quantifies the proportion of generated functions that Lizard can successfully parse. This indicates whether outputs are syntactically valid enough for automated analysis. & \ding{51} & \ding{51} & \ding{51} \\[6pt]
& \textit{Cyclomatic Complexity} (CyC) & Quantifies structural complexity through control flow analysis. Given a control flow graph with $E$ edges, $N$ nodes, and $Q$ connected components, the complexity $M$ is computed as $M = E + 2Q - N$. Higher values correspond to more branching logic. & \ding{51} & \ding{51} & \ding{51} \\[6pt]
& \textit{Lines of Code} (LoC) & Counts non-whitespace, non-comment source lines. This size metric correlates with maintenance burden and development effort. & \ding{51}  & \ding{51} & \ding{51} \\[6pt]
& \textit{Token Count} (Tok) & Enumerates lexical elements (keywords, identifiers, operators, literals) in the source. This language-neutral size measure is insensitive to formatting choices. & \ding{51} & \ding{51} & \ding{51} \\
\midrule
\multirow{12}{*}{\textbf{PMD}~\cite{pmd2025}}
& \textit{Best Practices} (BP) & Detects code patterns that deviate from recommended design approaches or create obstacles for future maintenance activities. & \ding{51} & & \\[6pt]
& \textit{Code Style} (CS) & Verifies compliance with formatting standards, particularly regarding bracket placement and identifier naming conventions. & \ding{51} & & \\[6pt]
& \textit{Design} & Reports structural concerns such as excessive nesting of conditional logic that impact long-term maintainability. & \ding{51} & & \\[6pt]
& \textit{Error Prone} (EP) & Identifies problematic constructs that exhibit poor clarity or have high potential for causing failures during execution. & \ding{51} & & \\[6pt]
& \textit{Multi-threading} (MT) & Detects concurrency problems that arise when managing parallel execution contexts, requiring appropriate coordination mechanisms. & \ding{51} & & \\[6pt]
& \textit{Performance} (Perf.) & Locates inefficient implementation choices that negatively affect execution speed and computational resource utilization. & \ding{51} & & \\
\midrule
\multirow{8}{*}{\textbf{Pylint}~\cite{PylintTeam2025}}
& \textit{Convention} (Conv) & Flags departures from PEP 8 style rules, encompassing spacing, indentation, identifier naming, and line formatting. These affect uniformity and scanability. & & \ding{51} & \\[6pt]
& \textit{Refactor} (Ref) & Identifies poor structural choices including duplicated logic, excessive nesting depth, and oversized functions. These suggest refactoring opportunities. & & \ding{51} & \\[6pt]
& \textit{Warning} (Warn) & Surfaces suspicious patterns like unreferenced imports, unused bindings, and name shadowing that may cause unexpected behavior. & & \ding{51} & \\[6pt]
& \textit{Error} (Err) & Reports likely defects, including malformed syntax, reference to undefined names, incorrect function signatures, and type violations that cause runtime crashes. & & \ding{51} & \\
\midrule
\multirow{8.5}{*}{\textbf{SonarCloud}~\cite{sonarsource2025sonarcloud}}
& \textit{Security Hotspots} (Sec) & Identifies security-sensitive code patterns requiring manual review. These include hard-coded credentials, weak cryptography usage, improper input validation, and unprotected resource access. & \ding{51} & \ding{51} &  \\[6pt]
& \textit{Reliability} (Rel) & Counts the total number of issues impacting code robustness detected through static analysis rules. & \ding{51} & \ding{51} &  \\[6pt]
& \textit{Maintainability} (Main) & Evaluates code quality by detecting issues that impact the ease of understanding, modifying, and extending code over time. & \ding{51} & \ding{51} &  \\[6pt]
& \textit{Cognitive Complexity} (CoC) & Measures the difficulty of understanding code's control flow structure by assigning penalties for nested constructs. & \ding{51} & \ding{51} &  \\
\midrule
\multirow{4}{*}{\textbf{Roslyn}~\cite{MicrosoftRoslyn2025}}
& \textit{Syntax Errors} & Reports malformed C\# constructs that block compilation, including incorrect statement structure, incompatible type usage, and disallowed operations. & & & \ding{51} \\[6pt]
& \textit{Maintainability Issues} & Identifies structural and stylistic problems that complicate code comprehension, modification, and evolution over time. & & & \ding{51} \\
\bottomrule
\end{tabular}
}
\end{table*}

Together, these tools provide a multifaceted assessment of the non-functional quality of model-generated code. By combining metrics that capture structural complexity, stylistic consistency, and maintainability, we obtain an objective characterization of code quality across programming languages. To ensure the robustness of our findings, we apply non-parametric statistical analysis to evaluate whether observed differences between configurations are statistically significant: the McNemar test for code generation \passatone, where outputs are evaluated as pass/fail, and the Wilcoxon signed-rank test~\cite{wilcoxon1945individual} for code translation, where outputs are assessed using continuous metrics. When conducting multiple comparisons, we apply the Holm--Bonferroni correction procedure to adjust $p$-values and account for multiple comparisons. This comprehensive evaluation framework allows us to assess how different fine-tuning strategies--such as \MT and \ST QLoRA--impact not only the correctness but also the practical usability and long-term maintainability of the generated software artifacts.

\subsubsection{Evaluating Natural Language in Software Engineering Application.}
\label{sec:eval-nl-quality}
We evaluate the performance of \MT and \ST QLoRA-optimized models in producing meaningful summaries for source code, using a set of established metrics widely employed in prior code summarization research~\cite{zhang2020retrieval, zhang2022survey, leclair2020improved, afrin2025resource, mastropaolo2024evaluating}:

\textbf{BLEU \cite{papineni2002bleu}} measures n-gram precision between generated and reference summaries, producing scores from 0 to 1 where higher values indicate greater similarity.

\textbf{METEOR \cite{meteor}} combines unigram precision and recall through harmonic mean, incorporating stemming and synonym matching for better human judgment correlation.

\textbf{ROUGE-L \cite{lin2004rouge}} assesses sequence-level similarity by measuring the longest common subsequence between summaries.

\textbf{chrF \cite{popovic2015chrf}} evaluates character-level similarity rather than token-level, offering robustness to minor word form variations.

\textbf{BERTScore \cite{zhang2019bertscore}} captures semantic similarity using contextualized BERT embeddings and token-level alignment, reporting F1-scores that go beyond surface-level lexical overlap.

\textbf{SIDE \cite{mastropaolo2024evaluating}} is a code-specific metric that uses contrastive learning to measure alignment between \java methods and their summaries, producing scores from -1 to 1 where values closer to 1 indicate stronger code-comment alignment.

We additionally employ the \textit{LLM-as-a-judge} methodology to evaluate summary quality, following the approach presented by Crupi \etal~\cite{crupi2025effectiveness}, who demonstrated that LLMs can effectively assess code summary quality with performance comparable to human evaluation of code documentation.

We replicate the experimental configuration that achieved the most reliable results in their study, which employs a zero-shot prompting approach to evaluate three dimensions of code summaries: (i) \textit{Content Adequacy}, which quantifies how completely the summary captures implementation details present in the documented code; (ii) \textit{Conciseness}, which assesses whether the summary contains redundant information beyond what is necessary for documentation; and (iii) \textit{Fluency}, which evaluates the grammatical correctness and clarity of the summary.

We instantiate this approach using \textbf{GPT-5 Mini}~\cite{OpenAI2025GPT5} with the evaluation prompt proposed by Crupi \etal~\cite{crupi2025effectiveness}.\footnote{The full prompt is included in our replication package~\cite{replication}.} To mitigate the inherent variability of LLM-based evaluations, we score each generated summary independently five times for all three criteria—\textit{content adequacy}, \textit{conciseness}, and \textit{fluency}—and report the arithmetic mean for each dimension as the final score, improving the stability and reproducibility of the evaluation.

Finally, we assess statistical significance using the Wilcoxon signed-rank test at $\alpha = 0.05$, applying Holm-Bonferroni $p$-value correction for multiple comparisons.
\rev{The reliability of GPT-5 Mini as a judge in our specific Java/Python summarization setting is empirically validated against human annotations in \secref{sec:human-validation}, where we report inter-annotator agreement and human--GPT-5 Mini correlation on a 200-pair stratified sample.}

\section{Results}

To ensure a clear and well-structured discussion, we present the results of our investigation by research question, with each RQ examining performance across the three code-related tasks: (i) code generation, (ii) code summarization, and (iii) code translation. For each task, we report both functional correctness metrics and quality attributes of generated outputs \ie code for generation and translation tasks, natural language for summarization.

To facilitate comparisons across multiple configurations, we adopt the following visual conventions. For functional correctness metrics (\eg \passatone, CodeBLEU, BLEU, \etc), the best-performing configuration within each model size is highlighted using \textbf{bold font}.
\texttt{For quality-related metrics}, we apply color coding to indicate the relative performance of artifacts obtained by specializing the underlying code model in a \MT setting via QLoRA-based optimization, compared to \ST QLoRA optimization. Specifically, \colorbox{green!20}{light green} denotes higher-quality artifacts under \MT QLoRA; \colorbox{blue!20}{light blue} indicates higher-quality artifacts under \ST QLoRA; and \colorbox{yellow!20}{light yellow} signifies no substantial difference between the two approaches. All results are available in our replication package~\cite{replication}.

\revlong{Beyond the metric-based comparisons reported below, Section~\ref{sec:qualitative_analysis} (\textit{Qualitative Analysis of \MTqlora in Code-Related Tasks}) presents one representative case per task --- code generation, translation, and summarization --- contrasting \MTqlora outputs against \STqlora and \MTfft outputs. Each case discusses the qualitative reasons behind the observed differences, including comparisons of correctness vs.\ readability for generated code and content adequacy for generated summaries.}

\revtwo{We clarify two conventions governing how the results are presented. First, the per-task discussions in each research question analyze the Qwen2.5-Coder models at the 0.5B, 1.5B, and 3B scales; the 7B results are reported separately in a dedicated \textit{Extension to 7B scale} paragraph at the end of each research question, so that the scale-dependent behavior is discussed in one place rather than interleaved throughout. Second, the CodeLlama-7B-Instruct replication is confined to RQ$_1$, where it is discussed alongside the 7B results in the same closing block at the end of \secref{sec:rq1}; RQ$_2$ and the ablation studies (\secref{sec:ablations}) use Qwen2.5-Coder only. Tables that include 7B or CodeLlama-7B row blocks are therefore read against the 0.5B--3B analysis in the surrounding text and against these closing paragraphs for the larger scale and the second family.}

\subsection{RQ1: \MT QLoRA \vs \ST QLoRA}
\label{sec:rq1}

This research question investigates whether \MT QLoRA fine-tuning across code generation, summarization, and translation can match or outperform \ST QLoRA training, focusing on whether cross-task knowledge transfer improves performance or whether competing objectives lead to degradation.

\subsubsection{Code Generation} 
\label{sec:rq1_codegen}

\begin{table}[h]
\centering
\scriptsize
\caption{RQ1: Comparison of \STqlora and \MTqlora for \python code generation on CoderEval \revtwo{across the Qwen2.5-Coder and CodeLlama (7B only) model families}.}
\label{tab:rq1_codegen_python}
\resizebox{\textwidth}{!}{%
\begin{tabular}{lc|c|cccc|cccc|cccc}
\toprule
\multirow{2}{*}{\textbf{Model}} & \multirow{2}{*}{\textbf{Method}} & 
\multirow{2}{*}{\textbf{Pass@1}} & 
\multicolumn{4}{c|}{\textbf{Lizard}} & 
\multicolumn{4}{c|}{\textbf{Pylint}} &
\multicolumn{4}{c}{\textbf{SonarCloud}} \\
\cmidrule(lr){4-7} \cmidrule(lr){8-11} \cmidrule(l){12-15}
& & & \textbf{LoC} & \textbf{Tok} & \textbf{DR} & \textbf{CyC} & 
\textbf{Err} & \textbf{Warn} & \textbf{Conv} & \textbf{Ref} &
\textbf{Sec} & \textbf{Rel} & \textbf{Main} & \textbf{CoC} \\
\midrule
\multirow{2}{*}{\textit{QwenCoder-0.5B}} 
& \STqlora & \textbf{15.79\%} & 1,931 & 12,904 & 100\% & 757 & 202 & 1,463 & 648 & 55 & 2 & 3 & 53 & 547 \\
& \MTqlora & \textbf{15.79\%} & 1,952 $\uparrow$ & 13,327 $\uparrow$ & \cellcolor{yellow!20}100\% & \cellcolor{blue!20}812 & \cellcolor{blue!20}251 & \cellcolor{blue!20}1,593 & \cellcolor{green!20}642 & \cellcolor{blue!20}56 & 2 & \cellcolor{blue!20}7 & \cellcolor{blue!20}90 & \cellcolor{blue!20}634 \\
\midrule
\multirow{2}{*}{\textit{QwenCoder-1.5B}} 
& \STqlora & 16.84\% & 2,017 & 13,460 & 100\% & 748 & 308 & 1,664 & 670 & 70 & 1 & 11 & 45 & 641 \\
& \MTqlora & \textbf{18.95\%} & 2,296 $\uparrow$ & 14,531 $\uparrow$ & \cellcolor{yellow!20}100\% & \cellcolor{blue!20}882 & \cellcolor{blue!20}315 & \cellcolor{blue!20}1,767 & \cellcolor{blue!20}677 & \cellcolor{blue!20}76 & 1 & \cellcolor{green!20}5 & \cellcolor{blue!20}72 & \cellcolor{blue!20}771 \\
\midrule
\multirow{2}{*}{\textit{QwenCoder-3B}} 
& \STqlora & 20.53\% & 1,945 & 12,195 & 100\% & 709 & 226 & 1,773 & 675 & 56 & 1 & 0 & 160 & 680 \\
& \MTqlora & \textbf{21.05\%} & 1,759 $\downarrow$ & 11,773 $\downarrow$ & \cellcolor{yellow!20}100\% & \cellcolor{green!20}677 & \cellcolor{blue!20}269 & \cellcolor{green!20}1,521 & \cellcolor{green!20}673 & \cellcolor{yellow!20}56 & 2 & \cellcolor{blue!20}5 & \cellcolor{green!20}106 & \cellcolor{green!20}549 \\
\midrule
\multirow{2}{*}{\rev{\textit{QwenCoder-7B}}}
& \rev{\STqlora} & \rev{\textbf{22.63\%}} & \rev{2,029} & \rev{13,229} & \rev{\cellcolor{yellow!20}100\%} & \rev{757} & \rev{271} & \rev{1,738} & \rev{701} & \rev{65} & \rev{1} & \rev{0} & \rev{39} & \rev{734} \\
& \rev{\MTqlora} & \rev{20.53\%} & \rev{1,957 $\downarrow$}  & \rev{12,424 $\downarrow$} & \rev{100\%} & \rev{\cellcolor{green!20}732} & \rev{\cellcolor{blue!20}287} & \rev{\cellcolor{blue!20}1,768} & \rev{\cellcolor{green!20}683} & \rev{\cellcolor{green!20}51} & \rev{3} & \rev{\cellcolor{blue!20}3} & \rev{\cellcolor{blue!20}59} & \rev{\cellcolor{green!20}727} \\
\midrule[\heavyrulewidth]
\multirow{2}{*}{\revtwo{\textit{CodeLlama-7B}}}
& \revtwo{\STqlora} & \revtwo{\textbf{21.58\%}} & \revtwo{1,996} & \revtwo{14,593} & \revtwo{100\%} & \revtwo{794} & \revtwo{231} & \revtwo{1,204} & \revtwo{598} & \revtwo{50} & \revtwo{0} & \revtwo{3} & \revtwo{62} & \revtwo{500} \\
& \revtwo{\MTqlora} & \revtwo{\textbf{21.58\%}} & \revtwo{2,048 $\uparrow$} & \revtwo{14,585 $\downarrow$} & \revtwo{\cellcolor{blue!20}98.4\%} & \revtwo{\cellcolor{blue!20}841} & \revtwo{\cellcolor{blue!20}248} & \revtwo{\cellcolor{blue!20}1,450} & \revtwo{\cellcolor{blue!20}672} & \revtwo{\cellcolor{green!20}46} & \revtwo{0} & \revtwo{\cellcolor{blue!20}5} & \revtwo{\cellcolor{blue!20}70} & \revtwo{\cellcolor{green!20}497} \\
\bottomrule
\end{tabular}
}
\end{table}

\begin{table}[h]
\centering
\scriptsize
\caption{RQ1: Comparison of \STqlora and \MTqlora for \java code generation on CoderEval \revtwo{across the Qwen2.5-Coder and CodeLlama (7B only) model families}.}
\label{tab:rq1_codegen_java}
\resizebox{\textwidth}{!}{%
\begin{tabular}{lc|c|cccc|cccccc|cccc}
\toprule
\multirow{2}{*}{\textbf{Model}} & \multirow{2}{*}{\textbf{Method}} & 
\multirow{2}{*}{\textbf{Pass@1}} & 
\multicolumn{4}{c|}{\textbf{Lizard}} & 
\multicolumn{6}{c|}{\textbf{PMD}} &
\multicolumn{4}{c}{\textbf{SonarCloud}} \\
\cmidrule(lr){4-7} \cmidrule(lr){8-13} \cmidrule(l){14-17}
& & & \textbf{LoC} & \textbf{Tok} & \textbf{DR} & \textbf{CyC} & 
\textbf{Best Prac.} & \textbf{CS} & \textbf{Design} & \textbf{EP} & \textbf{MT} & \textbf{Perf.} &
\textbf{Sec} & \textbf{Rel} & \textbf{Main} & \textbf{CoC} \\
\midrule
\multirow{2}{*}{\textit{QwenCoder-0.5B}} 
& \STqlora & \textbf{21.74\%} & 1,523 & 10,820 & 96.70\% & 589 & 74 & 1,177 & 128 & 33 & 4 & 19 & 1 & 3 & 525 & 497 \\
& \MTqlora & 20.11\% & 1,933 $\uparrow$ & 12,750 $\uparrow$ & \cellcolor{yellow!20}96.70\% & \cellcolor{blue!20}697 & \cellcolor{blue!20}76 & \cellcolor{green!20}1,123 & \cellcolor{blue!20}139 & \cellcolor{yellow!20}33 & \cellcolor{green!20}3 & \cellcolor{green!20}10 & 0 & \cellcolor{green!20}0 & \cellcolor{blue!20}531 & \cellcolor{green!20}494 \\
\midrule
\multirow{2}{*}{\textit{QwenCoder-1.5B}} 
& \STqlora & \textbf{25.54\%} & 1,717 & 12,389 & 96.70\% & 697 & 82 & 1,227 & 125 & 32 & 4 & 9 & 0 & 3 & 541 & 526 \\
& \MTqlora & 19.02\% & 1,490 $\downarrow$ & 10,313 $\downarrow$ & \cellcolor{yellow!20}96.20\% & \cellcolor{green!20}589 & \cellcolor{green!20}75 & \cellcolor{green!20}1,183 & \cellcolor{blue!20}127 & \cellcolor{green!20}31 & \cellcolor{blue!20}5 & \cellcolor{blue!20}11 & 0 & \cellcolor{green!20}2 & \cellcolor{green!20}530 & \cellcolor{green!20}442 \\
\midrule
\multirow{2}{*}{\textit{QwenCoder-3B}} 
& \STqlora & 29.89\% & 1,556 & 10,866 & 98.90\% & 589 & 79 & 1,185 & 128 & 40 & 3 & 11 & 0 & 3 & 545 & 491 \\
& \MTqlora & \textbf{32.07\%} & 1,580 $\uparrow$ & 10,920 $\uparrow$ & \cellcolor{yellow!20}98.40\% & \cellcolor{green!20}558 & \cellcolor{blue!20}86 & \cellcolor{green!20}1,184 & \cellcolor{yellow!20}128 & \cellcolor{green!20}32 & \cellcolor{blue!20}4 & \cellcolor{green!20}7 & 0 & \cellcolor{green!20}2 & \cellcolor{green!20}541 & \cellcolor{green!20}469 \\
\midrule
\multirow{2}{*}{\rev{\textit{QwenCoder-7B}}}
& \rev{\STqlora} & \rev{30.43\%} & \rev{1,714} & \rev{11,998} & \rev{98.9\%} & \rev{635} & \rev{88} & \rev{1,230} & \rev{156} & \rev{34} & \rev{3} & \rev{13} & \rev{0} & \rev{6} & \rev{545} & \rev{491} \\
& \rev{\MTqlora} & \rev{\textbf{33.70\%}} & \rev{1,799 $\uparrow$} & \rev{12,375 $\uparrow$} & \rev{\cellcolor{yellow!20}98.9\%} & \rev{\cellcolor{blue!20}663} & \rev{\cellcolor{green!20}87} & \rev{\cellcolor{blue!20}1,238} & \rev{\cellcolor{green!20}137} & \rev{\cellcolor{green!20}33} & \rev{\cellcolor{blue!20}5} & \rev{\cellcolor{green!20}11} & \rev{0} & \rev{\cellcolor{green!20}0} & \rev{\cellcolor{green!20}532} & \rev{\cellcolor{blue!20}576} \\ 
\midrule[\heavyrulewidth]
\multirow{2}{*}{\revtwo{\textit{CodeLlama-7B}}}
& \revtwo{\STqlora} & \revtwo{\textbf{33.15\%}} & \revtwo{1,472} & \revtwo{10,205} & \revtwo{95.7\%} & \revtwo{596} & \revtwo{78} & \revtwo{1,153} & \revtwo{117} & \revtwo{27} & \revtwo{1} & \revtwo{8} & \revtwo{0} & \revtwo{1} & \revtwo{506} & \revtwo{415} \\
& \revtwo{\MTqlora} & \revtwo{25.00\%} & \revtwo{1,290 $\downarrow$} & \revtwo{9,804 $\downarrow$} & \revtwo{\cellcolor{green!20}97.8\%} & \revtwo{\cellcolor{green!20}509} & \revtwo{\cellcolor{green!20}26} & \revtwo{\cellcolor{green!20}891} & \revtwo{\cellcolor{blue!20}132} & \revtwo{\cellcolor{blue!20}40} & \revtwo{\cellcolor{yellow!20}1} & \revtwo{\cellcolor{yellow!20}8} & \revtwo{0} & \revtwo{\cellcolor{green!20}0} & \revtwo{\cellcolor{green!20}420} & \revtwo{\cellcolor{green!20}314} \\
\bottomrule
\end{tabular}
}
\end{table}

Before delving into specific patterns, we examine the overall behavior of \MT QLoRA for code generation. From a general standpoint, we observe that \MT QLoRA demonstrates competitive \passatone scores, with performance advantages becoming increasingly pronounced as model size increases. 

For \textbf{\python code generation} (\tabref{tab:rq1_codegen_python}), considering \textbf{functional correctness as measured by \passatone}, \MT QLoRA achieves better performance in the 1.5B (18.95\% \vs\ 16.84\%, $\uparrow$12.5\%) and 3B (21.05\% \vs\ 20.53\%, $\uparrow$2.5\%) configurations, while maintaining identical performance in the 0.5B variant (15.79\%).  

For \textbf{\java code generation} (\tabref{tab:rq1_codegen_java}), again focusing on \textbf{\passatone functional correctness}, the 3B \MT model attains the highest score at 32.07\%, outperforming the \ST model (29.89\%) by $\uparrow$7.3\%. However, smaller \MT models underperform in Java: the 1.5B variant achieves 19.02\% compared to Single-task's 25.54\%, corresponding to a $\downarrow$25.5\% relative decline. 


While functional correctness is a fundamental requirement, it captures only one dimension of the practical value of automated techniques that generate code-related artifacts. Additional factors--including readability, maintainability, \textit{Conciseness}, and adherence to coding standards--are equally important in determining real-world usability. Indeed, code that passes test cases may still suffer from excessive complexity, poor maintainability, or violations of established quality guidelines, all of which can negatively affect long-term software evolution and maintenance \cite{mastropaolo2023robustness, afrin2025quantization, siddiq2024quality, khoury2023secure}.

To provide a more comprehensive evaluation, we complement functional correctness with aggregated metrics derived from static analysis tools. To give readers a clear and interpretable overview of code quality, we first aggregate results at the model level, which allows us to compare overall trends between \ST and \MT QLoRA independently of specific configurations. This aggregation enables a high-level assessment of differences in structural complexity and coding standard compliance. We then examine whether variations in model size and parameter count affect these quality metrics.

We begin by examining results for \python code generation. Aggregating across all model sizes (0.5B, 1.5B, and 3B), \ST QLoRA produces a total of 2{,}214 Cyclomatic Complexity issues, 7{,}810 Pylint warnings, and 2{,}144 SonarCloud quality issues. In comparison, \MT QLoRA yields 2{,}371 Cyclomatic Complexity issues ($\uparrow$7.1\%), 7{,}896 Pylint warnings ($\uparrow$1.1\%), and 2{,}244 SonarCloud issues ($\uparrow$4.7\%). 

For \java code generation, again aggregating across the 0.5B--3B scales, \ST QLoRA produces 1{,}875 Cyclomatic Complexity issues, 4{,}360 PMD warnings, and 3{,}135 SonarCloud issues. In contrast, \MT QLoRA generates 1{,}844 Cyclomatic Complexity issues ($\downarrow$1.7\%), 4{,}257 PMD warnings ($\downarrow$2.4\%), and 3{,}011 SonarCloud issues ($\downarrow$4.0\%). Overall, \python exhibits slightly higher aggregate complexity and quality issues under \MT fine-tuning, whereas \java shows modest improvements in both structural complexity and coding standard compliance.

This observed behavior of \MT models across aggregate metrics prompts a deeper examination of specific quality dimensions. Turning to Cyclomatic Complexity patterns, \python \MT shows $\uparrow$7.1\% higher CyC than \ST (2,371 \vs\ 2,214), indicating more complex control flow structures. However, this increase is not uniform across model sizes: the 0.5B model shows 812 \vs\ 757 CyC ($\uparrow$7.3\%), the 1.5B shows 882 \vs\ 748 ($\uparrow$17.9\%), but the 3B model reverses the trend with 677 \vs\ 709 ($\downarrow$4.5\%). For Java, \MT consistently produces simpler code: 1,844 total CyC versus Single-task's 1,875 ($\downarrow$1.7\%). Breaking down by model size reveals distinct patterns: the 0.5B model shows 697 \vs\ 589 ($\uparrow$18.3\%), indicating higher complexity in the smaller \MT variant, while the 1.5B shows 589 \vs\ 697 ($\downarrow$15.5\%) and the 3B shows 558 \vs\ 589 ($\downarrow$5.3\%), both demonstrating lower complexity. This suggests that joint training affects structural complexity differently across languages and model sizes, with larger models producing simpler implementations. The reversal in the 3B \python model is particularly noteworthy, as it suggests that sufficient model capacity enables \MT training to produce code that is not only more functionally correct, but also structurally simpler, aligning with recent findings that emphasize the role of model scale in enabling effective cross-task knowledge transfer without sacrificing code quality \cite{wei2022emergent, liu2024mftcoder}.

For Pylint metrics in \python, the 3B \MT model produces fewer issues than the \ST model (2,519 \vs\ 2,730, $\downarrow$7.7\%), aligning with its superior functional correctness and reduced complexity. While Errors increase slightly (269 \vs\ 226, $\uparrow$19.0\%), Warnings decrease substantially (1,521 \vs\ 1,773, $\downarrow$14.2\%). SonarCloud maintainability reinforces this pattern: the 3B \MT model achieves 106 issues compared to the \ST model's 160 ($\downarrow$33.8\%), while smaller models show higher issue counts (0.5B: 90 \vs\ 53, $\uparrow$69.8\%; 1.5B: 72 \vs\ 45, $\uparrow$60.0\%).

For Java, \textit{PMD} analysis reveals that the \MT QLoRA-optimized Qwen-Coder models produce code with better style conformance and design patterns across most model sizes. In particular, the 0.5B model generates 1,384 total PMD issues compared to the \ST model's 1,435 ($\downarrow$3.6\%), with improvements in Code Style (1,123 \vs\ 1,177, $\downarrow$4.6\%), Multi-threading (3 \vs\ 4, $\downarrow$25.0\%), and Performance (10 \vs\ 19, $\downarrow$47.4\%), a slight increase in Best Practices (76 \vs\ 74, $\uparrow$2.7\%), and Design and Error Prone remaining comparable. The 1.5B model shows 1,432 issues compared to 1,479 ($\downarrow$3.2\%), with reductions in Best Practices (75 \vs\ 82, $\downarrow$8.5\%), Code Style (1,183 \vs\ 1,227, $\downarrow$3.6\%), and Error Prone (31 \vs\ 32, $\downarrow$3.1\%), alongside slight increases in Design (127 \vs\ 125, $\uparrow$1.6\%), Performance (11 \vs\ 9, $\uparrow$22.2\%), and Multi-threading (5 \vs\ 4, $\uparrow$25.0\%). The 3B model demonstrates 1,441 issues compared to 1,446 ($\downarrow$0.3\%), with improvements in Code Style (1,184 \vs\ 1,185, $\downarrow$0.1\%), Error Prone (32 \vs\ 40, $\downarrow$20.0\%), and Performance (7 \vs\ 11, $\downarrow$36.4\%), while Multi-threading (4 \vs\ 3, $\uparrow$33.3\%) and Best Practices (86 \vs\ 79, $\uparrow$8.9\%) show slight increases.

SonarCloud maintainability results for \java show broadly similar patterns across both training variants. The \ST QLoRA-optimized model generates code affected by 525, 541, and 545 issues for the 0.5B, 1.5B, and 3B models, respectively, while the \MT variant generates 531 ($\uparrow$1.1\%), 530 ($\downarrow$2.0\%), and 541 ($\downarrow$0.7\%) issues. This pattern indicates that \MT training benefits increase with model capacity. Smaller models struggle to balance multiple objectives, producing code that violates more conventions and exhibits more code smells. In contrast, larger models successfully leverage knowledge gained from diverse tasks to generate cleaner, more maintainable code with better adherence to style guidelines and design patterns.


Examining code size metrics from \textit{Lizard} reveals that \MT models generate longer code in smaller configurations but more concise implementations in larger models. In the context of \python code generation, \MT QLoRA-based optimization with models featuring fewer than one billion parameters (\ie, Qwen-Coder 0.5B) produces 1,952 LoC compared to the \ST variant's 1,931 ($\uparrow$1.1\%), while the \MT 1.5B model generates 2,296 LoC compared to 2,017 for its \ST counterpart ($\uparrow$13.8\%). However, \MT 3B reverses this trend with 1,759 LoC versus Single-task's 1,945 ($\downarrow$9.6\%), while simultaneously achieving lower complexity (677 \vs\ 709 CyC, $\downarrow$4.5\%), slightly higher \passatone (21.05\% \vs\ 20.53\%, $\uparrow$2.5\%), fewer maintainability issues (106 \vs\ 160, $\downarrow$33.8\%), and reduced cognitive complexity (549 \vs\ 680, $\downarrow$19.3\%). 

For Java, model size plays a central role in how \MT training affects code length. The \MT 0.5B configuration generates longer implementations than its \ST counterpart (1{,}933 \vs\ 1{,}523 LoC, $\uparrow$26.9\%), 
whereas the 1.5B \MT model produces more compact code (1{,}490 \vs\ 1{,}717, $\downarrow$13.2\%) and the 3B models are essentially equivalent in length (1{,}580 \vs\ 1{,}556 LoC, $\uparrow$1.5\%).
A similar capacity-dependent pattern emerges when examining token counts.

For \python, \MT models generate slightly more tokens at smaller model sizes (13{,}327 \vs\ 12{,}904 for 0.5B, $\uparrow$3.3\%; 14{,}531 \vs\ 13{,}460 for 1.5B, $\uparrow$8.0\%), but fewer tokens in the 3B configuration (11{,}773 \vs\ 12{,}195, $\downarrow$3.5\%). \java exhibits an analogous trend: the \MT 0.5B model produces substantially more tokens (12{,}750 \vs\ 10{,}820, $\uparrow$17.8\%), the 1.5B configuration generates fewer tokens (10{,}313 \vs\ 12{,}389, $\downarrow$16.8\%), and the 3B models show nearly identical token counts (10{,}920 \vs\ 10{,}866, $\uparrow$0.5\%).

To determine whether the empirically observed differences also translate into statistically significant effects, we compare \ST and \MT QLoRA configurations using McNemar tests on \passatone outcomes and Wilcoxon signed-rank tests on quality metrics, with Holm--Bonferroni correction applied ($\alpha = 0.05$). To complement hypothesis testing, we report effect size measures to quantify the magnitude of observed differences. Specifically, odds ratios (ORs) from the McNemar tests capture the relative likelihood of \ST QLoRA fine-tuning outperforming \MT QLoRA in generating more functionally correct code, while Cliff's Delta from the Wilcoxon tests measures the degree of separation between quality metric distributions.



Across both \python and \java, no statistically significant differences are observed at any of the 0.5B--3B scales, indicating that empirical variations across configurations fall within the range expected from random fluctuation. Consequently, both \MT and \ST QLoRA-based training equip the underlying models with comparable knowledge, resulting in generated code that does not differ substantially from either a quantitative or qualitative perspective. Although larger models may exhibit practical advantages along certain performance and code quality dimensions, these differences are not statistically distinguishable under our testing framework. For transparency and reproducibility, we provide the complete set of test statistics, effect sizes, and adjusted $p$-values in the replication package \cite{replication}. \\

\begin{summarybox}[Key Findings]
\noindent \MT QLoRA is competitive with \ST QLoRA for code generation, with benefits that generally increase with model capacity. For \python, \MT matches 0.5B and improves \passatone at 1.5B and 3B. For \java, \MT peaks at 3B but underperforms at 1.5B, indicating scale-sensitive trade-offs. Static analysis suggests smaller \MT models can incur higher complexity or maintainability issues, while larger \MT models better balance correctness and quality. 
Statistical tests (McNemar for \passatone; Wilcoxon with Holm--Bonferroni for quality) find no significant differences across scales, implying practical but not statistically distinguishable gaps. \revtwo{These findings cover the 0.5B--3B scales; at 7B the two languages diverge, with \MTqlora ahead on \java and \STqlora ahead on \python, and the \java advantage does not reproduce for CodeLlama (see the end of this section).}
\end{summarybox}

\subsubsection{Code Translation} 
\label{sec:rq1_codetrans}

We evaluate code translation performance using CodeBLEU across both Java$\rightarrow$C\# and C\#$\rightarrow$Java directions. Tables~\ref{tab:rq1_codetrans_java2csharp} and~\ref{tab:rq1_codetrans_csharp2java} compare \MT QLoRA against \ST QLoRA across the 0.5B--3B scales, reporting both functional correctness and non-functional quality metrics.


We begin by evaluating the ability of \ST instruction-tuned code models to perform Java$\rightarrow$C\# and C\#$\rightarrow$Java translation, and then examine whether the empirical benefits previously observed for \MT training in code generation also carry over to translation tasks. In doing so, we find that, unlike code generation, translation performance varies substantially across both translation directions and model sizes, indicating a more complex interaction between task structure and model capacity.

For Java$\rightarrow$C\#, this variability is immediately apparent. The 0.5B \MT model attains a CodeBLEU score of 68.92\%, compared to 75.77\% for the \ST baseline, corresponding to a relative decrease of $\downarrow$9.0\%. A similar relative degradation is observed at the largest scale, where the 3B \MT configuration achieves 62.88\% versus 66.83\% for \ST ($\downarrow$5.9\%). In contrast, the intermediate 1.5B model exhibits a modest relative improvement under \MT QLoRA, reaching 73.08\% compared to 72.27\% for \ST ($\uparrow$1.1\%), suggesting a potential capacity sweet spot for this translation direction.

This capacity-dependent pattern persists for C\#$\rightarrow$Java translation. The 0.5B \MT model again underperforms relative to the \ST baseline (68.37\% \vs\ 75.36\%), corresponding to a relative decrease of $\downarrow$9.3\%. At the intermediate scale, however, \MT training yields a relative improvement of $\uparrow$2.0\% (70.72\% \vs\ 69.35\%), while at 3B the two approaches converge, with \MT achieving 71.22\% compared to 72.81\% for \ST ($\downarrow$2.2\%).

To examine whether these functional trends are accompanied by systematic differences in code quality--as we did for code generation (\secref{sec:rq1_codegen})--we aggregate metrics from static analysis tools. For Java$\rightarrow$C\#, where Lizard and Roslyn provide comprehensive coverage, \ST QLoRA produces 4,575 total Cyclomatic Complexity (CyC) issues and 4,919 Roslyn-reported issues across the 0.5B--3B scales. 
\MT QLoRA yields a marginal reduction in CyC issues (4,533; $\downarrow$0.9\%) but a higher number of Roslyn issues overall (5,183; $\uparrow$5.4\%).
In contrast, for C\#$\rightarrow$Java translation--again aggregating across the 0.5B--3B scales and evaluated using Lizard, PMD, and SonarCloud--\MT QLoRA-based fine-tuning consistently reduces aggregate quality issues, including fewer CyC issues ($\downarrow$1.0\%), fewer PMD violations ($\downarrow$0.6\%), and fewer SonarCloud-reported issues ($\downarrow$2.2\%). 


The observed behavior across aggregate metrics motivates a closer examination of capacity-dependent effects. Disaggregating Cyclomatic Complexity by model size reveals indeed some latent patterns that could go unnoticed. For Java$\rightarrow$C\#, \MT QLoRA demonstrates progressive improvements with increased capacity: the 0.5B model shows 1,515 CyC compared to 1,528 under \ST ($\downarrow$0.9\%), the 1.5B configuration shows 1,563 versus 1,561 ($\uparrow$0.1\%), and the 3B model achieves the largest reduction with 1,455 versus 1,486 ($\downarrow$2.1\%). For C\#$\rightarrow$Java, the pattern differs: the 0.5B model shows 1,594 versus 1,586 CyC ($\uparrow$0.5\%), the 1.5B model shows 1,566 versus 1,647 ($\downarrow$4.9\%), and the 3B configuration shows 1,612 versus 1,588 ($\uparrow$1.5\%). The particularly strong reduction observed for the 1.5B configuration aligns with its competitive CodeBLEU performance, suggesting that this capacity achieves an effective balance between similarity to reference implementations and structural simplicity.

A closer inspection of language-specific diagnostics further highlights non-monotonic quality trends. For Roslyn metrics in Java$\rightarrow$C\# translation, \MT QLoRA exhibits substantial variability across model sizes. The 0.5B model produces 1,091 total Roslyn issues compared to 850 under Single-task, representing a $\uparrow$28.4\% increase despite lower cyclomatic complexity. While syntax errors decrease from 283 to 256 ($\downarrow$9.5\%), maintainability issues increase substantially from 567 to 835 ($\uparrow$47.3\%). In contrast, the 1.5B model shows pronounced improvements, producing 905 total issues versus 2,724 under \ST ($\downarrow$66.8\%), with reductions in both syntax errors (313 \vs\ 356, $\downarrow$12.1\%) and maintainability issues (592 \vs\ 2,368, $\downarrow$75.0\%). The 3B model reverses this trend, generating 3,187 total issues versus 1,345 ($\uparrow$137.0\%), driven by increases in syntax errors (126 \vs\ 81, $\uparrow$55.6\%) and maintainability issues (3,061 \vs\ 1,264, $\uparrow$142.2\%). These results indicate that the 1.5B configuration achieves the most balanced outcome for Java$\rightarrow$C\# translation, while smaller and larger models exhibit different quality trade-offs.

PMD analysis for C\#$\rightarrow$Java translation shows mixed results across different model sizes. For the smallest 0.5B model, \MT QLoRA generates code with 4,793 total PMD issues compared to Single-task's 4,859 ($\downarrow$1.4\%), showing improvements in all categories: Best Practices (400 \vs\ 416, $\downarrow$3.8\%), Code Style (3,953 \vs\ 3,973, $\downarrow$0.5\%), Design (234 \vs\ 247, $\downarrow$5.3\%), Error Prone (54 \vs\ 61, $\downarrow$11.5\%), Multi-threading (12 \vs\ 15, $\downarrow$20.0\%), and Performance (140 \vs\ 147, $\downarrow$4.8\%). However, the 1.5B model shows the opposite trend, with \MT producing more issues (4,798 \vs\ 4,613, $\uparrow$4.0\%) across most categories, despite achieving higher CodeBLEU scores. This suggests a trade-off where the model prioritizes functional accuracy over code style. The largest 3B model demonstrates significant improvements with \MT training, reducing total issues from 1,358 to 1,171 ($\downarrow$13.8\%), with particularly large gains in Code Style (1,072 \vs\ 1,214, $\downarrow$11.7\%), Design (45 \vs\ 83, $\downarrow$45.8\%), Multi-threading (5 \vs\ 9, $\downarrow$44.4\%), and Performance (1 \vs\ 7, $\downarrow$85.7\%). Best Practices remain stable, while Error Prone violations increase slightly (11 \vs\ 8, $\uparrow$37.5\%).

SonarCloud metrics confirm these patterns for C\#$\rightarrow$Java translation. The 3B \MT model produces code affected by a grand total of 435 total quality issues compared to Single-task's 563 ($\downarrow$22.7\%), with substantial reductions in maintainability (278 \vs\ 366, $\downarrow$24.0\%) and cognitive complexity (157 \vs\ 192, $\downarrow$18.2\%). Notably, reliability issues are completely eliminated (0 \vs\ 5, $\downarrow$100\%). For smaller models, the results are less consistent: the 0.5B model shows modest improvement with 1,723 issues versus 1,760 ($\downarrow$2.1\%), while the 1.5B model actually produces more issues (1,785 \vs\ 1,708, $\uparrow$4.5\%), aligning with the PMD findings.

Finally, we examine code size metrics using Lizard. For Java$\rightarrow$C\#, \MT QLoRA consistently produces more compact implementations: 6,218 LoC versus 6,368 at 0.5B ($\downarrow$2.4\%), 6,369 versus 6,519 at 1.5B ($\downarrow$2.3\%), and 5,890 versus 6,189 at 3B ($\downarrow$4.8\%). Token counts follow the same trend, decreasing from 40,746 to 39,509 at 0.5B ($\downarrow$3.0\%), from 42,299 to 40,516 at 1.5B ($\downarrow$4.2\%), and from 39,845 to 37,716 at 3B ($\downarrow$5.3\%). For C\#$\rightarrow$Java, code size trends are mixed: 0.5B shows a slight increase in LoC (5,807 \vs\ 5,777, $\uparrow$0.5\%), 1.5B shows a reduction (5,745 \vs\ 5,860, $\downarrow$2.0\%), and 3B shows a small increase (5,742 \vs\ 5,687, $\uparrow$1.0\%). Token counts similarly increase at 0.5B ($\uparrow$0.8\%) but decrease at 1.5B ($\downarrow$3.3\%) and remain nearly unchanged at 3B ($\uparrow$0.7\%). Taken together, these results suggest that increased capacity allows \MT models to better leverage cross-task knowledge to produce concise and higher-quality translations, particularly for Java$\rightarrow$C\# and large-scale C\#$\rightarrow$Java settings.

Ultimately, to reliably determine whether the empirical differences observed in code translation reflect statistically meaningful effects rather than random variation, we conduct Wilcoxon signed-rank tests with Holm--Bonferroni correction ($\alpha=0.05$), comparing \ST QLoRA against \MT QLoRA across the 0.5B--3B scales and both translation directions. Complete statistical results are available in our replication package \cite{replication}. Java$\rightarrow$C\# exhibits statistically significant CodeBLEU differences at all three of these scales, whereas for C\#$\rightarrow$Java significance is limited to the 0.5B configuration. Quality metrics occasionally yield significant results, but effect sizes remain small, indicating that while some differences are statistically detectable, their practical impact is limited. \\

\noindent\begin{minipage}{\linewidth}
\begin{summarybox}[Key Findings]
\noindent Code translation under \MT QLoRA is direction- and scale-dependent relative to \ST QLoRA, with no consistent dominance. Across Java$\rightarrow$C\# and C\#$\rightarrow$Java, performance varies by model size, often favoring the intermediate configuration. Static analysis mirrors this non-monotonicity, where improvements in complexity or conciseness can co-occur with regressions in target-language diagnostics (\eg Roslyn for C\#). Wilcoxon tests with Holm--Bonferroni correction detect significant CodeBLEU differences for Java$\rightarrow$C\# across scales, while effects for C\#$\rightarrow$Java are more limited and typically small in magnitude.

\end{summarybox}
\end{minipage}
\begin{table}[h]
\centering
\scriptsize
\caption{RQ1: Comparison of \STqlora and \MTqlora for Java$\rightarrow$C\# code translation on CodeXGLUE \revtwo{across the Qwen2.5-Coder and CodeLlama (7B only) model families}.}
\label{tab:rq1_codetrans_java2csharp}
\resizebox{0.80\textwidth}{!}{%
\begin{tabular}{lc|c|cccc|cc}
\toprule
\multirow{2}{*}{\textbf{Model}} & \multirow{2}{*}{\textbf{Method}} & 
\multirow{2}{*}{\textbf{CodeBLEU}} & 
\multicolumn{4}{c|}{\textbf{Lizard}} & 
\multicolumn{2}{c}{\textbf{Roslyn}} \\
\cmidrule(lr){4-7} \cmidrule(l){8-9}
& & & \textbf{LoC} & \textbf{Tok} & \textbf{DR} & \textbf{CyC} & 
\textbf{Syntax Errors} & \textbf{Maintainability} \\
\midrule
\multirow{2}{*}{\textit{QwenCoder-0.5B}} 
& \STqlora & \textbf{75.77\%} & 6,368 & 40,746 & 99.1\% & 1,528 & 283 & 567 \\
& \MTqlora & 68.92\% & 6,218 $\downarrow$ & 39,509 $\downarrow$ & \cellcolor{blue!20}98.7\% & \cellcolor{green!20}1,515 & \cellcolor{green!20}256 & \cellcolor{blue!20}835 \\
\midrule
\multirow{2}{*}{\textit{QwenCoder-1.5B}} 
& \STqlora & 72.27\% & 6,519 & 42,299 & 95.5\% & 1,561 & 356 & 2,368 \\
& \MTqlora & \textbf{73.08\%} & 6,369 $\downarrow$ & 40,516 $\downarrow$ & \cellcolor{green!20}98.5\% & \cellcolor{blue!20}1,563 & \cellcolor{green!20}313 & \cellcolor{green!20}592 \\
\midrule
\multirow{2}{*}{\textit{QwenCoder-3B}} 
& \STqlora & \textbf{66.83\%} & 6,189 & 39,845 & 97.7\% & 1,486 & 81 & 1,264 \\
& \MTqlora & 62.88\% & 5,890 $\downarrow$ & 37,716 $\downarrow$ & \cellcolor{blue!20}93.6\% & \cellcolor{green!20}1,455 & \cellcolor{blue!20}126 & \cellcolor{blue!20}3,061 \\
\midrule
\multirow{2}{*}{\rev{\textit{QwenCoder-7B}}}
& \rev{\STqlora} & \rev{\textbf{77.70\%}} & \rev{6,805} & \rev{43,290} & \rev{99.7\%} & \rev{1,662} & \rev{292} & \rev{440} \\
& \rev{\MTqlora} & \rev{55.37\%} & \rev{5,641 $\downarrow$} & \rev{36,386 $\downarrow$} & \rev{\cellcolor{blue!20}88.1\%} & \rev{\cellcolor{green!20}1,340} & \rev{\cellcolor{green!20}123} & \rev{\cellcolor{blue!20}3,056} \\
\midrule[\heavyrulewidth]
\multirow{2}{*}{\revtwo{\textit{CodeLlama-7B}}}
& \revtwo{\STqlora} & \revtwo{\textbf{82.22\%}} & \revtwo{6,502} & \revtwo{41,895} & \revtwo{99.8\%} & \revtwo{1,604} & \revtwo{320} & \revtwo{726} \\
& \revtwo{\MTqlora} & \revtwo{79.45\%} & \revtwo{6,338 $\downarrow$} & \revtwo{41,032 $\downarrow$} & \revtwo{\cellcolor{yellow!20}99.7\%} & \revtwo{\cellcolor{green!20}1,569} & \revtwo{\cellcolor{green!20}298} & \revtwo{\cellcolor{green!20}645} \\
\bottomrule
\end{tabular}
}
\end{table}

\begin{table}[h]
\centering
\scriptsize
\caption{RQ1: Comparison of \STqlora and \MTqlora for C\#$\rightarrow$Java code translation on CodeXGLUE \revtwo{across the Qwen2.5-Coder and CodeLlama (7B only) model families}.}
\label{tab:rq1_codetrans_csharp2java}
\resizebox{\textwidth}{!}{%
\begin{tabular}{lc|c|cccc|cccccc|cccc}
\toprule
\multirow{2}{*}{\textbf{Model}} & \multirow{2}{*}{\textbf{Method}} & 
\multirow{2}{*}{\textbf{CodeBLEU}} & 
\multicolumn{4}{c|}{\textbf{Lizard}} & 
\multicolumn{6}{c|}{\textbf{PMD}} &
\multicolumn{4}{c}{\textbf{SonarCloud}} \\
\cmidrule(lr){4-7} \cmidrule(lr){8-13} \cmidrule(l){14-17}
& & & \textbf{LoC} & \textbf{Tok} & \textbf{DR} & \textbf{CyC} & 
\textbf{Best Prac.} & \textbf{CS} & \textbf{Design} & \textbf{EP} & \textbf{MT} & \textbf{Perf.} &
\textbf{Sec} & \textbf{Rel} & \textbf{Main} & \textbf{CoC} \\
\midrule
\multirow{2}{*}{\textit{QwenCoder-0.5B}} 
& \STqlora & \textbf{75.36\%} & 5,777 & 34,898 & 99.6\% & 1,586 & 416 & 3,973 & 247 & 61 & 15 & 147 & 0 & 22 & 1,071 & 667 \\
& \MTqlora & 68.37\% & 5,807 $\uparrow$ & 35,170 $\uparrow$ & \cellcolor{yellow!20}99.6\% & \cellcolor{blue!20}1,594 & \cellcolor{green!20}400 & \cellcolor{green!20}3,953 & \cellcolor{green!20}234 & \cellcolor{green!20}54 & \cellcolor{green!20}12 & \cellcolor{green!20}140 & 0 & \cellcolor{green!20}13 & \cellcolor{green!20}1,056 & \cellcolor{green!20}654 \\
\midrule
\multirow{2}{*}{\textit{QwenCoder-1.5B}} 
& \STqlora & 69.35\% & 5,860 & 36,265 & 97.8\% & 1,647 & 399 & 3,786 & 196 & 58 & 16 & 158 & 0 & 16 & 1,013 & 679 \\
& \MTqlora & \textbf{70.72\%} & 5,745 $\downarrow$ & 35,074 $\downarrow$ & \cellcolor{green!20}99.6\% & \cellcolor{green!20}1,566 & \cellcolor{blue!20}400 & \cellcolor{blue!20}3,939 & \cellcolor{blue!20}213 & \cellcolor{blue!20}61 & \cellcolor{blue!20}22 & \cellcolor{blue!20}163 & 0 & \cellcolor{blue!20}19 & \cellcolor{blue!20}1,088 & \cellcolor{green!20}678 \\
\midrule
\multirow{2}{*}{\textit{QwenCoder-3B}} 
& \STqlora & \textbf{72.81\%} & 5,687 & 34,622 & 99.4\% & 1,588 & 37 & 1,214 & 83 & 8 & 9 & 7 & 0 & 5 & 366 & 192 \\
& \MTqlora & 71.22\% & 5,742 $\uparrow$ & 34,862 $\uparrow$ & \cellcolor{yellow!20}99.5\% & \cellcolor{blue!20}1,612 & \cellcolor{yellow!20}37 & \cellcolor{green!20}1,072 & \cellcolor{green!20}45 & \cellcolor{blue!20}11 & \cellcolor{green!20}5 & \cellcolor{green!20}1 & 0 & \cellcolor{green!20}0 & \cellcolor{green!20}278 & \cellcolor{green!20}157 \\
\midrule
\multirow{2}{*}{\rev{\textit{QwenCoder-7B}}}
& \rev{\STqlora} & \rev{\textbf{71.26\%}} & \rev{5,650} & \rev{35,657} & \rev{99.7\%} & \rev{1,646} & \rev{404} & \rev{3,965} & \rev{232} & \rev{62} & \rev{16} & \rev{134} & \rev{1} & \rev{3} & \rev{457} & \rev{209} \\
& \rev{\MTqlora} & \rev{49.83\%} & \rev{4,882 $\downarrow$} & \rev{32,027 $\downarrow$} & \rev{\cellcolor{blue!20}90.8\%} & \rev{\cellcolor{green!20}1,383} & \rev{\cellcolor{green!20}20} & \rev{\cellcolor{green!20}383} & \rev{\cellcolor{green!20}10} & \rev{\cellcolor{green!20}12} & \rev{N/A} & \rev{\cellcolor{green!20}6} & \rev{0} & \rev{\cellcolor{green!20}1} & \rev{\cellcolor{green!20}168} & \rev{\cellcolor{green!20}3} \\
\midrule[\heavyrulewidth]
\multirow{2}{*}{\revtwo{\textit{CodeLlama-7B}}}
& \revtwo{\STqlora} & \revtwo{\textbf{82.48\%}} & \revtwo{5,725} & \revtwo{34,970} & \revtwo{99.9\%} & \revtwo{1,607} & \revtwo{347} & \revtwo{4,050} & \revtwo{222} & \revtwo{64} & \revtwo{12} & \revtwo{35} & \revtwo{0} & \revtwo{15} & \revtwo{1,136} & \revtwo{686} \\
& \revtwo{\MTqlora} & \revtwo{79.98\%} & \revtwo{5,859 $\uparrow$} & \revtwo{35,291 $\uparrow$} & \revtwo{\cellcolor{yellow!20}100\%} & \revtwo{\cellcolor{blue!20}1,614} & \revtwo{\cellcolor{green!20}346} & \revtwo{\cellcolor{blue!20}4,106} & \revtwo{\cellcolor{blue!20}234} & \revtwo{\cellcolor{yellow!20}64} & \revtwo{\cellcolor{blue!20}16} & \revtwo{\cellcolor{blue!20}39} & \revtwo{0} & \revtwo{\cellcolor{blue!20}16} & \revtwo{\cellcolor{green!20}1,120} & \revtwo{\cellcolor{blue!20}734} \\
\bottomrule
\end{tabular}
}
\end{table}

\subsubsection{Code Summarization} 
\label{sec:rq1_codesum}

To provide a holistic view of the real capabilities of QLoRA-based optimization, we include results for the code-to-natural language task, \rev{completing the evaluation across the three studied code-related tasks}, in the context of the \rq{1}.

Unlike the previous tasks, summarization outputs free-form text rather than executable code, necessitating evaluation metrics that capture both lexical overlap and semantic similarity. We therefore report a diverse set of similarity-based metrics, including BLEU, METEOR, Rouge-L, and chrF, which assess surface-level match to reference summaries, alongside contextual metrics such as BERTScore, which measures semantic similarity using contextualized embeddings. Additionally, for \java, we report SIDE, the novel code-aware metric presented by Mastropaolo \etal \cite{mastropaolo2024evaluating} that quantifies semantic alignment between code and summaries via contrastive learning.

\begin{table*}[h]
\centering
\scriptsize
\caption{RQ1: Comparison of \STqlora and \MTqlora for code summarization on CoderEval \revtwo{across the Qwen2.5-Coder and CodeLlama (7B only) model families}. SIDE applies to \java only.}
\label{tab:rq1_codesum}
\resizebox{0.75\textwidth}{!}{%
\begin{tabular}{c|c|c|cccccc}
\toprule
\textbf{Language} & \textbf{Model} & \textbf{Method} & \textbf{BLEU} & \textbf{METEOR} & \textbf{Rouge-L} & \textbf{chrF} & \textbf{BERTScore} & \textbf{SIDE} \\
\midrule
\multirow{11}{*}{\java}
& \multirow{2}{*}{\textit{QwenCoder-0.5B}} 
& \STqlora & 7.10\% & 21.91\% & 33.32\% & \textbf{28.66\%} & \textbf{66.56\%} & \textbf{92.38\%} \\
& & \MTqlora & \textbf{7.18\%} & \textbf{21.80\%} & \textbf{34.09\%} & 27.71\% & 66.31\% & 91.75\% \\
\cmidrule{2-9}
& \multirow{2}{*}{\textit{QwenCoder-1.5B}} 
& \STqlora & \textbf{8.48\%} & \textbf{23.43\%} & \textbf{35.60\%} & \textbf{30.21\%} & \textbf{67.01\%} & 91.96\% \\
& & \MTqlora & 8.12\% & 22.14\% & 34.15\% & 29.23\% & 66.78\% & \textbf{92.21\%} \\
\cmidrule{2-9}
& \multirow{2}{*}{\textit{QwenCoder-3B}} 
& \STqlora & \textbf{8.73\%} & \textbf{23.95\%} & \textbf{35.32\%} & \textbf{30.58\%} & \textbf{67.03\%} & \textbf{91.91\%} \\
& & \MTqlora & 7.63\% & 22.94\% & 34.93\% & 29.31\% & 66.40\% & 91.58\% \\
\cmidrule{2-9}
& \multirow{2}{*}{\rev{\textit{QwenCoder-7B}}}
& \rev{\STqlora} & \rev{8.19\%} & \rev{\textbf{24.15\%}} & \rev{\textbf{35.68\%}} & \rev{30.67\%} & \rev{66.90\%} & \rev{89.21\%} \\
& & \rev{\MTqlora} & \rev{\textbf{8.37\%}} & \rev{22.75\%} & \rev{35.07\%} & \rev{\textbf{30.73\%}} & \rev{\textbf{67.24\%}} & \rev{\textbf{90.49\%}} \\
\cmidrule[\heavyrulewidth]{2-9}
& \multirow{2}{*}{\revtwo{\textit{CodeLlama-7B}}}
& \revtwo{\STqlora} & \revtwo{8.50\%} & \revtwo{24.33\%} & \revtwo{37.56\%} & \revtwo{29.61\%} & \revtwo{67.76\%} & \revtwo{\textbf{91.32\%}} \\
& & \revtwo{\MTqlora} & \revtwo{\textbf{10.96\%}} & \revtwo{\textbf{27.09\%}} & \revtwo{\textbf{38.94\%}} & \revtwo{\textbf{31.99\%}} & \revtwo{\textbf{68.47\%}} & \revtwo{91.08\%} \\
\midrule
\multirow{11}{*}{\python}
& \multirow{2}{*}{\textit{QwenCoder-0.5B}} 
& \STqlora & 14.29\% & 25.03\% & 37.16\% & 34.89\% & 63.83\% & -- \\
& & \MTqlora & \textbf{19.43\%} & \textbf{29.53\%} & \textbf{40.67\%} & \textbf{39.16\%} & \textbf{65.48\%} & -- \\
\cmidrule{2-9}
& \multirow{2}{*}{\textit{QwenCoder-1.5B}} 
& \STqlora & 15.12\% & 26.84\% & 39.87\% & 36.88\% & 65.00\% & -- \\
& & \MTqlora & \textbf{22.90\%} & \textbf{33.07\%} & \textbf{44.50\%} & \textbf{42.28\%} & \textbf{66.84\%} & -- \\
\cmidrule{2-9}
& \multirow{2}{*}{\textit{QwenCoder-3B}} 
& \STqlora & 23.31\% & 33.22\% & 44.93\% & 41.75\% & 67.07\% & -- \\
& & \MTqlora & \textbf{29.90\%} & \textbf{40.06\%} & \textbf{47.43\%} & \textbf{48.20\%} & \textbf{68.59\%} & -- \\
\cmidrule{2-9}
& \multirow{2}{*}{\rev{\textit{QwenCoder-7B}}}
& \rev{\STqlora} & \rev{16.49\%} & \rev{28.42\%} & \rev{39.18\%} & \rev{36.67\%} & \rev{64.23\%} & -- \\
& & \rev{\MTqlora} & \rev{\textbf{30.80\%}} & \rev{\textbf{41.32\%}} & \rev{\textbf{51.12\%}} & \rev{\textbf{50.26\%}} & \rev{\textbf{70.15\%}} & -- \\
\cmidrule[\heavyrulewidth]{2-9}
& \multirow{2}{*}{\revtwo{\textit{CodeLlama-7B}}}
& \revtwo{\STqlora} & \revtwo{27.98\%} & \revtwo{37.09\%} & \revtwo{\textbf{49.65\%}} & \revtwo{47.60\%} & \revtwo{\textbf{69.28\%}} & -- \\
& & \revtwo{\MTqlora} & \revtwo{\textbf{30.89\%}} & \revtwo{\textbf{38.58\%}} & \revtwo{47.05\%} & \revtwo{\textbf{48.80\%}} & \revtwo{68.46\%} & -- \\
\bottomrule
\end{tabular}
}
\end{table*}

\tabref{tab:rq1_codesum} reports the performance of \MT QLoRA and \ST QLoRA across the 0.5B--3B scales for both \java and \python; the 7B and CodeLlama-7B results are discussed at the end of this section. Because CoderEval standardizes task requirements, the evaluated summarization instances exhibit comparable complexity across languages, enabling controlled cross-language analysis. Observed performance differences therefore, cannot be attributed to variations in task difficulty, but instead could be the result of the interplay between the training strategy and intrinsic characteristics of each programming language. However, we emphasize that our objective is not to draw broad generalizations from these observations, but rather to identify recurring patterns that shed light on subtle effects of QLoRA-based \MT training and motivate further investigation into language-specific training dynamics.

We begin with \python method-level code summarization, where \MT QLoRA demonstrates clear and consistent advantages. Across all model sizes and evaluation metrics, the Qwen-Coder model fine-tuned via \MT QLoRA outperforms its \ST counterpart, with performance gains that remain substantial across scales. All improvements are reported as \textit{relative gains} over the \ST baseline. At 0.5B parameters, \MT QLoRA achieves $\uparrow$36.0\% in BLEU (19.43\% \vs\ 14.29\%), $\uparrow$18.0\% in METEOR (29.53\% \vs\ 25.03\%), $\uparrow$9.4\% in ROUGE-L (40.67\% \vs\ 37.16\%), $\uparrow$12.2\% in chrF (39.16\% \vs\ 34.89\%), and $\uparrow$2.6\% in BERTScore (65.48\% \vs\ 63.83\%). These advantages persist and strengthen at 1.5B, where \MT QLoRA attains $\uparrow$51.5\% in BLEU (22.90\% \vs\ 15.12\%), $\uparrow$23.2\% in METEOR (33.07\% \vs\ 26.84\%), $\uparrow$11.6\% in ROUGE-L (44.50\% \vs\ 39.87\%), $\uparrow$14.6\% in chrF (42.28\% \vs\ 36.88\%), and $\uparrow$2.8\% in BERTScore (66.84\% \vs\ 65.00\%). At 3B, gains remain consistent, with BLEU reaching $\uparrow$28.3\% (29.90\% \vs\ 23.31\%), METEOR $\uparrow$20.6\% (40.06\% \vs\ 33.22\%), ROUGE-L $\uparrow$5.6\% (47.43\% \vs\ 44.93\%), chrF $\uparrow$15.4\% (48.20\% \vs\ 41.75\%), and BERTScore $\uparrow$2.3\% (68.59\% \vs\ 67.07\%).

\java summarization, in contrast, exhibits a distinctly different pattern under \MT training, with \ST QLoRA demonstrating consistent advantages that become more pronounced at larger scales. At 0.5B parameters, results are mixed: BLEU shows a marginal gain of $\uparrow$1.1\% (7.18\% \vs\ 7.10\%) and ROUGE-L improves by $\uparrow$2.3\% (34.09\% \vs\ 33.32\%) under \MT training. However, declines emerge in METEOR ($\downarrow$0.5\%), chrF ($\downarrow$3.3\%), BERTScore ($\downarrow$0.4\%), and SIDE ($\downarrow$0.7\%), indicating no clear advantage for \MT QLoRA at this scale.

At 1.5B parameters, \ST QLoRA establishes a clearer advantage across most metrics. Relative to \MT QLoRA, \ST achieves higher BLEU ($\uparrow$4.4\%; 8.48\% \vs\ 8.12\%), METEOR ($\uparrow$5.8\%; 23.43\% \vs\ 22.14\%), ROUGE-L ($\uparrow$4.2\%; 35.60\% \vs\ 34.15\%), chrF ($\uparrow$3.4\%; 30.21\% \vs\ 29.23\%), and BERTScore ($\uparrow$0.3\%; 67.01\% \vs\ 66.78\%). The code-aware SIDE metric shows a modest improvement of $\uparrow$0.3\% (92.21\% \vs\ 91.96\%) under \MT training, representing the only metric where \MT QLoRA maintains an edge at this scale.

At 3B parameters, the \ST advantage becomes most pronounced. \ST QLoRA outperforms \MT QLoRA across all metrics: BLEU by $\uparrow$14.4\% (8.73\% \vs\ 7.63\%), METEOR by $\uparrow$4.4\% (23.95\% \vs\ 22.94\%), ROUGE-L by $\uparrow$1.1\% (35.32\% \vs\ 34.93\%), chrF by $\uparrow$4.3\% (30.58\% \vs\ 29.31\%), BERTScore by $\uparrow$0.9\% (67.03\% \vs\ 66.40\%), and SIDE by $\uparrow$0.4\% (91.91\% \vs\ 91.58\%).

Overall, the metric-based evaluation reveals clear language-dependent differences in how \MT training scales. For \python, the benefits of \MT QLoRA are consistent across all model sizes, with substantial improvements in both lexical overlap and semantic similarity metrics. For \java, the pattern reverses: \ST QLoRA maintains an advantage that strengthens with model capacity, suggesting that task-specific training may be more effective for \java summarization, particularly at larger scales.

While these findings provide strong evidence based on automatic similarity metrics, such metrics are inherently limited in their ability to capture higher-level qualitative properties of summarization, including content adequacy, conciseness, and fluency. To address this limitation, we complement our analysis with an LLM-as-a-judge evaluation, in which a large language model is prompted to assess generated summaries along these qualitative dimensions. By using an LLM as a proxy for human judgment, this approach enables a more human-centric evaluation of summary quality that goes beyond surface-level similarity and has been shown to approximate expert assessments in prior work~\cite{crupi2025effectiveness}. To motivate our choice of GPT-5 Mini as the LLM judge, we follow recent work showing that lightweight, well-calibrated models can reliably approximate human judgments when properly validated. In software engineering contexts, GPT-5 Mini has been adopted as an LLM-based evaluator under established LLM-as-a-judge paradigms, where it serves as a strong proxy for human assessment using zero-shot prompting \cite{vitale2025toward}. Moreover, comparative validation studies across multiple candidate judge models report that GPT-5 Mini achieves the highest agreement with expert human judgments, outperforming larger models such as GPT-4.1 and GPT-4o~\cite{burleigh2025beyond}. Together, these findings indicate that GPT-5 Mini offers a favorable balance between evaluation reliability and computational efficiency, making it well suited for scalable, human-centric assessment of generated software artifacts.


Using the evaluation data described in Section~\ref{sec:codesumm}, we use zero-shot prompting with GPT-5 Mini as the judge. The evaluator scores each generated summary along three complementary dimensions--content Adequacy, conciseness, and fluency--using a five-point Likert scale ranging from 1 (Very poor) to 5 (Very good). The complete prompt used for this evaluation is shown in \figref{fig:llm_judge_prompt}.

\begin{figure}[t]
\centering
\fcolorbox{olive}{gray!10}{
\begin{minipage}{0.7\linewidth}
\small
\vspace{2pt}
\colorbox{olive!20}{\parbox{\dimexpr\linewidth-2\fboxsep}{\textbf{\textsf{Code Summary Evaluation Prompt}}}}
\vspace{4pt}

You will be provided with a \texttt{\{lang\}} function (``Function'') and a textual summary of it (``Comment''). The goal of the Comment is to document the functionality implemented in the Function. Your role is to evaluate the Comment across three criteria, providing as output for each of them a rating and a rationale as described in the following.

\vspace{4pt}
\textbf{\# Evaluation Criteria}\\
* \textit{Content adequacy}: the extent to which the comment summarizes all information that can be inferred from the source code.\\
* \textit{\textit{Conciseness}}: the extent to which the comment contains unnecessary information.\\
* \textit{Fluency}: the extent to which the comment is easy to understand.

\vspace{4pt}
For each criterion, provide a score on a scale from 1 to 5:\\
1. Very poor \hfill 2. Poor \hfill 3. Fair \hfill 4. Good \hfill 5. Very good

\vspace{4pt}
\textbf{\# Function}\\
\texttt{\{code\}}

\vspace{4pt}
\textbf{\# Comment}\\
\texttt{\{summary\}}

\vspace{4pt}
Please provide your ratings in this exact format:\\
\texttt{Content adequacy: [score]}\\
\texttt{\textit{Conciseness}: [score]}\\
\texttt{Fluency: [score]}
\vspace{2pt}
\end{minipage}
}
\caption{Zero-shot prompt used for LLM-as-a-judge evaluation of code summaries.}
\label{fig:llm_judge_prompt}
\end{figure}


We evaluate summaries generated by Qwen-Coder under two QLoRA fine-tuning strategies: (i) \MT training across multiple code-related tasks, and (ii) \ST training exclusively for code summarization. We compare these model-generated summaries against human-written references curated by Crupi \etal~\cite{crupi2025effectiveness} as reliable quality baselines. 

Complete distributions of GPT-5 mini quality scores across all configurations are presented in Appendix~\ref{appendix:llm_judge} (\figref{fig:llm_judge_java} and \figref{fig:llm_judge_python}).

Across both languages, \MT QLoRA achieves summary quality broadly comparable to \ST QLoRA across all three evaluation dimensions. For \java, both configurations attain Content Adequacy scores corresponding to approximately 73--78\% of the human-written baseline, indicating that the generated summaries capture a substantial portion of the information conveyed by developer-authored documentation. Differences between \MT and \ST QLoRA are minimal across all model sizes, typically within 0.1 points. \textit{\textit{Conciseness}} scores remain consistently high for both configurations, ranging from 4.92 to 4.99, and often surpass the human baseline of 4.62, suggesting that automated summaries tend to avoid unnecessary verbosity. Fluency scores are likewise close to human-written quality, with both configurations achieving values between 4.22 and 4.41, demonstrating that the generated summaries are generally clear, well-formed, and easy to understand. 

Consistent with these observations, Wilcoxon signed-rank tests with Holm--Bonferroni correction ($\alpha = 0.05$) reveal no statistically significant differences between training strategies for \java across any metric or model size ($p \geq 0.16$).

For \python, we observe similar qualitative trends, though with slightly lower Content Adequacy relative to \java. Human-written summaries establish a baseline of 2.64, while both QLoRA configurations achieve approximately 69--74\% of this level across model sizes. Differences between \MT and \ST QLoRA remain modest, with \MT models closely tracking \ST performance and, in the 3B configuration, slightly exceeding it in content adequacy. Conciseness scores for \python summaries remain high for both configurations (4.59--4.73), approaching the human baseline (4.80), while Fluency scores (3.71--3.81) fall somewhat below human quality (4.54) but remain consistent across training strategies. Statistical analysis supports these trends (complete results available in our replication package \cite{replication}): no significant differences are observed at 0.5B ($p \geq 0.05$), whereas at larger scales (1.5B and 3B), \MT QLoRA yields statistically significant improvements across all metrics ($p < 0.01$), indicating that \MT training provides measurable benefits for \python summarization at higher model capacities.

\vspace{10pt}

\begin{summarybox}[Key Findings]
\noindent Code summarization exhibits strong language-dependent effects under \MT training. For \python, \MT QLoRA consistently improves similarity-based metrics over \ST QLoRA across all model sizes, with gains ranging from 2.3\% to 51.5\% depending on metric and scale. 
For \java, the pattern reverses across 0.5B--3B: \ST QLoRA maintains an advantage that strengthens over this range, with \MT QLoRA showing declines of up to 14.4\% relative to \ST at 3B\revtwo{, though this advantage does not persist at 7B, where the two configurations split across metrics, nor for CodeLlama, where \MTqlora leads on most metrics}.
LLM-as-a-judge results reveal no significant quality differences for \java, whereas \python shows significant \MT advantages at larger scales under Wilcoxon tests with Holm--Bonferroni correction. 
Overall, \MT QLoRA reliably benefits \python summarization\revtwo{, including at 7B and for CodeLlama,} while \ST QLoRA proves more effective for \java \revtwo{over the 0.5B--3B range}.
\end{summarybox}

\rev{\noindent\textbf{Extension to 7B scale:}}
\rev{To address concerns regarding generalization beyond our primary 0.5B--3B range, we additionally evaluated \MTqlora and \STqlora at the 7B scale across all three tasks. \revtwo{The 7B results confirm our central RQ$_1$ findings for code generation and \python code summarization, while revealing a clear limitation for code translation.} For code generation, the two languages diverge at 7B: \MTqlora outperforms \STqlora on \java (\passatone 33.70\% \vs\ 30.43\%, $\uparrow$10.7\%), while \STqlora outperforms \MTqlora on \python (22.63\% \vs\ 20.53\%, $\uparrow$10.2\% for \STqlora) --- a reversal relative to 3B, where \MTqlora led in both languages. For code summarization, the language-dependent pattern persists: \MTqlora continues to dominate \python summarization (BLEU 30.80\% \vs\ 16.49\%, $\uparrow$86.8\%), while \java summarization remains split between the two configurations across metrics.
\revtwo{For code translation, however, the pattern does not hold: \STqlora outperforms \MTqlora in both directions by a substantially larger margin than at 3B ($\uparrow$40.3\% on \java$\rightarrow$C\#, $\uparrow$43.0\% on C\#$\rightarrow$\java); we discuss this further in \secref{sec:threats} (Threats to Validity). The 7B extension therefore confirms our conclusions for code generation and \python code summarization while revealing a clear limitation for translation; characterizing the cause of this degradation is left to future work.}
}

\revtwo{To assess whether these patterns generalize beyond a single model family, we replicate the RQ$_1$ comparison using CodeLlama-7B-Instruct while keeping the training protocol, hyperparameters, and evaluation procedure fixed, such that model family is the only varying factor. Because the two model families differ in tokenization and training data, we do not compare their absolute scores. Instead, we examine whether the same task- and language-dependent patterns observed in RQ$_1$ persist across model families. On code translation, \STqlora outperforms \MTqlora in both directions (CodeBLEU 82.22\% \vs\ 79.45\% on \java$\rightarrow$C\#, 82.48\% \vs\ 79.98\% on C\#$\rightarrow$\java), consistent with the \STqlora translation advantage. On \python code summarization, \MTqlora outperforms \STqlora on most metrics (BLEU 30.89\% \vs\ 27.98\%, METEOR 38.58\% \vs\ 37.09\%, chrF 48.80\% \vs\ 47.60\%), while \STqlora leads on Rouge-L and BERTScore, broadly consistent with the \MTqlora summarization advantage, and \java summarization again favors \MTqlora on most metrics, consistent with the mixed outcome observed at other scales. Code generation is where the two families differ: for CodeLlama, the two configurations are comparable on \python (\passatone 21.58\% for both) and \STqlora outperforms \MTqlora on \java (33.15\% \vs\ 25.00\%), rather than exhibiting the \java \MTqlora advantage observed for Qwen2.5-Coder at this scale.}

\revtwo{Turning to non-functional quality, the CodeLlama results are broadly consistent with the functional picture, with one notable exception. For \java code generation (\tabref{tab:rq1-cg-java}), \MTqlora produces markedly cleaner code than \STqlora despite its lower \passatone: PMD violations fall from 1,384 to 1,098 ($\downarrow$20.7\%), SonarCloud issues from 922 to 734 ($\downarrow$20.4\%), and Cyclomatic Complexity from 596 to 509 ($\downarrow$14.6\%), alongside more compact implementations (1,290 \vs\ 1,472 LoC, $\downarrow$12.4\%). For \python code generation (\tabref{tab:rq1-cg-py}), the two configurations are closer, with \MTqlora showing slightly higher complexity (841 \vs\ 794 CyC, $\uparrow$5.9\%) and Pylint counts (2,416 \vs\ 2,083, $\uparrow$16.0\%) at identical \passatone. For code translation, quality differences are small in both directions: \MTqlora yields fewer Roslyn-detected issues on \java$\rightarrow$C\# (943 \vs\ 1,046, $\downarrow$9.8\%), while on C\#$\rightarrow$\java the two configurations are close, with \MTqlora slightly higher on PMD (4,805 \vs\ 4,730, $\uparrow$1.6\%) and SonarCloud (1,870 \vs\ 1,837, $\uparrow$1.8\%). The \java code-generation case is the most notable: \MTqlora trades functional correctness for structurally simpler and more standards-compliant code, a correctness--quality divergence not observed for Qwen2.5-Coder at this scale. None of the CodeLlama \STqlora--\MTqlora differences reaches statistical significance, so these are directional observations. Overall, the translation and summarization findings are reproduced for CodeLlama, while code generation differs between the two families in both correctness and quality.}

\subsection{RQ2: \MT QLoRA \vs \MT Full Fine-Tuning}
\label{sec:rq2}

Having established the comparative trends between \ST QLoRA and \MT QLoRA, we next examine whether the observed \MT advantages persist when \MT QLoRA is compared against \MT full fine-tuning (\MTfft) across multiple code-related objectives. Unlike QLoRA, which updates only low-rank adaptation parameters, \MTfft updates all model parameters and therefore provides a reference point for assessing the trade-offs between efficiency and overall effectiveness. We structure this analysis by task, considering code generation (\secref{sec:rq2_codegen}), code translation (\secref{sec:rq2_codetrans}), and code summarization (\secref{sec:rq2_codesum}).

\subsubsection{Code Generation}
\label{sec:rq2_codegen}

We summarize the code generation results in \tabref{tab:rq2_codegen_python} and \tabref{tab:rq2_codegen_java} for \python and \java, respectively. Across both languages and all model scales, \MT QLoRA matches or exceeds \MTfft in most configurations. 

For \python, \MT QLoRA achieves higher \passatone scores across all model scales, increasing from 14.74\% to 15.79\% at 0.5B ($\uparrow$7.1\%), from 15.79\% to 18.95\% at 1.5B ($\uparrow$20.0\%), and from 17.37\% to 21.05\% at 3B ($\uparrow$21.2\%). The performance gap widens as model capacity increases, with the largest relative improvement observed at 3B. 

For Java, \MT QLoRA similarly outperforms \MTfft at 0.5B (20.11\% \vs\ 19.02\%, $\uparrow$5.7\%) and demonstrates substantial advantage at 3B (32.07\% \vs\ 26.63\%, $\uparrow$20.4\%). The 1.5B configuration represents the only case where \MTfft slightly exceeds \MT QLoRA (20.11\% \vs\ 19.02\%, $\downarrow$5.4\%). 

Comparing \MT QLoRA performance across languages, we notice that the same model exhibits distinct behaviors when generating \java versus \python code under the same setting configuration. In particular, we notice that in terms of correctness (\passatone) gains are more pronounced and uniformly increasing with scale for \python, while \java exhibits a mild regression at 1.5B before recovering strongly at 3B. These differences may stem from intrinsic language characteristics, such as syntactic verbosity, type system rigidity, and the degree of structural constraint imposed during generation, which can interact differently with shared \MT representations. In addition to that, variations in the distribution of the training corpora may amplify scale-dependent effects.

\begin{table}[h]
\centering
\scriptsize
\caption{RQ2: Comparison of \MTfft and \MTqlora for \python code generation on CoderEval.}
\label{tab:rq2_codegen_python}
\resizebox{\textwidth}{!}{%
\begin{tabular}{lc|c|cccc|cccc|cccc}
\toprule
\multirow{2}{*}{\textbf{Model}} & \multirow{2}{*}{\textbf{Method}} & 
\multirow{2}{*}{\textbf{Pass@1}} & 
\multicolumn{4}{c|}{\textbf{Lizard}} & 
\multicolumn{4}{c|}{\textbf{Pylint}} &
\multicolumn{4}{c}{\textbf{SonarCloud}} \\
\cmidrule(lr){4-7} \cmidrule(lr){8-11} \cmidrule(l){12-15}
& & & \textbf{LoC} & \textbf{Tok} & \textbf{DR} & \textbf{CyC} & 
\textbf{Err} & \textbf{Warn} & \textbf{Conv} & \textbf{Ref} &
\textbf{Sec} & \textbf{Rel} & \textbf{Main} & \textbf{CoC} \\
\midrule
\multirow{2}{*}{\textit{QwenCoder-0.5B}} 
& \MTfft & \textbf{14.74\%} & 2,468 & 16,917 & 98.40\% & 928 & 352 & 1,809 & 725 & 68 & 3 & 28 & 133 & 569 \\
& \MTqlora & \textbf{15.79\%} & 1,952 $\downarrow$ & 13,327 $\downarrow$ & \cellcolor{green!20}100\% & \cellcolor{green!20}812 & \cellcolor{green!20}251 & \cellcolor{green!20}1,593 & \cellcolor{green!20}642 & \cellcolor{green!20}56 & \cellcolor{green!20}2 & \cellcolor{green!20}7 & \cellcolor{green!20}90 & \cellcolor{blue!20}634 \\
\midrule
\multirow{2}{*}{\textit{QwenCoder-1.5B}} 
& \MTfft & 15.79\% & 2,374 & 16,385 & 100\% & 928 & 350 & 1,596 & 650 & 66 & 1 & 6 & 51 & 762 \\
& \MTqlora & \textbf{18.95\%} & 2,296 $\downarrow$ & 14,531 $\downarrow$ & \cellcolor{yellow!20}100\% & \cellcolor{green!20}882 & \cellcolor{green!20}315 & \cellcolor{blue!20}1,767 & \cellcolor{blue!20}677 & \cellcolor{blue!20}76 & 1 & \cellcolor{green!20}5 & \cellcolor{blue!20}72 & \cellcolor{blue!20}771 \\
\midrule
\multirow{2}{*}{\textit{QwenCoder-3B}} 
& \MTfft & 17.37\% & 2,706 & 18,038 & 100\% & 996 & 351 & 1,762 & 579 & 75 & 2 & 0 & 135 & 686 \\
& \MTqlora & \textbf{21.05\%} & 1,759 $\downarrow$ & 11,773 $\downarrow$ & \cellcolor{yellow!20}100\% & \cellcolor{green!20}677 & \cellcolor{green!20}269 & \cellcolor{green!20}1,521 & \cellcolor{blue!20}673 & \cellcolor{green!20}56 & 2 & \cellcolor{blue!20}5 & \cellcolor{green!20}106 & \cellcolor{green!20}549 \\
\midrule
\multirow{2}{*}{\rev{\textit{QwenCoder-7B}}}
& \rev{\MTfft} & \rev{14.74\%} & \rev{1,993} & \rev{14,253} & \rev{99.5\%} & \rev{782} & \rev{372} & \rev{1,409} & \rev{707} & \rev{54} & \rev{1} & \rev{4} & \rev{85} & \rev{647} \\
& \rev{\MTqlora} & \rev{\textbf{20.53\%}} & \rev{1,957 $\downarrow$} & \rev{12,424 $\downarrow$} & \rev{\cellcolor{green!20}100\%} & \rev{\cellcolor{green!20}732} & \rev{\cellcolor{green!20}287} & \rev{\cellcolor{blue!20}1,768} & \rev{\cellcolor{green!20}683} & \rev{\cellcolor{green!20}51} & \rev{3} & \rev{\cellcolor{green!20}3} & \rev{\cellcolor{green!20}59} & \rev{\cellcolor{blue!20}727} \\
\bottomrule
\end{tabular}
}
\end{table}
\begin{table}[h]
\centering
\scriptsize
\caption{RQ2: Comparison of \MTfft and \MTqlora for \java code generation on CoderEval.}
\label{tab:rq2_codegen_java}
\resizebox{\textwidth}{!}{%
\begin{tabular}{lc|c|cccc|cccccc|cccc}
\toprule
\multirow{2}{*}{\textbf{Model}} & \multirow{2}{*}{\textbf{Method}} & 
\multirow{2}{*}{\textbf{Pass@1}} & 
\multicolumn{4}{c|}{\textbf{Lizard}} & 
\multicolumn{6}{c|}{\textbf{PMD}} &
\multicolumn{4}{c}{\textbf{SonarCloud}} \\
\cmidrule(lr){4-7} \cmidrule(lr){8-13} \cmidrule(l){14-17}
& & & \textbf{LoC} & \textbf{Tok} & \textbf{DR} & \textbf{CyC} & 
\textbf{Best Prac.} & \textbf{CS} & \textbf{Design} & \textbf{EP} & \textbf{MT} & \textbf{Perf.} &
\textbf{Sec} & \textbf{Rel} & \textbf{Main} & \textbf{CoC} \\
\midrule
\multirow{2}{*}{\textit{QwenCoder-0.5B}} 
& \MTfft & 19.02\% & 1,838 & 12,339 & 98.40\% & 649 & 90 & 1,183 & 131 & 58 & 4 & 15 & 0 & 5 & 570 & 629 \\
& \MTqlora & \textbf{20.11\%} & 1,933 $\uparrow$ & 12,750 $\uparrow$ & \cellcolor{blue!20}96.70\% & \cellcolor{blue!20}697 & \cellcolor{green!20}76 & \cellcolor{green!20}1,123 & \cellcolor{blue!20}139 & \cellcolor{green!20}33 & \cellcolor{green!20}3 & \cellcolor{green!20}10 & 0 & \cellcolor{green!20}0 & \cellcolor{green!20}531 & \cellcolor{green!20}494 \\
\midrule
\multirow{2}{*}{\textit{QwenCoder-1.5B}} 
& \MTfft & \textbf{20.11\%} & 1,828 & 13,029 & 98.90\% & 676 & 94 & 1,288 & 119 & 49 & 9 & 17 & 1 & 22 & 569 & 576 \\
& \MTqlora & 19.02\% & 1,490 $\downarrow$ & 10,313 $\downarrow$ & \cellcolor{blue!20}96.20\% & \cellcolor{green!20}589 & \cellcolor{green!20}75 & \cellcolor{green!20}1,183 & \cellcolor{blue!20}127 & \cellcolor{green!20}31 & \cellcolor{green!20}5 & \cellcolor{green!20}11 & 0 & \cellcolor{green!20}2 & \cellcolor{green!20}530 & \cellcolor{green!20}442 \\
\midrule
\multirow{2}{*}{\textit{QwenCoder-3B}} 
& \MTfft & 26.63\% & 1,727 & 11,565 & 99.50\% & 607 & 80 & 1,200 & 131 & 45 & 3 & 27 & 0 & 1 & 548 & 501 \\
& \MTqlora & \textbf{32.07\%} & 1,580 $\downarrow$ & 10,920 $\downarrow$ & \cellcolor{blue!20}98.40\% & \cellcolor{green!20}558 & \cellcolor{blue!20}86 & \cellcolor{green!20}1,184 & \cellcolor{green!20}128 & \cellcolor{green!20}32 & \cellcolor{blue!20}4 & \cellcolor{green!20}7 & 0 & \cellcolor{blue!20}2 & \cellcolor{green!20}541 & \cellcolor{green!20}469 \\
\midrule
\multirow{2}{*}{\rev{\textit{QwenCoder-7B}}}
& \rev{\MTfft} & \rev{18.48\%} & \rev{2,161} & \rev{15,310} & \rev{100\%} & \rev{778} & \rev{94} & \rev{1,376} & \rev{149} & \rev{53} & \rev{2} & \rev{34} & \rev{0} & \rev{3} & \rev{586} & \rev{759} \\
& \rev{\MTqlora} & \rev{\textbf{33.70\%}} & \rev{1,799 $\downarrow$} & \rev{12,375 $\downarrow$} & \rev{\cellcolor{blue!20}98.9\%} & \rev{\cellcolor{green!20}663} & \rev{\cellcolor{green!20}87} & \rev{\cellcolor{green!20}1,238} & \rev{\cellcolor{green!20}137} & \rev{\cellcolor{green!20}33} & \rev{\cellcolor{blue!20}5} & \rev{\cellcolor{green!20}11} & \rev{0} & \rev{\cellcolor{green!20}0} & \rev{\cellcolor{green!20}532} & \rev{\cellcolor{green!20}576} \\
\bottomrule
\end{tabular}
}
\vspace{-10pt}
\end{table}

We move forward presenting the results of automated qualitative code analysis we conducted.

Starting from \python, \MT QLoRA consistently produces code with lower complexity than \MTfft, reducing CyC from 928 to 812 at 0.5B ($\downarrow$12.5\%), from 928 to 882 at 1.5B ($\downarrow$5.0\%), and from 996 to 677 at 3B ($\downarrow$32.0\%). For Java, the pattern varies across scales: at 0.5B, \MT QLoRA exhibits slightly higher complexity than \MTfft (697 \vs\ 649, $\uparrow$7.4\%), whereas at 1.5B and 3B it produces lower CyC values (589 \vs\ 676, $\downarrow$12.9\%; and 558 \vs\ 607, $\downarrow$8.1\%). These trends suggest that complexity-related benefits of parameter-efficient training become more consistent as model capacity increases.

Static analysis results further contextualize the functional correctness findings by contrasting \MT QLoRA with \MTfft across languages and scales. Overall, \MT QLoRA either matches or improves code quality and implementation efficiency relative to full fine-tuning, particularly at larger model sizes. For \python, \MT QLoRA reduces total Pylint violations by 5.8\% across the 0.5B--3B scales (from 8{,}383 to 7{,}896), with clear improvements at 0.5B ($\downarrow$13.9\%) and 3B ($\downarrow$9.0\%), alongside a mild regression at 1.5B. SonarCloud maintainability metrics follow a similar trend, showing substantial reductions at 0.5B ($\downarrow$32.3\%) and 3B ($\downarrow$21.5\%). In parallel, \MT QLoRA consistently produces more compact \python implementations, reducing lines of code and token counts at all scales, with the largest gains observed at 3B (35.0\% fewer lines of code and 34.7\% fewer tokens).

For \java, \MT QLoRA likewise improves static analysis outcomes relative to \MTfft, though with smaller and more uniform gains. Total PMD violations decrease by 6.3\% across the 0.5B--3B scales, with consistent reductions across all model sizes. 
Notably, improvements are concentrated in Error Prone and Performance categories, where violations drop by up to 43.1\% (at 0.5B) and 74.1\% (at 3B), respectively.
SonarCloud maintainability issues also decrease modestly but consistently across all configurations. Code size reductions for \java mirror the \python trend at medium and large scales--showing decreases of 18.5\% in lines of code at 1.5B and 8.5\% at 3B--while exhibiting a slight increase at 0.5B.

Overall, these results reinforce the emerging trend that QLoRA-based parameter-efficient \MT fine-tuning does not compromise, and in many cases can improve code quality and implementation compactness relative to \MTfft. Although both languages benefit from \MT QLoRA, \python exhibits earlier and more pronounced improvements, whereas \java follows a similar trajectory with steadier and more conservative gains, particularly as model capacity increases. This contrast suggests that the effectiveness of \MT parameter-efficient fine-tuning is modulated by language-specific characteristics, with larger models better able to leverage shared representations across tasks.



To assess statistical significance, we conduct McNemar tests for \passatone and Wilcoxon signed-rank tests for quality metrics, applying Holm--Bonferroni correction ($\alpha=0.05$) to compare \MT QLoRA against \MTfft. Across the 0.5B--3B scales and both languages, \MT QLoRA achieves statistical parity in functional correctness, supporting the hypothesis that parameter-efficient \MT fine-tuning can preserve \passatone performance relative to full fine-tuning. 

Beyond functional correctness, quality analyses reveal additional advantages of \MT QLoRA. For \python, larger model scales exhibit statistically significant improvements in code conciseness alongside reductions in complexity-related metrics. For \java, quality benefits are more pronounced at smaller scales, including significant reductions in code size and structural complexity.
\\

\begin{summarybox}[Key Findings:] 
\noindent Code generation under \MT QLoRA consistently matches or exceeds \MTfft across both \python and \java, with gains becoming more pronounced as model capacity increases. For \python, \MT QLoRA yields steadily increasing improvements in functional correctness and code quality across all scales, producing less complex, more compact, and cleaner implementations. For \java, improvements are more conservative and scale-dependent, with a mild regression at intermediate scale followed by clear gains at larger models. Statistical analysis confirms parity in functional correctness across all configurations, while quality metrics reveal additional advantages of \MT QLoRA, indicating that parameter-efficient \MT fine-tuning preserves correctness and can enhance code quality, particularly at larger scales.
\end{summarybox}






\subsubsection{Code Translation}
\label{sec:rq2_codetrans}

We evaluate code translation performance using CodeBLEU across both Java$\rightarrow$C\# and C\#$\rightarrow$Java directions. \tabref{tab:rq2_codetrans_java2csharp} and \tabref{tab:rq2_codetrans_csharp2java} compare \MT QLoRA against \MT FFT across the 0.5B--3B scales.

\begin{table}[h]
\centering
\scriptsize
\caption{RQ2: Comparison of \MTfft and \MTqlora for Java$\rightarrow$C\# code translation on CodeXGLUE.}
\label{tab:rq2_codetrans_java2csharp}
\resizebox{0.75\textwidth}{!}{%
\begin{tabular}{lc|c|cccc|cc}
\toprule
\multirow{2}{*}{\textbf{Model}} & \multirow{2}{*}{\textbf{Method}} & 
\multirow{2}{*}{\textbf{CodeBLEU}} & 
\multicolumn{4}{c|}{\textbf{Lizard}} & 
\multicolumn{2}{c}{\textbf{Roslyn}} \\
\cmidrule(lr){4-7} \cmidrule(l){8-9}
& & & \textbf{LoC} & \textbf{Tok} & \textbf{DR} & \textbf{CyC} & 
\textbf{Syntax Errors} & \textbf{Maintainability} \\
\midrule
\multirow{2}{*}{\textit{QwenCoder-0.5B}} 
& \MTfft & \textbf{74.91\%} & 6,230 & 40,104 & 99.1\% & 1,492 & 255 & 566 \\
& \MTqlora & 68.92\% & 6,218 $\downarrow$ & 39,509 $\downarrow$ & \cellcolor{blue!20}98.7\% & \cellcolor{blue!20}1,515 & \cellcolor{blue!20}256 & \cellcolor{blue!20}835 \\
\midrule
\multirow{2}{*}{\textit{QwenCoder-1.5B}} 
& \MTfft & \textbf{78.53\%} & 6,671 & 43,343 & 99.3\% & 1,688 & 302 & 494 \\
& \MTqlora & 73.08\% & 6,369 $\downarrow$ & 40,516 $\downarrow$ & \cellcolor{blue!20}98.5\% & \cellcolor{green!20}1,563 & \cellcolor{blue!20}313 & \cellcolor{blue!20}592 \\
\midrule
\multirow{2}{*}{\textit{QwenCoder-3B}} 
& \MTfft & \textbf{66.38\%} & 5,986 & 38,418 & 92.1\% & 1,492 & 232 & 1,829 \\
& \MTqlora & 62.88\% & 5,890 $\downarrow$ & 37,716 $\downarrow$ & \cellcolor{blue!20}93.6\% & \cellcolor{green!20}1,455 & \cellcolor{green!20}126 & \cellcolor{blue!20}3,061 \\
\midrule
\multirow{2}{*}{\rev{\textit{QwenCoder-7B}}}
& \rev{\MTfft} & \rev{\textbf{72.10\%}} & \rev{6,192} & \rev{40,182} & \rev{92.5\%} & \rev{1,474} & \rev{306} & \rev{691} \\
& \rev{\MTqlora} & \rev{55.37\%} & \rev{5,641 $\downarrow$} & \rev{36,386 $\downarrow$} & \rev{\cellcolor{blue!20}88.1\%} & \rev{\cellcolor{green!20}1,340} & \rev{\cellcolor{green!20}123} & \rev{\cellcolor{blue!20}3,056} \\
\bottomrule
\end{tabular}
}
\end{table}
\begin{table}[h]
\centering
\scriptsize
\caption{RQ2: Comparison of \MTfft and \MTqlora for C\#$\rightarrow$Java code translation on CodeXGLUE.}
\label{tab:rq2_codetrans_csharp2java}
\resizebox{\textwidth}{!}{%
\begin{tabular}{lc|c|cccc|cccccc|cccc}
\toprule
\multirow{2}{*}{\textbf{Model}} & \multirow{2}{*}{\textbf{Method}} & 
\multirow{2}{*}{\textbf{CodeBLEU}} & 
\multicolumn{4}{c|}{\textbf{Lizard}} & 
\multicolumn{6}{c|}{\textbf{PMD}} &
\multicolumn{4}{c}{\textbf{SonarCloud}} \\
\cmidrule(lr){4-7} \cmidrule(lr){8-13} \cmidrule(l){14-17}
& & & \textbf{LoC} & \textbf{Tok} & \textbf{DR} & \textbf{CyC} & 
\textbf{Best Prac.} & \textbf{CS} & \textbf{Design} & \textbf{EP} & \textbf{MT} & \textbf{Perf.} &
\textbf{Sec} & \textbf{Rel} & \textbf{Main} & \textbf{CoC} \\
\midrule
\multirow{2}{*}{\textit{QwenCoder-0.5B}} 
& \MTfft & \textbf{73.82\%} & 5,506 & 33,600 & 98.6\% & 1,536 & 398 & 3,862 & 188 & 57 & 14 & 141 & 0 & 17 & 1,049 & 604 \\
& \MTqlora & 68.37\% & 5,807 $\uparrow$ & 35,170 $\uparrow$ & \cellcolor{green!20}99.6\% & \cellcolor{blue!20}1,594 & \cellcolor{blue!20}400 & \cellcolor{blue!20}3,953 & \cellcolor{blue!20}234 & \cellcolor{green!20}54 & \cellcolor{green!20}12 & \cellcolor{green!20}140 & 0 & \cellcolor{green!20}13 & \cellcolor{blue!20}1,056 & \cellcolor{blue!20}654 \\
\midrule
\multirow{2}{*}{\textit{QwenCoder-1.5B}} 
& \MTfft & 70.07\% & 5,784 & 35,926 & 98.9\% & 1,633 & 402 & 3,894 & 189 & 60 & 17 & 152 & 0 & 12 & 1,046 & 625 \\
& \MTqlora & \textbf{70.72\%} & 5,745 $\downarrow$ & 35,074 $\downarrow$ & \cellcolor{green!20}99.6\% & \cellcolor{green!20}1,566 & \cellcolor{green!20}400 & \cellcolor{blue!20}3,939 & \cellcolor{blue!20}213 & \cellcolor{blue!20}61 & \cellcolor{blue!20}22 & \cellcolor{blue!20}163 & 0 & \cellcolor{blue!20}19 & \cellcolor{blue!20}1,088 & \cellcolor{blue!20}678 \\
\midrule
\multirow{2}{*}{\textit{QwenCoder-3B}} 
& \MTfft & \textbf{76.98\%} & 5,822 & 35,831 & 97.8\% & 1,614 & 378 & 3,060 & 204 & 42 & 12 & 162 & 0 & 12 & 777 & 641 \\
& \MTqlora & 71.22\% & 5,742 $\downarrow$ & 34,862 $\downarrow$ & \cellcolor{green!20}99.5\% & \cellcolor{green!20}1,612 & \cellcolor{green!20}37 & \cellcolor{green!20}1,072 & \cellcolor{green!20}45 & \cellcolor{green!20}11 & \cellcolor{green!20}5 & \cellcolor{green!20}1 & 0 & \cellcolor{green!20}0 & \cellcolor{green!20}278 & \cellcolor{green!20}157 \\
\midrule
\multirow{2}{*}{\rev{\textit{QwenCoder-7B}}}
& \rev{\MTfft} & \rev{\textbf{70.54\%}} & \rev{5,682} & \rev{35,191} & \rev{99.3\%} & \rev{1,586} & \rev{399} & \rev{3,971} & \rev{200} & \rev{60} & \rev{15} & \rev{144} & \rev{0} & \rev{11} & \rev{1,118} & \rev{683} \\
& \rev{\MTqlora} & \rev{49.83\%} & \rev{4,882 $\downarrow$} & \rev{32,027 $\downarrow$} & \rev{\cellcolor{blue!20}90.8\%} & \rev{\cellcolor{green!20}1,383} & \rev{\cellcolor{green!20}20} & \rev{\cellcolor{green!20}383} & \rev{\cellcolor{green!20}10} & \rev{12} & \rev{N/A} & \rev{\cellcolor{green!20}6} & \rev{0} & \rev{\cellcolor{green!20}1} & \rev{\cellcolor{green!20}168} & \rev{\cellcolor{green!20}3} \\
\bottomrule
\end{tabular}
}
\end{table}

\MT QLoRA shows mixed results compared to \MT FFT in code translation, with performance varying by direction and model size. In the Java$\rightarrow$C\# direction, \MT QLoRA falls short across the board: the 0.5B model achieves 68.92\% versus FFT's 74.91\% ($\downarrow$8.0\%), the 1.5B model reaches 73.08\% versus 78.53\% ($\downarrow$6.9\%), and the 3B model attains 62.88\% versus 66.38\% ($\downarrow$5.3\%). The C\#$\rightarrow$Java direction reveals a more nuanced picture: while the 0.5B model lags behind at 68.37\% versus 73.82\% ($\downarrow$7.4\%) and the 3B model similarly underperforms at 71.22\% versus 76.98\% ($\downarrow$7.5\%), the 1.5B configuration bucks this trend, slightly edging out \MT FFT with 70.72\% versus 70.07\% ($\uparrow$0.9\%). This lone success at 1.5B stands in contrast to the otherwise consistent advantage of \MT FFT in translation tasks--a pattern that differs markedly from code generation, where \MT QLoRA matched or exceeded \MT FFT performance throughout.

When we look at quality metrics across static analysis tools, the patterns echo what we observed for correctness. In the Java$\rightarrow$C\# direction, where Lizard and Roslyn assess code quality, \MT FFT generates 4,672 Cyclomatic Complexity issues and 3,678 Roslyn issues (combining syntax errors and maintainability concerns) across the 0.5B--3B scales. \MT QLoRA produces slightly fewer CyC issues at 4,533 ($\downarrow$3.0\%), but Roslyn flags jump considerably to 5,183 ($\uparrow$40.9\%). The reverse direction, C\#$\rightarrow$Java, tells a different story. Across the same scales, while \MT FFT produces 4,783 CyC issues, 13,232 PMD issues, and 4,783 SonarCloud quality issues, \MT QLoRA shows improvements across the board: 4,772 CyC issues ($\downarrow$0.2\%), 10,762 PMD issues ($\downarrow$18.7\%), and 3,943 SonarCloud quality issues ($\downarrow$17.6\%). In essence, \MT QLoRA struggles with Roslyn-detected issues when translating to C\#, but delivers notable quality gains in PMD and SonarCloud metrics when translating to Java.

Cyclomatic Complexity trends shift depending on model size and translation direction. For Java$\rightarrow$C\#, the results are mixed: the 0.5B model produces 1,515 CyC issues versus FFT's 1,492 ($\uparrow$1.5\%), while the 1.5B model improves to 1,563 versus 1,688 ($\downarrow$7.4\%), and the 3B model shows 1,455 versus 1,492 ($\downarrow$2.5\%). The C\#$\rightarrow$Java direction follows a similar pattern: 0.5B generates 1,594 versus 1,536 ($\uparrow$3.8\%), 1.5B improves to 1,566 versus 1,633 ($\downarrow$4.1\%), and 3B achieves near parity at 1,612 versus 1,614 ($\downarrow$0.1\%). 

Language-specific analysis reveal pronounced differences in quality profiles. For Roslyn metrics in Java$\rightarrow$C\# translation, \MT QLoRA shows an increasing numbers of detected issues across all scales. Breaking down by components: 0.5B produces 1,091 total issues (256 syntax errors + 835 maintainability) compared to FFT's 821 (255 + 566), representing a $\uparrow$32.9\% increase. The 1.5B configuration generates 905 total issues (313 + 592) \vs\ 796 (302 + 494), yielding a $\uparrow$13.7\% increase. Notably, 3B produces 3,187 total issues (126 + 3,061) compared to 2,061 (232 + 1,829), representing a $\uparrow$54.6\% increase. While syntax errors decrease at 3B (126 \vs\ 232, $\downarrow$45.7\%), maintainability issues increase substantially (3,061 \vs\ 1,829, $\uparrow$67.4\%).

SonarCloud analysis for C\#$\rightarrow$Java reinforces these scale-dependent patterns. Aggregate quality issues across the 0.5B--3B scales decrease from 4,783 to 3,943 ($\downarrow$17.6\%), with particularly pronounced improvements in the 3B configuration. Breaking down by scale: 0.5B shows mixed results with total issues at 1,723 \vs\ 1,670 ($\uparrow$3.2\%), featuring increases in Maintainability (1,056 \vs\ 1,049, $\uparrow$0.7\%) and Cognitive Complexity (654 \vs\ 604, $\uparrow$8.3\%), while achieving reductions in Reliability (13 \vs\ 17, $\downarrow$23.5\%). The 1.5B configuration shows 1,785 \vs\ 1,683 ($\uparrow$6.1\%), with increases in Maintainability (1,088 \vs\ 1,046, $\uparrow$4.0\%), Cognitive Complexity (678 \vs\ 625, $\uparrow$8.5\%), and Reliability (19 \vs\ 12, $\uparrow$58.3\%). The 3B configuration achieves exceptional improvement with 435 \vs\ 1,430 ($\downarrow$69.6\%), demonstrating substantial reductions in Maintainability (278 \vs\ 777, $\downarrow$64.2\%), Reliability (0 \vs\ 12, $\downarrow$100\%), and Cognitive Complexity (157 \vs\ 641, $\downarrow$75.5\%). 


Code size metrics reveal mixed patterns across translation directions and scales. For Java$\rightarrow$C\#, \MT QLoRA generates comparable or reduced code: 0.5B produces 6,218 LoC \vs\ FFT's 6,230 ($\downarrow$0.2\%), 1.5B generates 6,369 \vs\ 6,671 ($\downarrow$4.5\%), and 3B produces 5,890 \vs\ 5,986 ($\downarrow$1.6\%). Token counts demonstrate consistent reductions: 0.5B uses 39,509 \vs\ 40,104 ($\downarrow$1.5\%), 1.5B uses 40,516 \vs\ 43,343 ($\downarrow$6.5\%), and 3B uses 37,716 \vs\ 38,418 ($\downarrow$1.8\%). For C\#$\rightarrow$Java, patterns vary: 0.5B produces 5,807 LoC \vs\ 5,506 ($\uparrow$5.5\%), 1.5B generates 5,745 \vs\ 5,784 ($\downarrow$0.7\%), and 3B produces 5,742 \vs\ 5,822 ($\downarrow$1.4\%). Token usage follows similar patterns: 0.5B uses 35,170 \vs\ 33,600 ($\uparrow$4.7\%), 1.5B uses 35,074 \vs\ 35,926 ($\downarrow$2.4\%), and 3B uses 34,862 \vs\ 35,831 ($\downarrow$2.7\%).

Following the approach used for code generation, we apply Wilcoxon signed-rank tests with Holm--Bonferroni correction ($\alpha=0.05$) to assess whether the quality of code produced by \MT QLoRA differs significantly from that produced by \MTfft.

For the Java$\rightarrow$C\# translation direction, the results are unambiguous: across all three model sizes, statistically significant differences in CodeBLEU scores are observed ($p<0.05$), consistently favoring \MTfft. These differences align with the previously observed gaps in functional correctness and static quality metrics, with corresponding effect sizes (Cliff's delta) indicating a \textit{negligible}-magnitude effect.

In contrast, the C\#$\rightarrow$Java direction exhibits a scale-dependent pattern. At 0.5B, \MTfft maintains a statistically significant advantage ($p<0.01$) with negligible effect size. This advantage disappears at 1.5B, where the two approaches achieve statistical parity, indicating no significant difference in performance. Notably, at 3B, the relationship reverses: \MT QLoRA demonstrates statistically significant superiority across both functional correctness and quality metrics ($p<0.01$). This result is consistent with the substantial improvements observed at this scale in static analysis outcomes and code conciseness, with effect sizes (Cliff's delta) suggesting a \textit{small-to-medium}-magnitude effect.\\

\begin{summarybox}[Key Findings:]
\noindent Code translation under \MT QLoRA exhibits strong scale- and direction-dependent behavior. For Java$\rightarrow$C\#, quality improvements in syntax are offset by increased maintainability issues, whereas for C\#$\rightarrow$Java, larger models achieve clear gains in both correctness and quality. 
Overall, across 0.5B--3B the benefits of parameter-efficient \MT fine-tuning emerge primarily at larger scales, underscoring the role of model capacity in balancing translation quality and efficiency\revtwo{; at 7B, however, this trend does not continue, and \MTfft outperforms \MTqlora in both directions}.
\end{summarybox}

\subsubsection{Code Summarization}
\label{sec:rq2_codesum}

We evaluate code summarization performance by comparing \MT QLoRA against \MT Full Fine-Tuning (\MTfft) using similarity-based metrics. \tabref{tab:rq2_codesum} reports results for both \java and \python across the 0.5B--3B scales.

\begin{table}[h]
\centering
\scriptsize
\caption{RQ2: Comparison of \MTfft and \MTqlora for code summarization on CoderEval. SIDE applies to \java only.}
\label{tab:rq2_codesum}
\resizebox{0.75\textwidth}{!}{%
\begin{tabular}{c|c|c|cccccc}
\toprule
\textbf{Lang} & \textbf{Model} & \textbf{Method} & \textbf{BLEU} & \textbf{METEOR} & \textbf{Rouge-L} & \textbf{chrF} & \textbf{BERTScore} & \textbf{SIDE} \\
\midrule
\multirow{9}{*}{\java}
& \multirow{2}{*}{\textit{QwenCoder-0.5B}} 
& \MTfft & \textbf{7.70\%} & 21.36\% & 33.29\% & \textbf{28.76\%} & \textbf{71.25\%} & 90.44\% \\
& & \MTqlora & 7.18\% & \textbf{21.80\%} & \textbf{34.09\%} & 27.71\% & 66.31\% & \textbf{91.75\%} \\
\cmidrule{2-9}
& \multirow{2}{*}{\textit{QwenCoder-1.5B}} 
& \MTfft & 7.43\% & 21.86\% & 33.73\% & 27.66\% & \textbf{69.79\%} & 89.68\% \\
& & \MTqlora & \textbf{8.12\%} & \textbf{22.14\%} & \textbf{34.15\%} & \textbf{29.23\%} & 66.78\% & \textbf{92.21\%} \\
\cmidrule{2-9}
& \multirow{2}{*}{\textit{QwenCoder-3B}} 
& \MTfft & \textbf{7.90}\% & 21.85\% & 33.59\% & 28.80\% & 65.90\% & 90.53\% \\
& & \MTqlora & 7.63\% & \textbf{22.94\%} & \textbf{34.93\%} & \textbf{29.31\%} & \textbf{66.40\%} & \textbf{91.58\%} \\
\cmidrule{2-9}
& \multirow{2}{*}{\rev{\textit{QwenCoder-7B}}}
& \rev{\MTfft} & \rev{6.77\%} & \rev{21.51\%} & \rev{33.54\%} & \rev{28.15\%} & \rev{65.96\%} & \rev{\textbf{90.90\%}} \\
& & \rev{\MTqlora} & \rev{\textbf{8.37\%}} & \rev{\textbf{22.75\%}} & \rev{\textbf{35.07\%}} & \rev{\textbf{30.73\%}} & \rev{\textbf{67.24\%}} & \rev{90.49\%} \\
\midrule
\multirow{9}{*}{\python}
& \multirow{2}{*}{\textit{QwenCoder-0.5B}} 
& \MTfft & 9.53\% & 20.28\% & 31.55\% & 29.14\% & 60.66\% & -- \\
& & \MTqlora & \textbf{19.43\%} & \textbf{29.53\%} & \textbf{40.67\%} & \textbf{39.16\%} & \textbf{65.48\%} & -- \\
\cmidrule{2-9}
& \multirow{2}{*}{\textit{QwenCoder-1.5B}} 
& \MTfft & 12.67\% & 24.36\% & 34.72\% & 33.28\% & 62.01\% & -- \\
& & \MTqlora & \textbf{22.90\%} & \textbf{33.07\%} & \textbf{44.50\%} & \textbf{42.28\%} & \textbf{66.84\%} & -- \\
\cmidrule{2-9}
& \multirow{2}{*}{\textit{QwenCoder-3B}} 
& \MTfft & 22.54\% & 32.91\% & 43.41\% & 42.27\% & 67.17\% & -- \\
& & \MTqlora & \textbf{29.90\%} & \textbf{40.06\%} & \textbf{47.43\%} & \textbf{48.20\%} & \textbf{68.59\%} & -- \\
\cmidrule{2-9}
& \multirow{2}{*}{\rev{\textit{QwenCoder-7B}}}
& \rev{\MTfft} & \rev{9.76\%} & \rev{20.46\%} & \rev{31.48\%} & \rev{28.53\%} & \rev{60.28\%} & -- \\
& & \rev{\MTqlora} & \rev{\textbf{30.80\%}} & \rev{\textbf{41.32\%}} & \rev{\textbf{51.12\%}} & \rev{\textbf{50.26\%}} & \rev{\textbf{70.15\%}} & -- \\
\bottomrule
\end{tabular}
}
\end{table}

\MT QLoRA exhibits clear language-dependent behavior when compared to \MT FFT. For \python, \MT QLoRA consistently outperforms \MTfft across all metrics and model scales. At 0.5B, \MT QLoRA achieves 19.43\% BLEU compared to \MTfft's 9.53\% ($\uparrow$103.9\%), 29.53\% METEOR \vs\ 20.28\% ($\uparrow$45.6\%), 40.67\% Rouge-L \vs\ 31.55\% ($\uparrow$28.9\%), 39.16\% chrF \vs\ 29.14\% ($\uparrow$34.4\%), and 65.48\% BERTScore \vs\ 60.66\% ($\uparrow$7.9\%). These advantages persist at 1.5B, where \MT QLoRA attains 22.90\% BLEU \vs\ 12.67\% ($\uparrow$80.7\%), 33.07\% METEOR \vs\ 24.36\% ($\uparrow$35.8\%), 44.50\% Rouge-L \vs\ 34.72\% ($\uparrow$28.2\%), 42.28\% chrF \vs\ 33.28\% ($\uparrow$27.0\%), and 66.84\% BERTScore \vs\ 62.01\% ($\uparrow$7.8\%). At 3B, \MT QLoRA continues to outperform, with BLEU reaching 29.90\% \vs\ 22.54\% ($\uparrow$32.7\%), METEOR 40.06\% \vs\ 32.91\% ($\uparrow$21.7\%), Rouge-L 47.43\% \vs\ 43.41\% ($\uparrow$9.3\%), chrF 48.20\% \vs\ 42.27\% ($\uparrow$14.0\%), and BERTScore 68.59\% \vs\ 67.17\% ($\uparrow$2.1\%).

For Java, differences between \MT QLoRA and \MT FFT are mixed across configurations, with each approach showing advantages on different metrics. At 0.5B, \MT QLoRA shows heterogeneous outcomes: METEOR improves by $\uparrow$2.1\% (21.80\% \vs\ 21.36\%), Rouge-L by $\uparrow$2.4\% (34.09\% \vs\ 33.29\%), and SIDE by $\uparrow$1.4\% (91.75\% \vs\ 90.44\%). However, \MTfft achieves higher BLEU ($\uparrow$7.2\%; 7.70\% \vs\ 7.18\%), chrF ($\uparrow$3.8\%; 28.76\% \vs\ 27.71\%), and notably BERTScore ($\uparrow$7.4\%; 71.25\% \vs\ 66.31\%). At 1.5B, \MT QLoRA exhibits improvements on most metrics: BLEU of 8.12\% \vs\ 7.43\% ($\uparrow$9.3\%), METEOR of 22.14\% \vs\ 21.86\% ($\uparrow$1.3\%), Rouge-L of 34.15\% \vs\ 33.73\% ($\uparrow$1.2\%), chrF of 29.23\% \vs\ 27.66\% ($\uparrow$5.7\%), and SIDE of 92.21\% \vs\ 89.68\% ($\uparrow$2.8\%), while \MTfft maintains an advantage in BERTScore (69.79\% \vs\ 66.78\%; $\uparrow$4.5\%). At 3B, the pattern continues with \MT QLoRA achieving higher METEOR ($\uparrow$5.0\%; 22.94\% \vs\ 21.85\%), Rouge-L ($\uparrow$4.0\%; 34.93\% \vs\ 33.59\%), chrF ($\uparrow$1.8\%; 29.31\% \vs\ 28.80\%), BERTScore ($\uparrow$0.8\%; 66.40\% \vs\ 65.90\%), and SIDE ($\uparrow$1.2\%; 91.58\% \vs\ 90.53\%), while \MTfft shows a slight edge in BLEU ($\uparrow$3.5\%; 7.90\% \vs\ 7.63\%). Overall, while \MT QLoRA demonstrates consistent advantages on overlap-based metrics (METEOR, Rouge-L) and the code-aware SIDE metric for \java, \MTfft tends to achieve comparable or better performance on lexical precision metrics, particularly BERTScore at smaller scales.

Scaling behavior further highlights these language-dependent effects. For \python, \MT QLoRA benefits remain substantial across model capacities, though the relative advantage over \MTfft narrows at larger scales as both approaches improve. From 0.5B to 3B under \MT QLoRA, BLEU improves from 19.43\% to 29.90\% ($\uparrow$53.9\%), METEOR from 29.53\% to 40.06\% ($\uparrow$35.7\%), Rouge-L from 40.67\% to 47.43\% ($\uparrow$16.6\%), chrF from 39.16\% to 48.20\% ($\uparrow$23.1\%), and BERTScore from 65.48\% to 68.59\% ($\uparrow$4.7\%). \MT FFT also exhibits strong scaling, with BLEU improving from 9.53\% to 22.54\% ($\uparrow$136.5\%), yet \MT QLoRA maintains its advantage across all scales. For Java, both approaches show relatively stable performance across model sizes, with differences between training strategies remaining within a narrow range and neither approach establishing consistent dominance across all metrics.

To complement similarity-based evaluation with qualitative assessment, we employ the LLM-as-a-judge methodology described in RQ$_1$ (\secref{sec:rq1_codesum}). Complete distributions of GPT-5 mini quality scores across all configurations are reported in the Appendix~\ref{appendix:llm_judge}. Statistical significance is assessed using Wilcoxon signed-rank tests with Holm--Bonferroni correction ($\alpha=0.05$), comparing \MT QLoRA against \MTfft for both functional and quality metrics.

For \java, Content Adequacy scores indicate that \MT QLoRA maintains summary quality comparable to \MTfft across all model sizes. \MT QLoRA achieves scores of 2.178 \vs\ 2.139 at 0.5B ($\uparrow$1.8\%), 2.224 \vs\ 2.198 at 1.5B ($\uparrow$1.2\%), and 2.293 \vs\ 2.228 at 3B ($\uparrow$2.9\%), corresponding to approximately 73--77\% of the human baseline (2.960). Conciseness scores remain consistently high for \MT QLoRA (4.925--4.980), marginally exceeding both \MTfft (4.925--4.954) and the human baseline (4.620), while Fluency scores show small, scale-dependent variations. Importantly, none of these differences are statistically significant across metrics or model sizes, indicating performance parity between \MT QLoRA and full fine-tuning for \java summarization.

For \python, qualitative assessment reveals a different pattern. Content Adequacy scores for \MT QLoRA and \MTfft remain close across all model sizes (1.818 \vs\ 1.806 at 0.5B, 1.862 \vs\ 1.894 at 1.5B, and 1.938 \vs\ 1.962 at 3B), corresponding to 69--74\% of the human baseline (2.636). Conciseness scores are consistently high for both approaches (4.624--4.730 for \MT QLoRA and 4.626--4.670 for \MTfft), approaching the human baseline (4.800), while Fluency scores remain comparable and below human quality but maintain acceptable readability. Despite the similarity in absolute scores, statistical testing reveals significant \MT QLoRA advantages across all qualitative metrics and model scales ($p<0.05$)\footnote{Complete statistical results are provided in the replication package \cite{replication}.}.

Finally, an interesting pattern emerges: QLoRA-based \MT optimization can improve automatic metrics without proportionally changing human assessments. The 0.5B model illustrates this clearly--\MT QLoRA increases BLEU by 103.9\% over \MTfft (19.43\% \vs\ 9.53\%), while Content Adequacy remains flat (1.818 \vs\ 1.806). \MTqlora appears to enhance lexical alignment in ways that automatic metrics detect but human evaluators don't perceive as proportional quality improvements.  A detailed breakdown of statistical results and metric-level analyses is provided in the Appendix \ref{appendix:llm_judge}. \\


\begin{summarybox}[Key Findings:]
\noindent Code summarization under \MT QLoRA shows clear language-dependent effects. \MT QLoRA yields substantial improvements for \python across all similarity-based metrics, with gains ranging from 2.1\% to 103.9\% depending on metric and scale. For \java, results are more nuanced: \MT QLoRA demonstrates advantages on overlap-based metrics (METEOR, Rouge-L) and the code-aware SIDE metric, while \MTfft achieves comparable or better performance on lexical precision metrics like BERTScore. LLM-as-a-judge evaluation confirms that \MT QLoRA preserves content adequacy, conciseness, and fluency for both languages. Overall, \MTqlora strongly benefits \python summarization while achieving competitive performance with full fine-tuning for Java.
\end{summarybox}

\rev{\noindent\textbf{Extension to 7B scale:}
    \revtwo{The 7B extension confirms the QLoRA-over-full-fine-tuning advantage for code generation and \python summarization, while revealing a clear limitation for translation, rather than uniformly reinforcing the smaller-scale pattern.} For code generation, \MTqlora outperforms \MTfft in both languages: on \java, \MTqlora reaches Pass@1 33.70\% \vs\ 18.48\% for \MTfft ($\uparrow$82.4\%); on \python, \MTqlora reaches 20.53\% \vs\ 14.74\% for \MTfft ($\uparrow$39.3\%). For code summarization, \MTqlora preserves its \python advantage and remains comparable on \java: on \python, \MTqlora reaches BLEU 30.80\% \vs\ 9.76\% for \MTfft ($\uparrow$215.6\%); on \java, the two configurations remain close across metrics, with \MTqlora ahead on overlap-based scores (BLEU 8.37\% \vs\ 6.77\%, $\uparrow$23.6\%; chrF 30.73\% \vs\ 28.15\%, $\uparrow$9.2\%). 
    \revtwo{For code translation, the gap widens substantially at 7B compared to 0.5B--3B: \MTfft outperforms \MTqlora by $\uparrow$30.2\% on Java$\rightarrow$C\# (72.10\% vs. 55.37\%) and $\uparrow$41.6\% on C\#$\rightarrow$Java (70.54\% vs. 49.83\%; Tables~\ref{tab:rq2_codetrans_java2csharp} and~\ref{tab:rq2_codetrans_csharp2java}), in contrast to within-8\% gaps at smaller scales. Overall, the 7B extension reinforces the central RQ2 finding for code generation and \python summarization, where \MTqlora matches or substantially exceeds \MTfft at scale, while revealing a clear limitation for translation at this scale; we regard characterizing this translation degradation as an open question warranting dedicated investigation (\secref{sec:conclusion})}.
}

\begin{table*}[h]
\centering
\scriptsize
\caption{\rev{Efficiency comparison of \MTfft and \MTqlora across model scales. Percentages are computed against the full base model parameter count.}}
\label{tab:efficiency-comparison}
\resizebox{0.7\textwidth}{!}{%
\begin{tabular}{l|cc|cc|c}
\toprule
\multirow{2}{*}{\rev{\textbf{Model}}} &
\multicolumn{2}{c|}{\rev{\textbf{Peak GPU Memory (GB)}}} &
\multicolumn{2}{c|}{\rev{\textbf{Trainable Parameters}}} &
\multirow{2}{*}{\rev{\textbf{Param Reduction}}} \\
\cmidrule(lr){2-3} \cmidrule(lr){4-5}
& \rev{\textbf{\MTfft}} & \rev{\textbf{\MTqlora}} & \rev{\textbf{\MTfft}} & \rev{\textbf{\MTqlora}} & \\
\midrule
\rev{\textit{QwenCoder-1.5B}} & \rev{13.04} & \rev{10.48} & \rev{1.54B (100\%)} & \rev{9.2M (0.60\%)}  & \rev{167$\times$} \\
\rev{\textit{QwenCoder-3B}}   & \rev{23.28} & \rev{15.68} & \rev{3.09B (100\%)} & \rev{15.0M (0.49\%)} & \rev{206$\times$} \\
\rev{\textit{QwenCoder-7B}}   & \rev{39.92} & \rev{24.15} & \rev{7.62B (100\%)} & \rev{20.2M (0.27\%)} & \rev{377$\times$} \\
\bottomrule
\end{tabular}
}
\end{table*}

\rev{\noindent\textbf{Efficiency comparison.}} 
\rev{Having compared \MTqlora and \MTfft on performance and quality outcomes across all three model scales --- including the 7B extension reported above --- we now turn to the efficiency side of the same comparison. Building on the adapter configuration specified in \tabref{tab:qlora_config}, \tabref{tab:efficiency-comparison} reports the corresponding training-cost profile of the two paradigms at all three model scales.
Two things stand out. \MTqlora reduces the trainable parameter count from $167\times$ at 1.5B to $377\times$ at 7B, with the reduction strengthening as the model grows: at 7B, fewer than $0.3\%$ of the model's parameters are updated during training, and the remaining $99.7\%$ are held fixed in 4-bit quantized form. On memory, \MTfft peak memory grows from 13.04~GB at 1.5B to 39.92~GB at 7B, while \MTqlora stays substantially lower across the entire range, peaking at 24.15~GB at 7B. Combined with the parameter reduction, this profile makes \MTqlora practical to train at the scales evaluated in this study (1.5B--7B) under more constrained hardware budgets than \MTfft.}

\subsection{\rev{Ablation Studies}}
\label{sec:ablations}

\rev{The RQ$_1$ and RQ$_2$ findings rest on two design choices: the joint training of all three tasks, and the proportional sampling that follows from the dataset sizes. We isolate each through a dedicated ablation: a balanced-sampling comparison (Section~\ref{sec:balanced-ablation}) that disentangles the contribution of training-data composition, and a pairwise task-pair analysis (Section~\ref{sec:pairwise-ablation}) that disentangles the contribution of individual task interactions.}

\subsubsection{\rev{\textbf{Balanced vs. Proportional Task Sampling}}}
\label{sec:balanced-ablation}

\rev{The \MT analyses in RQ$_1$ and RQ$_2$ use proportional sampling, in which the model encounters examples from each task with probability proportional to the task's share of the combined training corpus. Because code translation contributes $\sim$20K instances versus $\sim$416K each for generation and summarization, the model sees translation examples roughly twenty times less often than generation or summarization examples per epoch. To disentangle whether the translation-direction weaknesses observed under \MTqlora reflect genuine cross-task interference or simply this exposure imbalance, we re-train the 3B \MTqlora configuration with \emph{balanced} sampling: each task contributes an equal number of training instances per epoch, matched to the smallest task ($\sim$20K instances). All other training hyperparameters --- adapter rank, learning rate schedule, optimizer, and early stopping --- are identical to the proportional baseline, isolating the effect of sampling strategy. Results are reported by task in \tabref{tab:balanced-sampling-cg-python}, \tabref{tab:balanced-sampling-cg-java}, \tabref{tab:balanced-sampling-ct-java2csharp}, \tabref{tab:balanced-sampling-ct-csharp2java}, and \tabref{tab:balanced-sampling}.}


\begin{table}[h]
\centering
\scriptsize
\caption{\rev{Ablation: Comparison of proportional and balanced task sampling for \MTqlora at the 3B scale on CoderEval for Python code generation.}}
\label{tab:balanced-sampling-cg-python}
\resizebox{0.9\textwidth}{!}{%
\begin{tabular}{lc|c|cccc|cccc|cccc}
\toprule
\multirow{2}{*}{\rev{\textbf{Model}}} & \multirow{2}{*}{\rev{\textbf{Method}}} & 
\multirow{2}{*}{\rev{\textbf{Pass@1}}} & 
\multicolumn{4}{c|}{\rev{\textbf{Lizard}}} & 
\multicolumn{4}{c|}{\rev{\textbf{Pylint}}} &
\multicolumn{4}{c}{\rev{\textbf{SonarCloud}}} \\
\cmidrule(lr){4-7} \cmidrule(lr){8-11} \cmidrule(l){12-15}
& & & \rev{\textbf{LoC}} & \rev{\textbf{Tok}} & \rev{\textbf{DR}} & \rev{\textbf{CyC}} & 
\rev{\textbf{Err}} & \rev{\textbf{Warn}} & \rev{\textbf{Conv}} & \rev{\textbf{Ref}} &
\rev{\textbf{Sec}} & \rev{\textbf{Rel}} & \rev{\textbf{Main}} & \rev{\textbf{CoC}} \\
\midrule
\multirow{2}{*}{\rev{\textit{Qwen-3B}}} 
& \rev{\MTqlora\ (proportional)} & \rev{21.05\%} & \rev{1,759} & \rev{11,773} & \rev{100\%} & \rev{677} & \rev{269} & \rev{1,521} & \rev{673} & \rev{56} & \rev{2} & \rev{5} & \rev{106} & \rev{549} \\
& \rev{\MTqlora\ (balanced)} & \rev{\textbf{22.11\%}} & \rev{1,725 $\downarrow$} & \rev{11,643 $\downarrow$} & \cellcolor{yellow!20}\rev{100\%} & \cellcolor{green!20}\rev{630} & \cellcolor{green!20}\rev{211} & \cellcolor{green!20}\rev{1,339} & \cellcolor{green!20}\rev{644} & \cellcolor{green!20}\rev{39} & \cellcolor{blue!20}\rev{3} & \cellcolor{green!20}\rev{2} & \cellcolor{green!20}\rev{72} & \cellcolor{green!20}\rev{537} \\
\bottomrule
\end{tabular}
}
\end{table}

\begin{table}[h]
\centering
\scriptsize
\caption{\rev{Ablation: Comparison of proportional and balanced task sampling for \MTqlora at the 3B scale on CoderEval for \java code generation.}}
\label{tab:balanced-sampling-cg-java}
\resizebox{0.9\textwidth}{!}{%
\begin{tabular}{lc|c|cccc|cccccc|cccc}
\toprule
\multirow{2}{*}{\rev{\textbf{Model}}} & \multirow{2}{*}{\rev{\textbf{Method}}} & 
\multirow{2}{*}{\rev{\textbf{Pass@1}}} & 
\multicolumn{4}{c|}{\rev{\textbf{Lizard}}} & 
\multicolumn{6}{c|}{\rev{\textbf{PMD}}} &
\multicolumn{4}{c}{\rev{\textbf{SonarCloud}}} \\
\cmidrule(lr){4-7} \cmidrule(lr){8-13} \cmidrule(l){14-17}
& & & \rev{\textbf{LoC}} & \rev{\textbf{Tok}} & \rev{\textbf{DR}} & \rev{\textbf{CyC}} & 
\rev{\textbf{Best Prac.}} & \rev{\textbf{CS}} & \rev{\textbf{Design}} & \rev{\textbf{EP}} & \rev{\textbf{MT}} & \rev{\textbf{Perf.}} &
\rev{\textbf{Sec}} & \rev{\textbf{Rel}} & \rev{\textbf{Main}} & \rev{\textbf{CoC}} \\
\midrule
\multirow{2}{*}{\rev{\textit{Qwen-3B}}} 
& \rev{\MTqlora\ (proportional)} & \rev{\textbf{32.07\%}} & \rev{1,580} & \rev{10,920} & \rev{98.40\%} & \rev{558} & \rev{86} & \rev{1,184} & \rev{128} & \rev{32} & \rev{4} & \rev{7} & \rev{0} & \rev{2} & \rev{541} & \rev{469} \\
& \rev{\MTqlora\ (balanced)} & \rev{30.98\%} & \rev{1,675 $\uparrow$} & \rev{11,216 $\uparrow$} & \cellcolor{green!20}\rev{99.5\%} & \cellcolor{blue!20}\rev{607} & \cellcolor{green!20}\rev{81} & \cellcolor{blue!20}\rev{1,213} & \cellcolor{green!20}\rev{127} & \cellcolor{green!20}\rev{31} & \cellcolor{yellow!20}\rev{4} & \cellcolor{blue!20}\rev{15} & \rev{0} & \cellcolor{blue!20}\rev{3} & \cellcolor{green!20}\rev{540} & \cellcolor{green!20}\rev{459} \\
\bottomrule
\end{tabular}
}
\end{table}

\begin{table}[h]
\centering
\scriptsize
\caption{\rev{Ablation: Comparison of proportional and balanced task sampling for \MTqlora at the 3B scale on CodeXGLUE for Java$\rightarrow$C\# code translation.}}
\label{tab:balanced-sampling-ct-java2csharp}
\resizebox{0.85\textwidth}{!}{%
\begin{tabular}{lc|c|cccc|cc}
\toprule
\multirow{2}{*}{\rev{\textbf{Model}}} & \multirow{2}{*}{\rev{\textbf{Method}}} & 
\multirow{2}{*}{\rev{\textbf{CodeBLEU}}} & 
\multicolumn{4}{c|}{\rev{\textbf{Lizard}}} & 
\multicolumn{2}{c}{\rev{\textbf{Roslyn}}} \\
\cmidrule(lr){4-7} \cmidrule(l){8-9}
& & & \rev{\textbf{LoC}} & \rev{\textbf{Tok}} & \rev{\textbf{DR}} & \rev{\textbf{CyC}} & 
\rev{\textbf{Syntax Errors}} & \rev{\textbf{Maintainability}} \\
\midrule
\multirow{2}{*}{\rev{\textit{Qwen-3B}}} 
& \rev{\MTqlora\ (proportional)} & \rev{62.88\%} & \rev{5,890} & \rev{37,716} & \rev{93.6\%} & \rev{1,455} & \rev{126} & \rev{3,061} \\
& \rev{\MTqlora\ (balanced)} & \rev{\textbf{66.12\%}} & \rev{6,373 $\uparrow$} & \rev{40,974 $\uparrow$} & \cellcolor{green!20}\rev{98\%} & \cellcolor{blue!20}\rev{1,552} & \cellcolor{green!20}\rev{33} & \cellcolor{green!20}\rev{1,712} \\
\bottomrule
\end{tabular}
}
\end{table}

\begin{table}[h]
\centering
\scriptsize
\caption{\rev{Ablation: Comparison of proportional and balanced task sampling for \MTqlora at the 3B scale on CodeXGLUE for C\#$\rightarrow$Java code translation.}}
\label{tab:balanced-sampling-ct-csharp2java}
\resizebox{0.9\textwidth}{!}{%
\begin{tabular}{lc|c|cccc|cccccc|cccc}
\toprule
\multirow{2}{*}{\rev{\textbf{Model}}} & \multirow{2}{*}{\rev{\textbf{Method}}} & 
\multirow{2}{*}{\rev{\textbf{CodeBLEU}}} & 
\multicolumn{4}{c|}{\rev{\textbf{Lizard}}} & 
\multicolumn{6}{c|}{\rev{\textbf{PMD}}} &
\multicolumn{4}{c}{\rev{\textbf{SonarCloud}}} \\
\cmidrule(lr){4-7} \cmidrule(lr){8-13} \cmidrule(l){14-17}
& & & \rev{\textbf{LoC}} & \rev{\textbf{Tok}} & \rev{\textbf{DR}} & \rev{\textbf{CyC}} & 
\rev{\textbf{Best Prac.}} & \rev{\textbf{CS}} & \rev{\textbf{Design}} & \rev{\textbf{EP}} & \rev{\textbf{MT}} & \rev{\textbf{Perf.}} &
\rev{\textbf{Sec}} & \rev{\textbf{Rel}} & \rev{\textbf{Main}} & \rev{\textbf{CoC}} \\
\midrule
\multirow{2}{*}{\rev{\textit{Qwen-3B}}} 
& \rev{\MTqlora\ (proportional)} & \rev{\textbf{71.22\%}} & \rev{5,742} & \rev{34,862} & \rev{99.5\%} & \rev{1,612} & \rev{37} & \rev{1,072} & \rev{45} & \rev{11} & \rev{5} & \rev{1} & \rev{0} & \rev{0} & \rev{278} & \rev{157} \\
& \rev{\MTqlora\ (balanced)} & \rev{65.56\%} & \rev{5,172 $\downarrow$} & \rev{31,995 $\downarrow$} & \cellcolor{blue!20}\rev{95.3\%} & \cellcolor{green!20}\rev{1,456} & \cellcolor{blue!20}\rev{81} & \cellcolor{blue!20}\rev{1,213} & \cellcolor{blue!20}\rev{127} & \cellcolor{blue!20}\rev{31} & \cellcolor{green!20}\rev{4} & \cellcolor{blue!20}\rev{15} & \rev{0} & \rev{0} & \cellcolor{green!20}\rev{230} & \cellcolor{green!20}\rev{24} \\
\bottomrule
\end{tabular}
}
\end{table}

\begin{table}[h]
\centering
\scriptsize
\caption{\rev{Ablation: Comparison of proportional and balanced task sampling for \MTqlora at the 3B scale on CoderEval. SIDE applies to Java only.}}
\label{tab:balanced-sampling}
\resizebox{0.75\textwidth}{!}{%
\begin{tabular}{c|c|c|cccccc}
\toprule
\rev{\textbf{Lang}} & \rev{\textbf{Model}} & \rev{\textbf{Method}} & \rev{\textbf{BLEU}} & \rev{\textbf{METEOR}} & \rev{\textbf{Rouge-L}} & \rev{\textbf{chrF}} & \rev{\textbf{BERTScore}} & \rev{\textbf{SIDE}} \\
\midrule
\multirow{2}{*}{\rev{Java}}
& \multirow{2}{*}{\rev{\textit{Qwen-3B}}} 
& \rev{\MTqlora\ (proportional)} & \rev{7.63\%} & \rev{22.94\%} & \rev{34.93\%} & \rev{29.31\%} & \rev{66.40\%} & \rev{\textbf{91.58\%}} \\
& & \rev{\MTqlora\ (balanced)} & \rev{\textbf{8.31\%}} & \rev{\textbf{25.27\%}} & \rev{\textbf{36.14\%}} & \rev{\textbf{30.28\%}} & \rev{\textbf{67.34\%}} & \rev{90.53\%} \\
\midrule
\multirow{2}{*}{\rev{Python}}
& \multirow{2}{*}{\rev{\textit{Qwen-3B}}} 
& \rev{\MTqlora\ (proportional)} & \rev{29.90\%} & \rev{40.06\%} & \rev{47.43\%} & \rev{48.20\%} & \rev{68.59\%} & \rev{--} \\
& & \rev{\MTqlora\ (balanced)} & \rev{\textbf{35.43\%}} & \rev{\textbf{42.62\%}} & \rev{\textbf{52.61\%}} & \rev{\textbf{51.91\%}} & \rev{\textbf{71.44\%}} & \rev{--} \\
\bottomrule
\end{tabular}
}
\vspace{-10pt}
\end{table}

\rev{\textbf{Code Generation.} For code generation, the two strategies yield broadly comparable performance. \python \passatone improves slightly under balanced sampling (22.11\% vs.\ 21.05\%, $\uparrow$5.0\% relative; \tabref{tab:balanced-sampling-cg-python}), with corresponding reductions in Cyclomatic Complexity (630 vs.\ 677, $\downarrow$6.9\%), Pylint warnings (1,339 vs.\ 1,521, $\downarrow$12.0\%), and SonarCloud maintainability (72 vs.\ 106, $\downarrow$32.1\%). \java \passatone declines slightly (30.98\% vs.\ 32.07\%, $\downarrow$3.4\% relative; \tabref{tab:balanced-sampling-cg-java}), with Cyclomatic Complexity increasing modestly (607 vs.\ 558, $\uparrow$8.8\%) and other quality metrics remaining within a few percent of the proportional baseline. The roughly neutral generation outcome is expected given that code generation already contributes the largest single share of the proportional corpus; reducing its weight does not deprive the model of generation-specific learning to the extent that translation under-representation harms translation learning.}

\rev{\textbf{Code Translation.} The effect of data imbalance is most directly testable on translation, where the gap between task corpora is most pronounced. For \java$\rightarrow$C\# (\tabref{tab:balanced-sampling-ct-java2csharp}), balanced sampling improves CodeBLEU from 62.88\% to 66.12\% ($\uparrow$5.2\% relative), with substantial reductions in Roslyn-detected syntax errors (126 to 33, $\downarrow$73.8\%) and maintainability findings (3,061 to 1,712, $\downarrow$44.0\%). The detection rate rises from 93.6\% to 98.0\%, indicating that balanced exposure produces more analyzable C\# output. These improvements narrow the gap between \MTqlora and \STqlora on \java$\rightarrow$C\# from $-5.9\%$ to $-1.1\%$, confirming that data imbalance was a meaningful contributor to the \java$\rightarrow$C\# weakness observed in RQ$_1$.}


\rev{The C\#$\rightarrow$\java direction (\tabref{tab:balanced-sampling-ct-csharp2java}), however, exhibits the opposite pattern: balanced sampling \emph{reduces} CodeBLEU from 71.22\% to 65.56\% ($\downarrow$7.9\% relative), with detection rate dropping from 99.5\% to 95.3\%. PMD violations rise substantially across most categories (\eg Best Practices from 37 to 81, $\uparrow$118.9\%; Design from 45 to 127, $\uparrow$182.2\%; Performance from 1 to 15), while SonarCloud cognitive complexity drops sharply (157 to 24). This direction-asymmetric behavior indicates that the under-representation of translation data is not uniformly disadvantageous: \java appears as a target language in both code generation and code summarization training data, while C\# appears only in translation, so the proportional mixture's higher exposure to \java content during non-translation training is consistent with the stronger C\#$\rightarrow$\java performance observed under proportional sampling. We report this as an observation rather than a causal claim, since other explanations (\eg asymmetries in tokenizer coverage or in the test-set distribution) remain compatible with the data.}

\rev{\textbf{Code Summarization.} Summarization shows the largest gains under balanced sampling, despite already contributing the majority of the proportional training corpus. \tabref{tab:balanced-sampling} shows that \python summarization improves substantially across all metrics: BLEU rises from 29.90\% to 35.43\% ($\uparrow$18.5\%), METEOR from 40.06\% to 42.62\% ($\uparrow$6.4\%), Rouge-L from 47.43\% to 52.61\% ($\uparrow$10.9\%), chrF from 48.20\% to 51.91\% ($\uparrow$7.7\%), and BERTScore from 68.59\% to 71.44\% ($\uparrow$4.2\%). \java summarization improves more modestly but consistently: BLEU rises from 7.63\% to 8.31\% ($\uparrow$8.9\%), METEOR from 22.94\% to 25.27\% ($\uparrow$10.2\%), Rouge-L from 34.93\% to 36.14\% ($\uparrow$3.5\%), and chrF from 29.31\% to 30.28\% ($\uparrow$3.3\%), while SIDE remains essentially stable (90.53\% vs.\ 91.58\%, $\downarrow$1.1\%). These improvements suggest that absolute training-set size is not the binding constraint for summarization quality at this scale: even though the proportional mixture provides far more summarization examples in raw count, re-balancing the mixture to give translation a larger relative share yields better summarization metrics. We do not attribute this to a specific mechanism (\eg interference in the shared adapter), since the ablation varies only the sampling strategy and does not isolate the underlying cause.}

\rev{\textbf{Implications.} Taken together, these results show that data imbalance is a partial but incomplete explanation for the translation weakness observed under proportional \MTqlora in RQ$_1$. Balanced sampling clearly helps \java$\rightarrow$C\# translation (narrowing the gap with \STqlora from $-5.9\%$ to $-1.1\%$) and \python summarization (BLEU $+18.5\%$), confirming that under-exposure to translation data contributed to those specific weaknesses. However, the same intervention \emph{degrades} C\#$\rightarrow$\java translation by $7.9\%$, indicating that the proportional mixture also provided a benefit that balanced sampling removes --- most likely the broader exposure to \java content present in the generation and summarization corpora, though we report this as an observation rather than a verified causal mechanism. Translation is therefore the task most sensitive to data composition in our experiments, but not in a way that a single sampling strategy resolves uniformly: each strategy helps one direction and hurts the other. This sensitivity is consistent with the high translation variance observed across scales in RQ$_1$ and RQ$_2$. We retain proportional sampling as the primary configuration in RQ$_1$ and RQ$_2$ --- it remains the most widely used \MT training strategy in the SE literature~\cite{ciniselli2021empirical, mastropaolo2022using, mastropaolo2021studying} --- and report the balanced ablation explicitly so that the underlying trade-off, rather than a single best-strategy claim, is visible.}

\subsubsection{\rev{\textbf{Task-Pair Ablation: Positive and Negative Transfer}}}
\label{sec:pairwise-ablation}

\rev{The three-task analysis in Sections~\ref{sec:rq1} and~\ref{sec:rq2} establishes that \MTqlora is competitive with both \STqlora and \MTfft across all evaluated configurations, but it does not isolate which task interactions drive that behavior. To examine whether the observed transfer is uniformly beneficial or whether specific task combinations exhibit positive or negative interference, we trained three additional 3B QLoRA variants restricted to two-task subsets: code generation with translation (CG+CT), code generation with summarization (CG+CS), and summarization with translation (CS+CT). All other training hyperparameters --- adapter rank, learning rate schedule, optimizer, and early stopping --- match the three-task baseline reported in RQ$_1$, isolating the effect of task composition. We restrict the ablation to the 3B scale: the three-task evaluation already exhibits clear scale-dependent patterns at this size, and the 7B-scale results in RQ$_1$ and RQ$_2$ confirm that the broader findings hold at larger capacity. Results are reported by task, considering both functional correctness and code quality.}

\rev{\textbf{Code Generation.} Functional correctness reveals language-dependent transfer patterns. For \python (\tabref{tab:rq3_codegen_python}), \MTqlora (CG+CT) matches the three-task baseline at 21.05\% \passatone, while \MTqlora (CG+CS) drops to 19.47\% vs.\ 21.05\% ($\downarrow$7.5\%), suggesting mild negative interference when summarization replaces translation in the mixture. For \java (\tabref{tab:rq3_codegen_java}), both two-task variants outperform the three-task baseline: \MTqlora (CG+CT) reaches 36.41\% vs.\ 32.07\% ($\uparrow$13.5\%), and \MTqlora (CG+CS) reaches 34.23\% vs.\ 32.07\% ($\uparrow$6.7\%). The asymmetry is consistent with the broader RQ$_1$ pattern in which \java code generation is more responsive to multi-task composition than \python.}

\begin{table}[h]
\centering
\scriptsize
\caption{\rev{Ablation: Comparison of \MTqlora task-pair variants for Python code generation on CoderEval.}}
\label{tab:rq3_codegen_python}
\resizebox{0.9\textwidth}{!}{%
\begin{tabular}{lc|c|cccc|cccc|cccc}
\toprule
\multirow{2}{*}{\rev{\textbf{Model}}} & \multirow{2}{*}{\rev{\textbf{Method}}} & 
\multirow{2}{*}{\rev{\textbf{Pass@1}}} & 
\multicolumn{4}{c|}{\rev{\textbf{Lizard}}} & 
\multicolumn{4}{c|}{\rev{\textbf{Pylint}}} &
\multicolumn{4}{c}{\rev{\textbf{SonarCloud}}} \\
\cmidrule(lr){4-7} \cmidrule(lr){8-11} \cmidrule(l){12-15}
& & & \rev{\textbf{LoC}} & \rev{\textbf{Tok}} & \rev{\textbf{DR}} & \rev{\textbf{CyC}} & 
\rev{\textbf{Err}} & \rev{\textbf{Warn}} & \rev{\textbf{Conv}} & \rev{\textbf{Ref}} &
\rev{\textbf{Sec}} & \rev{\textbf{Rel}} & \rev{\textbf{Main}} & \rev{\textbf{CoC}} \\
\midrule
\multirow{3}{*}{\rev{\textit{Qwen-3B}}} 
& \rev{\MTqlora\ (3-task)} & \rev{\textbf{21.05\%}} & \rev{1,759} & \rev{11,773} & \rev{100\%} & \rev{677} & \rev{269} & \rev{1,521} & \rev{673} & \rev{56} & \rev{2} & \rev{5} & \rev{106} & \rev{549} \\
& \rev{\MTqlora\ (CG+CT)} & \rev{\textbf{21.05\%}} & \rev{1,761 $\uparrow$} & \rev{11,202 $\downarrow$} & \cellcolor{blue!20}\rev{99.5\%} & \cellcolor{green!20}\rev{656} & \cellcolor{green!20}\rev{231} & \cellcolor{green!20}\rev{1,457} & \cellcolor{blue!20}\rev{678} & \cellcolor{green!20}\rev{51} & \cellcolor{green!20}\rev{1} & \cellcolor{green!20}\rev{2} & \cellcolor{green!20}\rev{73} & \cellcolor{green!20}\rev{544} \\
& \rev{\MTqlora\ (CG+CS)} & \rev{19.47\%} & \rev{1,703 $\downarrow$} & \rev{10,461 $\downarrow$} & \cellcolor{yellow!20}\rev{100\%} & \cellcolor{green!20}\rev{632} & \cellcolor{green!20}\rev{239} & \cellcolor{green!20}\rev{1,377} & \cellcolor{green!20}\rev{666} & \cellcolor{blue!20}\rev{59} & \rev{2} & \cellcolor{green!20}\rev{2} & \cellcolor{green!20}\rev{35} & \cellcolor{green!20}\rev{510} \\
\bottomrule
\end{tabular}
}
\end{table}

\begin{table}[h]
\centering
\scriptsize
\caption{\rev{Ablation: Comparison of \MTqlora task-pair variants for Java code generation on CoderEval.}}
\label{tab:rq3_codegen_java}
\resizebox{0.9\textwidth}{!}{%
\begin{tabular}{lc|c|cccc|cccccc|cccc}
\toprule
\multirow{2}{*}{\rev{\textbf{Model}}} & \multirow{2}{*}{\rev{\textbf{Method}}} & 
\multirow{2}{*}{\rev{\textbf{Pass@1}}} & 
\multicolumn{4}{c|}{\rev{\textbf{Lizard}}} & 
\multicolumn{6}{c|}{\rev{\textbf{PMD}}} &
\multicolumn{4}{c}{\rev{\textbf{SonarCloud}}} \\
\cmidrule(lr){4-7} \cmidrule(lr){8-13} \cmidrule(l){14-17}
& & & \rev{\textbf{LoC}} & \rev{\textbf{Tok}} & \rev{\textbf{DR}} & \rev{\textbf{CyC}} & 
\rev{\textbf{Best Prac.}} & \rev{\textbf{CS}} & \rev{\textbf{Design}} & \rev{\textbf{EP}} & \rev{\textbf{MT}} & \rev{\textbf{Perf.}} &
\rev{\textbf{Sec}} & \rev{\textbf{Rel}} & \rev{\textbf{Main}} & \rev{\textbf{CoC}} \\
\midrule
\multirow{3}{*}{\rev{\textit{Qwen-3B}}} 
& \rev{\MTqlora\ (3-task)} & \rev{32.07\%} & \rev{1,580} & \rev{10,920} & \rev{98.40\%} & \rev{558} & \rev{86} & \rev{1,184} & \rev{128} & \rev{32} & \rev{4} & \rev{7} & \rev{0} & \rev{2} & \rev{541} & \rev{469} \\
& \rev{\MTqlora\ (CG+CT)} & \rev{\textbf{36.41\%}} & \rev{1,654 $\uparrow$} & \rev{11,704 $\uparrow$} & \cellcolor{yellow!20}\rev{98.9\%} & \cellcolor{blue!20}\rev{630} & \cellcolor{blue!20}\rev{92} & \cellcolor{blue!20}\rev{1,216} & \cellcolor{blue!20}\rev{135} & \cellcolor{blue!20}\rev{58} & \cellcolor{yellow!20}\rev{4} & \cellcolor{blue!20}\rev{13} & \rev{1} & \cellcolor{blue!20}\rev{4} & \cellcolor{blue!20}\rev{556} & \cellcolor{blue!20}\rev{558} \\
& \rev{\MTqlora\ (CG+CS)} & \rev{34.23\%} & \rev{1,555 $\downarrow$} & \rev{11,194 $\uparrow$} & \cellcolor{yellow!20}\rev{98.9\%} & \cellcolor{blue!20}\rev{593} & \cellcolor{green!20}\rev{75} & \cellcolor{blue!20}\rev{1,254} & \cellcolor{blue!20}\rev{141} & \cellcolor{blue!20}\rev{45} & \cellcolor{yellow!20}\rev{4} & \cellcolor{blue!20}\rev{11} & \rev{0} & \cellcolor{yellow!20}\rev{2} & \cellcolor{blue!20}\rev{548} & \cellcolor{blue!20}\rev{523} \\
\bottomrule
\end{tabular}
}
\end{table}

\rev{Code-quality outcomes diverge sharply between the two languages. For \python, both two-task variants reduce static-analysis violations relative to the baseline, with CG+CS yielding the largest improvements: Cyclomatic Complexity 632 vs.\ 677 ($\downarrow$6.6\%), total Pylint violations 2,341 vs.\ 2,519 ($\downarrow$7.1\%), SonarCloud Maintainability 35 vs.\ 106 ($\downarrow$67.0\%), and Cognitive Complexity 510 vs.\ 549 ($\downarrow$7.1\%). \MTqlora (CG+CT) shows more moderate gains while preserving \passatone\ parity: CyC 656 vs.\ 677 ($\downarrow$3.1\%), Pylint 2,417 vs.\ 2,519 ($\downarrow$4.0\%), and Maintainability 73 vs.\ 106 ($\downarrow$31.1\%). \python therefore admits a clean trade-off: CG+CT retains correctness, while CG+CS gives up correctness ($\downarrow$7.5\%) to obtain substantially cleaner code.}

\rev{For \java, the gains in correctness are accompanied by uniformly increased quality violations. \MTqlora (CG+CT) raises Cyclomatic Complexity to 630 vs.\ 558 ($\uparrow$12.9\%), PMD violations to 1,518 vs.\ 1,441 ($\uparrow$5.3\%), and SonarCloud issue count to 1,119 vs.\ 1,012 ($\uparrow$10.6\%). Within PMD, Error Prone (58 vs.\ 32, $\uparrow$81.3\%) and Performance (13 vs.\ 7, $\uparrow$85.7\%) show the largest increases. \MTqlora (CG+CS) follows the same direction with smaller magnitudes (CyC 593 vs.\ 558, $\uparrow$6.3\%; PMD 1,530 vs.\ 1,441, $\uparrow$6.2\%; SonarCloud 1,073 vs.\ 1,012, $\uparrow$6.0\%), with one exception --- Best Practices violations 75 vs.\ 86 ($\downarrow$12.8\%). The \java pattern is therefore distinct: \passatone\ improvements come at the cost of structurally more complex and stylistically less compliant code.}

\rev{Code-size metrics align with the quality patterns. For \python, CG+CS produces the most compact implementations (1,703 LoC vs.\ 1,759, $\downarrow$3.2\%; 10,461 tokens vs.\ 11,773, $\downarrow$11.1\%), consistent with its reduced complexity. For \java, CG+CT produces longer code (1,654 LoC vs.\ 1,580, $\uparrow$4.7\%; 11,704 tokens vs.\ 10,920, $\uparrow$7.2\%), consistent with the elaborate-solution interpretation of its correctness gains. Across both languages, the absolute magnitudes of the two-task improvements remain within the range of variation observed across model sizes in RQ$_1$, so we report the \java \passatone\ gains as practically observable but not systematically dominant.}

\rev{\textbf{Code Translation.} Translation results reveal effects an order of magnitude larger than those observed for generation. For \java$\rightarrow$C\# (\tabref{tab:rq3_codetrans_java2csharp}), \MTqlora (CG+CT) achieves a CodeBLEU of 73.71\% vs.\ the three-task baseline's 62.88\% ($\uparrow$17.2\%), whereas \MTqlora (CS+CT) collapses to 5.35\% vs.\ 62.88\% ($\downarrow$91.5\%). The C\#$\rightarrow$\java direction (\tabref{tab:rq3_codetrans_csharp2java}) shows the same asymmetry at smaller magnitude: CG+CT reaches 72.48\% vs.\ 71.22\% ($\uparrow$1.8\%), while CS+CT collapses to 7.35\% vs.\ 71.22\% ($\downarrow$89.7\%). The pattern is unambiguous --- pairing generation with translation yields positive transfer, and excluding generation breaks translation entirely.}


\begin{table}[h]
\centering
\scriptsize
\caption{\rev{Ablation: Comparison of \MTqlora task-pair variants for Java$\rightarrow$C\# code translation on CodeXGLUE.}}
\label{tab:rq3_codetrans_java2csharp}
\resizebox{0.85\textwidth}{!}{%
\begin{tabular}{lc|c|cccc|cc}
\toprule
\multirow{2}{*}{\rev{\textbf{Model}}} & \multirow{2}{*}{\rev{\textbf{Method}}} & 
\multirow{2}{*}{\rev{\textbf{CodeBLEU}}} & 
\multicolumn{4}{c|}{\rev{\textbf{Lizard}}} & 
\multicolumn{2}{c}{\rev{\textbf{Roslyn}}} \\
\cmidrule(lr){4-7} \cmidrule(l){8-9}
& & & \rev{\textbf{LoC}} & \rev{\textbf{Tok}} & \rev{\textbf{DR}} & \rev{\textbf{CyC}} & 
\rev{\textbf{Syntax Errors}} & \rev{\textbf{Maintainability}} \\
\midrule
\multirow{3}{*}{\rev{\textit{Qwen-3B}}} 
& \rev{\MTqlora\ (3-task)} & \rev{62.88\%} & \rev{5,890} & \rev{37,716} & \rev{93.6\%} & \rev{1,455} & \rev{126} & \rev{3,061} \\
& \rev{\MTqlora\ (CG+CT)} & \rev{\textbf{73.71\%}} & \rev{6,342 $\uparrow$} & \rev{40,752 $\uparrow$} & \cellcolor{green!20}\rev{98.8\%} & \cellcolor{blue!20}\rev{1,560} & \cellcolor{blue!20}\rev{302} & \cellcolor{green!20}\rev{820} \\
& \rev{\MTqlora\ (CS+CT)} & \rev{5.35\%} & \rev{845} & \rev{4,417} & \rev{19.8\%} & \rev{251} & \rev{68} & \rev{1,828} \\
\bottomrule
\end{tabular}
}
\end{table}

\begin{table}[h]
\centering
\scriptsize
\caption{\rev{Ablation: Comparison of \MTqlora task-pair variants for C\#$\rightarrow$Java code translation on CodeXGLUE.}}
\label{tab:rq3_codetrans_csharp2java}
\resizebox{0.9\textwidth}{!}{%
\begin{tabular}{lc|c|cccc|cccccc|cccc}
\toprule
\multirow{2}{*}{\rev{\textbf{Model}}} & \multirow{2}{*}{\rev{\textbf{Method}}} & 
\multirow{2}{*}{\rev{\textbf{CodeBLEU}}} & 
\multicolumn{4}{c|}{\rev{\textbf{Lizard}}} & 
\multicolumn{6}{c|}{\rev{\textbf{PMD}}} &
\multicolumn{4}{c}{\rev{\textbf{SonarCloud}}} \\
\cmidrule(lr){4-7} \cmidrule(lr){8-13} \cmidrule(l){14-17}
& & & \rev{\textbf{LoC}} & \rev{\textbf{Tok}} & \rev{\textbf{DR}} & \rev{\textbf{CyC}} & 
\rev{\textbf{Best Prac.}} & \rev{\textbf{CS}} & \rev{\textbf{Design}} & \rev{\textbf{EP}} & \rev{\textbf{MT}} & \rev{\textbf{Perf.}} &
\rev{\textbf{Sec}} & \rev{\textbf{Rel}} & \rev{\textbf{Main}} & \rev{\textbf{CoC}} \\
\midrule
\multirow{3}{*}{\rev{\textit{Qwen-3B}}} 
& \rev{\MTqlora\ (3-task)} & \rev{71.22\%} & \rev{5,742} & \rev{34,862} & \rev{99.5\%} & \rev{1,612} & \rev{37} & \rev{1,072} & \rev{45} & \rev{11} & \rev{5} & \rev{1} & \rev{0} & \rev{0} & \rev{278} & \rev{157} \\
& \rev{\MTqlora\ (CG+CT)} & \rev{\textbf{72.48\%}} & \rev{5,746 $\uparrow$} & \rev{34,654 $\downarrow$} & \cellcolor{yellow!20}\rev{99.2\%} & \cellcolor{green!20}\rev{1,583} & \cellcolor{blue!20}\rev{99} & \cellcolor{blue!20}\rev{2,453} & \cellcolor{blue!20}\rev{144} & \cellcolor{blue!20}\rev{44} & \cellcolor{blue!20}\rev{12} & \cellcolor{blue!20}\rev{21} & \rev{0} & \cellcolor{blue!20}\rev{15} & \cellcolor{blue!20}\rev{653} & \cellcolor{blue!20}\rev{404} \\
& \rev{\MTqlora\ (CS+CT)} & \rev{7.35\%} & \rev{850} & \rev{4,597} & \rev{20.9\%} & \rev{280} & \rev{2,050} & \rev{\textit{N/A}} & \rev{\textit{N/A}} & \rev{\textit{N/A}} & \rev{\textit{N/A}} & \rev{\textit{N/A}} & \rev{0} & \rev{0} & \rev{1,360} & \rev{0} \\
\bottomrule
\end{tabular}
}
\end{table}

\rev{Code-quality outcomes track the functional results but reveal an additional asymmetry. For \java$\rightarrow$C\#, \MTqlora (CG+CT) increases syntax errors to 302 vs.\ 126 ($\uparrow$139.7\%) yet substantially reduces Roslyn maintainability findings to 820 vs.\ 3,061 ($\downarrow$73.2\%); the model produces more syntactically rough but structurally cleaner C\#. For C\#$\rightarrow$\java, the trade-off inverts: CG+CT achieves only marginal CodeBLEU gains ($\uparrow$1.8\%) while raising PMD violations to 2,773 vs.\ 1,171 ($\uparrow$136.8\%) and SonarCloud issues to 1,072 vs.\ 435 ($\uparrow$146.4\%), with the largest growth in Error Prone (44 vs.\ 11, $\uparrow$300.0\%) and Performance (21 vs.\ 1, $\uparrow$2,000.0\%) categories. The CG+CT configuration therefore optimizes translation accuracy at the expense of target-language stylistic conformance, with the cost most visible when translating into \java.}

\rev{\MTqlora (CS+CT) outputs are functionally and statically inviable in both directions. Detection rates fall to 19.8\% (\java$\rightarrow$C\#) and 20.9\% (C\#$\rightarrow$\java) against the baseline's 93.6\% and 99.5\%, indicating that most generated outputs cannot be parsed by the static-analysis tools at all. Output sizes shrink correspondingly --- 845 LoC for \java$\rightarrow$C\# vs.\ 5,890 baseline ($\downarrow$85.7\%) and 850 LoC for C\#$\rightarrow$\java vs.\ 5,742 ($\downarrow$85.2\%) --- consistent with truncated or syntactically invalid translations rather than alternative valid solutions. The CS+CT failure is not a degradation but a removal of capability: code generation appears necessary for the model to produce parseable cross-language code at all.}

\rev{\textbf{Code Summarization.} The summarization picture differs from translation in both magnitude and direction. For \python (\tabref{tab:rq3_codesum}), restricting to two tasks produces sharply divergent outcomes: \MTqlora (CG+CS) substantially exceeds the three-task baseline (BLEU 35.79\% vs.\ 29.90\%, $\uparrow$19.7\%; METEOR 43.98\% vs.\ 40.06\%, $\uparrow$9.8\%; Rouge-L 53.79\% vs.\ 47.43\%, $\uparrow$13.4\%; chrF 53.65\% vs.\ 48.20\%, $\uparrow$11.3\%; BERTScore 72.00\% vs.\ 68.59\%, $\uparrow$5.0\%), while \MTqlora (CS+CT) underperforms it (BLEU 22.46\% vs.\ 29.90\%, $\downarrow$24.9\%; METEOR 32.53\% vs.\ 40.06\%, $\downarrow$18.8\%; chrF 41.55\% vs.\ 48.20\%, $\downarrow$13.8\%). Pairing summarization with code generation amplifies summarization quality; pairing it with translation introduces moderate negative interference --- a sign that code-to-code (CT) and code-to-NL (CS) training place structurally different demands on the shared adapter space.}


\begin{table}[h]
\centering
\scriptsize
\caption{\rev{Ablation: Comparison of \MTqlora task-pair variants for code summarization on CoderEval. SIDE applies to Java only.}}
\label{tab:rq3_codesum}
\resizebox{0.75\textwidth}{!}{%
\begin{tabular}{c|c|c|cccccc}
\toprule
\rev{\textbf{Lang}} & \rev{\textbf{Model}} & \rev{\textbf{Method}} & \rev{\textbf{BLEU}} & \rev{\textbf{METEOR}} & \rev{\textbf{Rouge-L}} & \rev{\textbf{chrF}} & \rev{\textbf{BERTScore}} & \rev{\textbf{SIDE}} \\
\midrule
\multirow{3}{*}{\rev{Java}}
& \multirow{3}{*}{\rev{\textit{Qwen-3B}}} 
& \rev{\MTqlora\ (3-task)} & \rev{7.63\%} & \rev{22.94\%} & \rev{34.93\%} & \rev{29.31\%} & \rev{66.40\%} & \rev{91.58\%} \\
& & \rev{\MTqlora\ (CG+CS)} & \rev{8.59\%} & \rev{23.72\%} & \rev{34.10\%} & \rev{\textbf{30.40\%}} & \rev{\textbf{67.06\%}} & \rev{\textbf{91.79\%}} \\
& & \rev{\MTqlora\ (CS+CT)} & \rev{\textbf{8.73\%}} & \rev{\textbf{24.28\%}} & \rev{\textbf{35.06\%}} & \rev{30.27\%} & \rev{66.96\%} & \rev{91.67\%} \\
\midrule
\multirow{3}{*}{\rev{Python}}
& \multirow{3}{*}{\rev{\textit{Qwen-3B}}} 
& \rev{\MTqlora\ (3-task)} & \rev{29.90\%} & \rev{40.06\%} & \rev{47.43\%} & \rev{48.20\%} & \rev{68.59\%} & \rev{--} \\
& & \rev{\MTqlora\ (CG+CS)} & \rev{\textbf{35.79\%}} & \rev{\textbf{43.98\%}} & \rev{\textbf{53.79\%}} & \rev{\textbf{53.65\%}} & \rev{\textbf{72.00\%}} & \rev{--} \\
& & \rev{\MTqlora\ (CS+CT)} & \rev{22.46\%} & \rev{32.53\%} & \rev{44.15\%} & \rev{41.55\%} & \rev{66.83\%} & \rev{--} \\
\bottomrule
\end{tabular}
}
\end{table}

\rev{For \java, both two-task variants modestly outperform the three-task baseline within the narrow band characteristic of \java summarization across all configurations in RQ$_1$ and RQ$_2$. \MTqlora (CS+CT) leads on lexical-overlap metrics (BLEU 8.73\% vs.\ 7.63\%, $\uparrow$14.4\%; METEOR 24.28\% vs.\ 22.94\%, $\uparrow$5.8\%; Rouge-L 35.06\% vs.\ 34.93\%, $\uparrow$0.4\%), while \MTqlora (CG+CS) leads on character-level and contextual metrics (chrF 30.40\% vs.\ 29.31\%, $\uparrow$3.7\%; BERTScore 67.06\% vs.\ 66.40\%, $\uparrow$1.0\%; SIDE 91.79\% vs.\ 91.58\%, $\uparrow$0.2\%). The three-task baseline is not the best on any \java metric, but no relative margin exceeds 14.4\% --- a signal that \java summarization is bounded by factors other than task selection within QLoRA's parameter budget.}

\begin{figure*}[ht]
    \centering
    \begin{subfigure}{0.95\textwidth}
        \includegraphics[width=\textwidth]{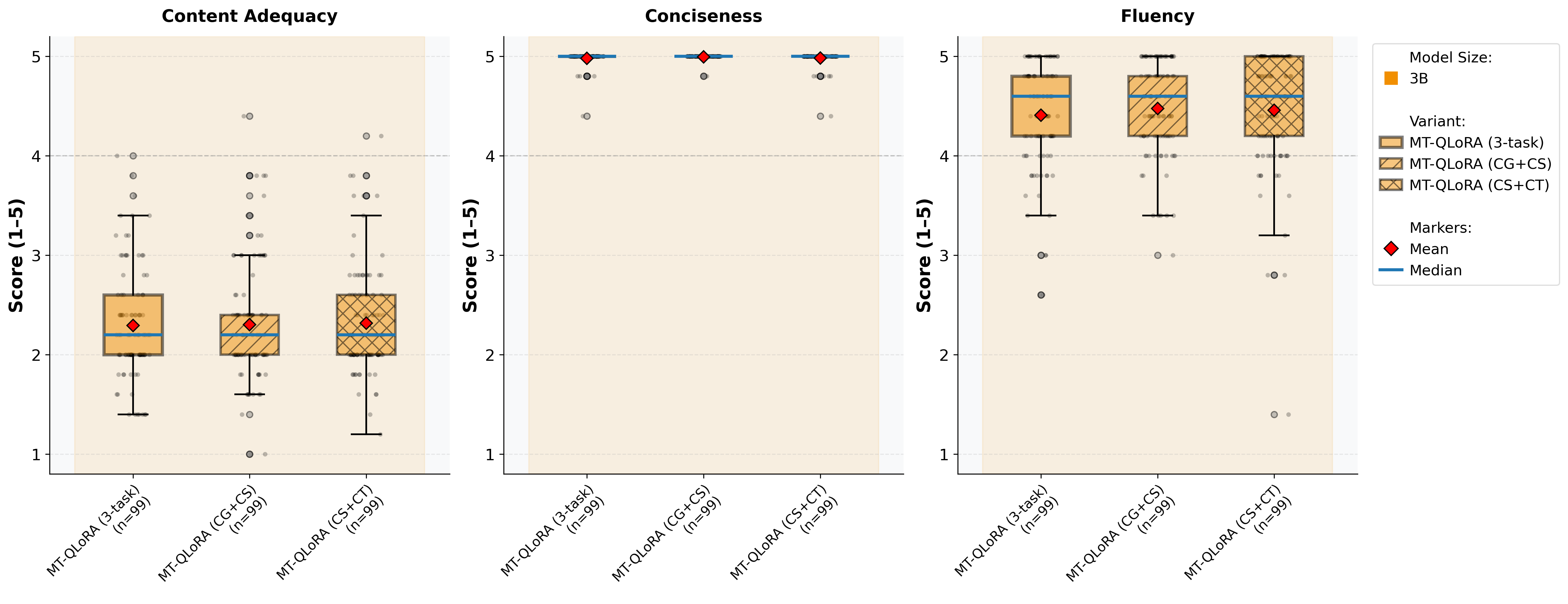}
        \caption{\java}
        \label{fig:llm_judge_java_pairwise}
    \end{subfigure}
    
    \begin{subfigure}{0.95\textwidth}
        \includegraphics[width=\textwidth]{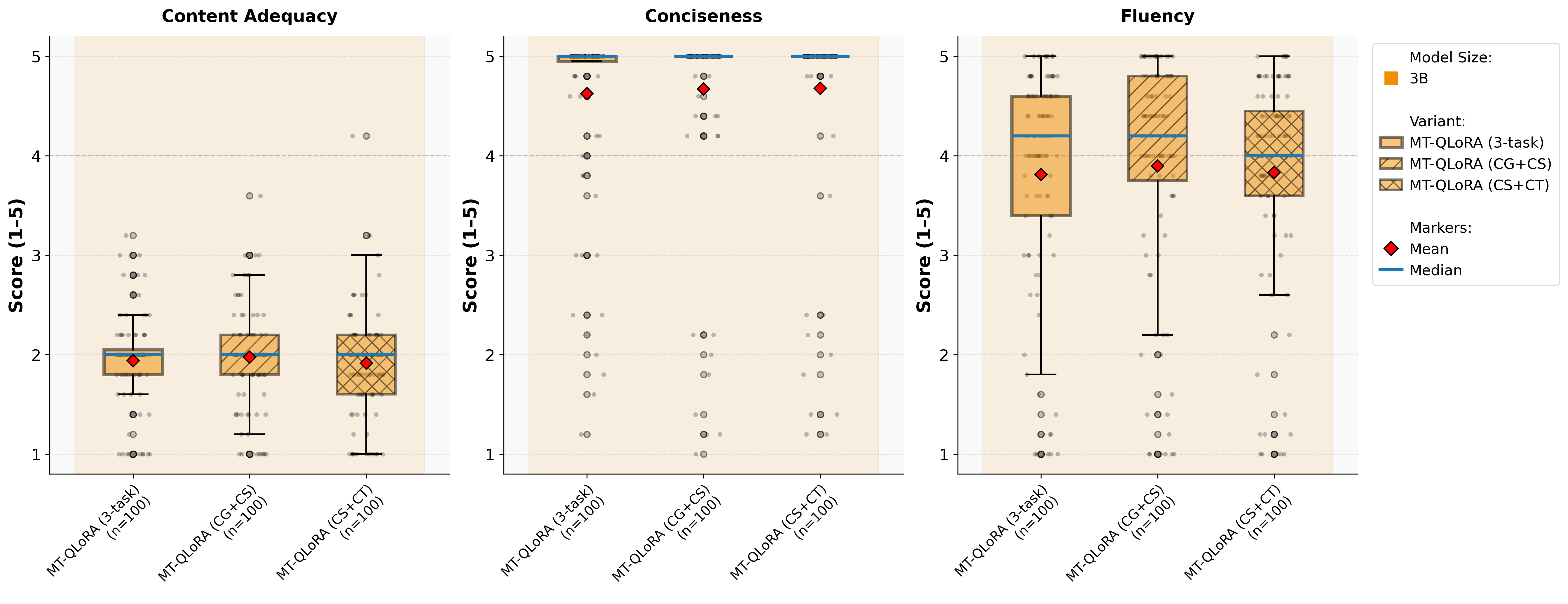}
        \caption{\python}
        \label{fig:llm_judge_python_pairwise}
    \end{subfigure}
    \caption{\rev{Ablation: LLM-as-judge boxplots for code summarization across Content Adequacy, Conciseness, and Fluency. Red diamonds indicate means; blue lines indicate medians.}}
    \label{fig:llm_judge_pairwise}
\end{figure*}

\rev{LLM-as-judge scores (\figref{fig:llm_judge_pairwise}) only partially track the lexical-overlap differences. For \python, Content Adequacy moves modestly in the direction the lexical metrics predict (1.974 vs.\ 1.938 baseline for CG+CS, $\uparrow$1.9\%; 1.914 vs.\ 1.938 for CS+CT, $\downarrow$1.2\%), but the human-aligned magnitude is far smaller than the corresponding BLEU swing ($\uparrow$19.7\% and $\downarrow$24.9\%). Conciseness (4.624 / 4.672 / 4.678 across baseline, CG+CS, CS+CT) and Fluency (3.812 / 3.898 / 3.834) are essentially flat. For \java, all three dimensions remain flat across configurations: Content Adequacy varies by less than 0.025 absolute points (2.293 / 2.301 / 2.317), Conciseness by less than 0.016 (4.980 / 4.996 / 4.984), and Fluency by less than 0.07 (4.408 / 4.475 / 4.459). The pattern reinforces an observation already reported in RQ$_2$: lexical-overlap metrics can move substantially without proportional changes in human-aligned quality scores, a known limitation of similarity-based summarization evaluation.}

\rev{The role of code generation in the multi-task mixture differs sharply between code-to-code and code-to-NL targets. For code translation, generation is load-bearing: \MTqlora (CS+CT) collapses by $\downarrow$91.5\% (\java$\rightarrow$C\#) and $\downarrow$89.7\% (C\#$\rightarrow$\java) in CodeBLEU, with detection rates below 21\% indicating loss of capability rather than degraded quality, while \MTqlora (CG+CT) yields the largest positive transfer observed in this study ($\uparrow$17.2\% CodeBLEU for \java$\rightarrow$C\#; $\uparrow$13.5\% \passatone\ for \java code generation). For code summarization, the picture inverts at the surface and softens in magnitude: \MTqlora (CG+CS) substantially exceeds the three-task baseline for \python ($\uparrow$19.7\% BLEU), while \MTqlora (CS+CT) introduces only moderate negative interference ($\downarrow$24.9\% BLEU) and leaves \java summarization broadly intact. Two implications follow. First, the three-task configuration is not strictly optimal for every target: for \python summarization specifically, the CG+CS pairing is meaningfully better, suggesting that translation contributes negative transfer to summarization when both share a constrained adapter space. Second, the cross-task transfer mechanism is target-specific. Code generation acts as a load-bearing task for translation but as one of two equivalent partners for summarization, and any practical recommendation about multi-task composition must condition on the deployment target rather than treat the three-task mixture as universally preferable.}

\subsection{\rev{Human Validation of GPT-5 Mini as a Judge}}
\label{sec:human-validation}

\rev{To assess the reliability of GPT-5 Mini as a judge, we conducted a human evaluation study on the same \java/\python code summarization setting used in this paper. We sampled 200 (code, summary) pairs in total --- 50 each across the four buckets defined by \{\STqlora, \MTqlora\} $\times$ {\java, \python} --- with summaries generated by the 3B-scale models, which represent the most representative scale in our experimental grid for this comparison. Two annotators (the authors) independently scored each summary on the same three dimensions used by GPT-5 Mini --- Content Adequacy, Conciseness, and Fluency --- using the 1--5 Likert scale described in Section~3.5.2 and shown in Fig.~1. Annotators were blind to the configuration (\STqlora or \MTqlora) that produced each summary, and the order of summaries within each bucket was randomized. We report two complementary forms of agreement: (i)~inter-annotator agreement, to establish that human judgment is itself reliable on this task, and (ii)~agreement between the mean human scores and GPT-5 Mini's mean scores (5-run mean), to assess whether GPT-5 Mini tracks human consensus.}

\rev{\textbf{Inter-annotator agreement.} We report quadratic-weighted Cohen's $\kappa$ as our inter-annotator agreement metric, which is appropriate for ordinal Likert-scale data. Pooled across all 200 paired observations, we obtain $\kappa_w = 0.51$ for Content Adequacy (moderate agreement), $\kappa_w = 0.38$ for Conciseness (fair agreement), and $\kappa_w = 0.63$ for Fluency (substantial agreement).}

\rev{\textbf{Agreement between GPT-5 Mini and the human consensus.} We compute Pearson and Spearman correlations between the mean human scores and GPT-5 Mini's mean scores on the same 200 pairs. Pearson $r$ is $0.633$ (CA), $0.827$ (Conciseness), and $0.877$ (Fluency); Spearman $\rho$ is $0.625$ (CA), $0.167$ (Conciseness), and $0.722$ (Fluency), all with $p < 0.0001$ except Conciseness Spearman ($p = 0.018$). For a more direct, instance-level view, $69.0\%$ of CA scores, $93.0\%$ of Conciseness scores, and $97.0\%$ of Fluency scores fall within one Likert point of the mean human value --- a level of practical agreement that supports the use of GPT-5 Mini as a scalable proxy for human consensus on Conciseness and Fluency, and as a calibration-shifted but rank-preserving proxy on Content Adequacy.}

\rev{\textbf{Content Adequacy calibration.} On the Content Adequacy dimension, GPT-5 Mini's mean scores are systematically lower than the mean human scores by approximately $0.8$ Likert points on average. We attribute this to a rubric-interpretation difference: the judge prompt asks GPT-5 Mini to evaluate ``the extent to which the comment summarizes \emph{all} information that can be inferred from the source code,'' which the model interprets as a strict completeness standard, awarding a 5 only when the comment is exhaustive; human annotators, in contrast, more readily awarded high scores to summaries that captured the function's main purpose without exhaustively listing every implementation detail. The shift is uniform rather than erratic, as evidenced by the Pearson and Spearman correlations on CA ($r = 0.633$, $\rho = 0.625$), both of which indicate that the LLM judge preserves the \emph{rank ordering} of summaries even when its absolute scores are calibrated more strictly than human consensus. Because the comparisons reported in RQ$_1$ and RQ$_2$ rely on \emph{relative} differences between configurations (\STqlora vs \MTqlora vs \MTfft) rather than absolute score levels, rank preservation is the property that matters for the conclusions drawn in this study.}

\rev{Taken together, the human evaluation supports the use of GPT-5 Mini as a judge for the comparisons reported in this paper: agreement with the mean human consensus on Fluency is substantial (within-1-point: $97.0\%$); on Conciseness it is strong in linear association and within-1-point ($93.0\%$); on Content Adequacy it is rank-preserving with a known calibration offset. The full per-instance data, including individual annotator scores, mean human scores, GPT-5 Mini mean scores, and per-instance differences, is released as part of our replication package to allow independent verification of these findings.}

\rev{Lastly, one further point worth discussing is bias. Recent work by Li \etal~\cite{li2025preference} on preference leakage shows that LLM judges systematically favor models trained on related data, with bias scaling by relatedness: $23.6\%$ when the judge and evaluated model are the same, $8.9\%$ within the same model family, $2.8\%$ for different series in the same family, and between $-0.1\%$ and $1.7\%$ on average for cross-family pairs. Since our evaluated model is Qwen2.5-Coder and our judge is GPT-5 Mini, our setup falls within the cross-family regime, where the residual self-preference bound is below $2\%$. To further reduce variance in the residual range, we report the mean across five independent runs per summary as described earlier in this section.}

\subsection{Qualitative Analysis of \MTqlora in Code-Related Tasks}
\label{sec:qualitative_analysis}

To better understand the quality characteristics behind the quantitative results, we conduct a qualitative analysis of model outputs across code generation, summarization, and translation tasks. We perform an instance-level overlap analysis that partitions evaluation instances based on which configuration(s) succeed, enabling us to examine both shared successes and \emph{unique} successes attributable to a specific training setup.

For \textbf{code generation}, we use \passatone to partition instances into those solved by both ST-QLoRA and \MTqlora and those uniquely solved by only one configuration. For \textbf{code translation}, we use Exact Match (EM) to isolate instances where only one configuration achieves a perfect prediction (EM$=1$). We then inspect representative examples from these partitions to characterize qualitative differences in correctness, completeness, and semantic preservation. For \textbf{code summarization}, we use Content Adequacy (CA) to identify instances where only \STqlora, only \MTqlora, or only \MTfft attains the highest CA score. 

\subsubsection{Code Generation}

Our analysis for code generation shows substantial agreement between \STqlora and \MTqlora, but instances where one succeeds and the other fails reveal clear qualitative differences on multi-step specifications.

\begin{figure*}[t]
\centering
\includegraphics[width=0.95\textwidth]{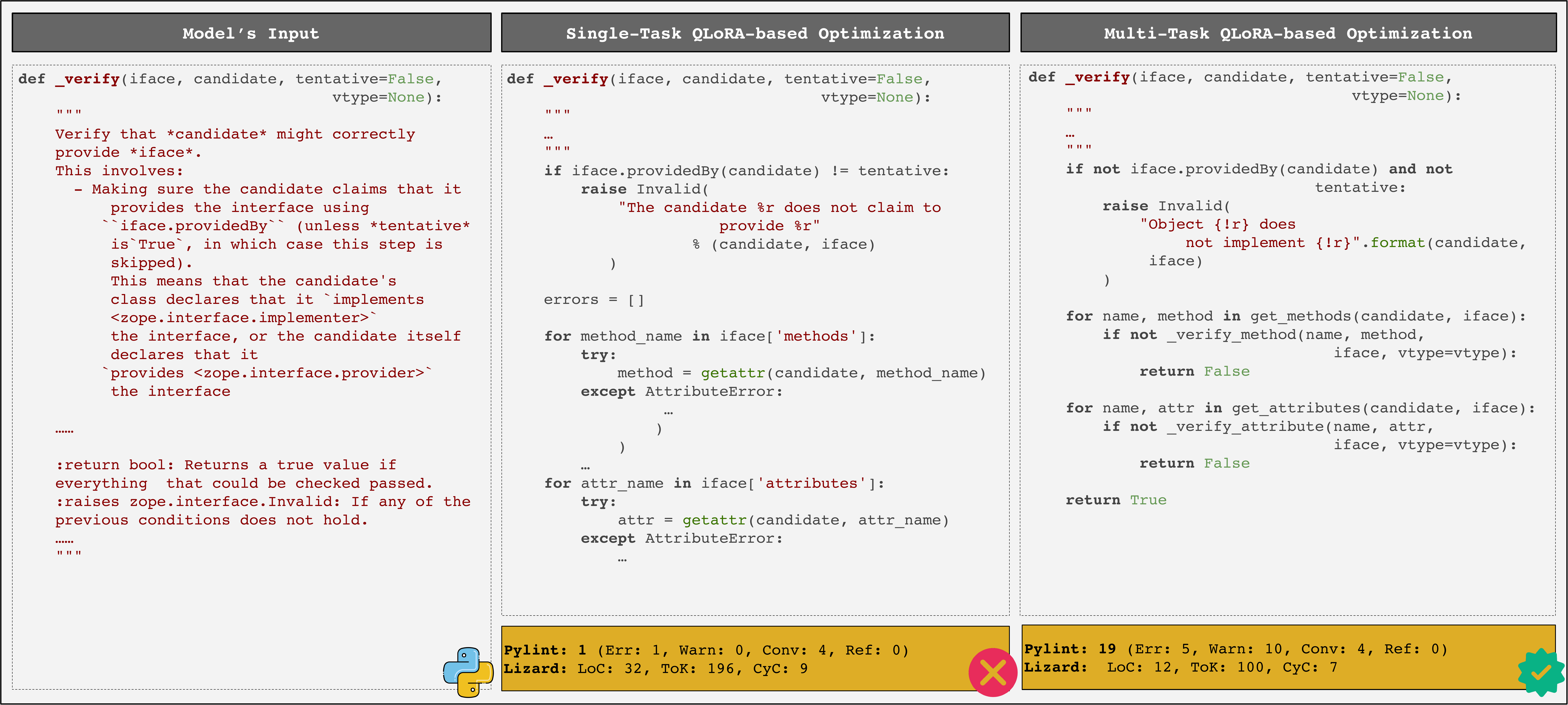}
\caption{Qualitative comparison of \python code generated by \STqlora and \MTqlora for an interface verification routine. \MTqlora (right, green checkmark) produces a passing solution while \STqlora (middle, red X) fails functional testing.}
\label{fig:codegen_example}
\end{figure*}

Figure~\ref{fig:codegen_example} presents a representative Python case where \MTqlora produces a correct, passing solution (green checkmark) while \STqlora fails (red X). The prompt specifies a verification routine that conditionally checks \texttt{iface.providedBy} (skipped under \texttt{tentative}), validates required methods (including signatures) and attributes, and raises \texttt{Invalid} on contract violations.

\MTqlora generates a concise, specification-aligned implementation: it correctly gates the \texttt{providedBy} check (\texttt{if not iface.providedBy(candidate) and not tentative}), raises an informative \texttt{Invalid} exception on failure, and cleanly verifies methods and attributes via helper routines with early exits before returning \texttt{True}. In contrast, \STqlora diverges from the intended control flow by using an incorrect gating condition (\texttt{iface.providedBy(candidate) != tentative}), introducing fragmented error accumulation, and devolving into incomplete branches (ellipsis placeholders), yielding a non-passing solution. As reported in Fig.~\ref{fig:codegen_example}, \STqlora also produces a longer and more complex output (LoC: 32, Tok: 196, CyC: 9) than \MTqlora (LoC: 12, Tok: 100, CyC: 7), without improving correctness. This example reflects a recurring pattern: \MTqlora more consistently preserves the contract and control-flow structure implied by docstrings, whereas \STqlora is more prone to semantic drift on multi-step behavioral requirements.

\subsubsection{Code Translation}

To characterize how structural complexity relates to the observed success patterns in translation, we analyze the source inputs using Lizard~\citep{yin2019lizard}. Specifically, we focused on cyclomatic complexity alongside complementary indicators of code size and structure (\eg token count and nesting depth) of the source code. This analysis helps determine whether \MTqlora's gains are concentrated in particular complexity regimes and provides context for common qualitative failure modes, such as missing control-flow logic, broken API consistency, or semantic drift.

\begin{figure*}[t]
\centering
\includegraphics[width=0.95\textwidth]{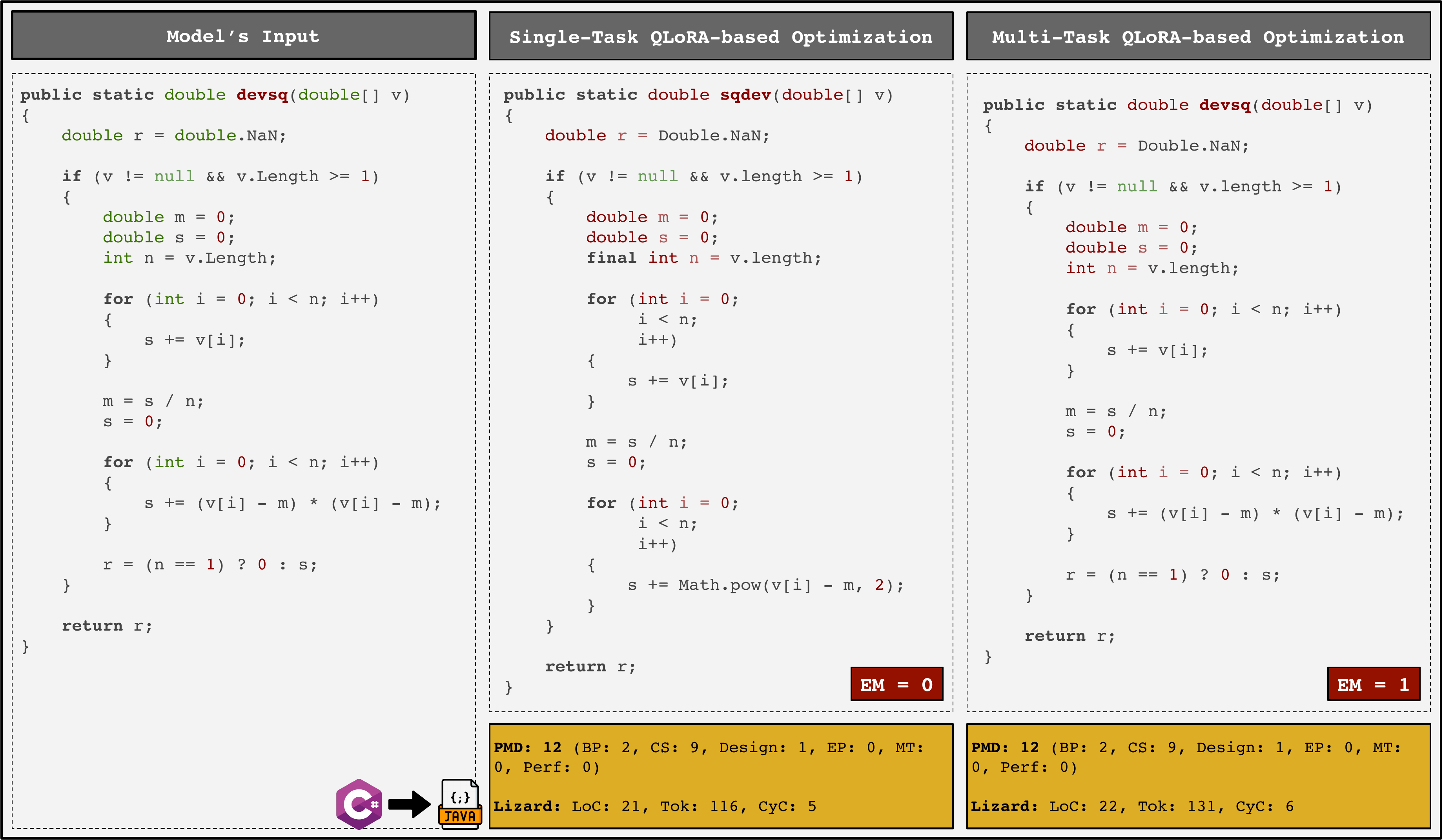}
\caption{Qualitative comparison of C\#$\rightarrow$\java translations generated by \STqlora and \MTqlora for a representative instance where \MTqlora achieves exact match (EM$=1$) while \STqlora fails (EM$=0$).}
\label{fig:translation_example}
\end{figure*}

Our overlap analysis reveals distinct patterns in the structural complexity of code uniquely translated by different model configurations. For C\#$\rightarrow$Java translation, \MTqlora uniquely achieves perfect translations on structurally more complex code compared to both ST-QLoRA (RQ1) and MT-FFT (RQ2). Across the instances it solves uniquely, \MTqlora more often handles inputs with higher cyclomatic complexity ($\approx 39$--$51$\%), larger code size ($\approx 30$--$33$\%), and deeper nesting ($\approx 28$--$32$\%), suggesting improved robustness to complex control flow and nested structures in this translation direction. In contrast, the Java$\rightarrow$C\# direction shows an opposite trend: ST-QLoRA and MT-FFT more often uniquely solve structurally complex inputs, whereas \MTqlora more often succeeds on comparatively simpler instances, suggesting direction-dependent translation difficulty and asymmetric transfer effects.


Figure~\ref{fig:translation_example} shows a representative C\#$\rightarrow$Java case where \MTqlora produces a perfect translation (EM$=$1) while \STqlora fails (EM$=$0). The source input exhibits moderate structural complexity (CyC$=$6), with nested control flow structures including multiple \texttt{for} loops and conditional logic. Although the \STqlora output remains syntactically plausible, it introduces semantic and interface-breaking changes: (1) it renames the method from \texttt{devsq} to \texttt{sqdev}, breaking API consistency; (2) it omits the assignment \texttt{r = (n == 1) ? 0 : s}, leaving \texttt{r} never updated from its initialization \texttt{Double.NaN} and thus returning \texttt{NaN} regardless of the computed accumulator \texttt{s}; and (3) it replaces the direct square \texttt{(v[i] - m) * (v[i] - m)} with \texttt{Math.pow(v[i] - m, 2)}, an unnecessary deviation from the original implementation style. In contrast, \MTqlora preserves the method signature, retains the complete conditional assignment logic, and maintains the original arithmetic structure. While both translations trigger comparable static analysis warnings (PMD: 12 issues), the EM discrepancy is driven by substantive semantic preservation rather than superficial style differences, illustrating \MTqlora's stronger robustness on nested control-flow translation in C\#$\rightarrow$Java.

\subsubsection{Code Summarization}

For summarization, we inspect instances where one configuration achieves higher Content Adequacy (CA) than the others, focusing on whether the generated summaries capture \emph{behaviorally salient} details (\eg filtering conditions, edge cases, or implicit constraints) rather than only paraphrasing surface-level operations. Across the overlap partitions, \MTqlora more consistently produces summaries that reflect such conditional behaviors, whereas \MTfft summaries are more likely to remain correct but underspecified, which reduces adequacy when key constraints are omitted.

\begin{figure*}[t]
\centering
\includegraphics[width=0.95\textwidth]{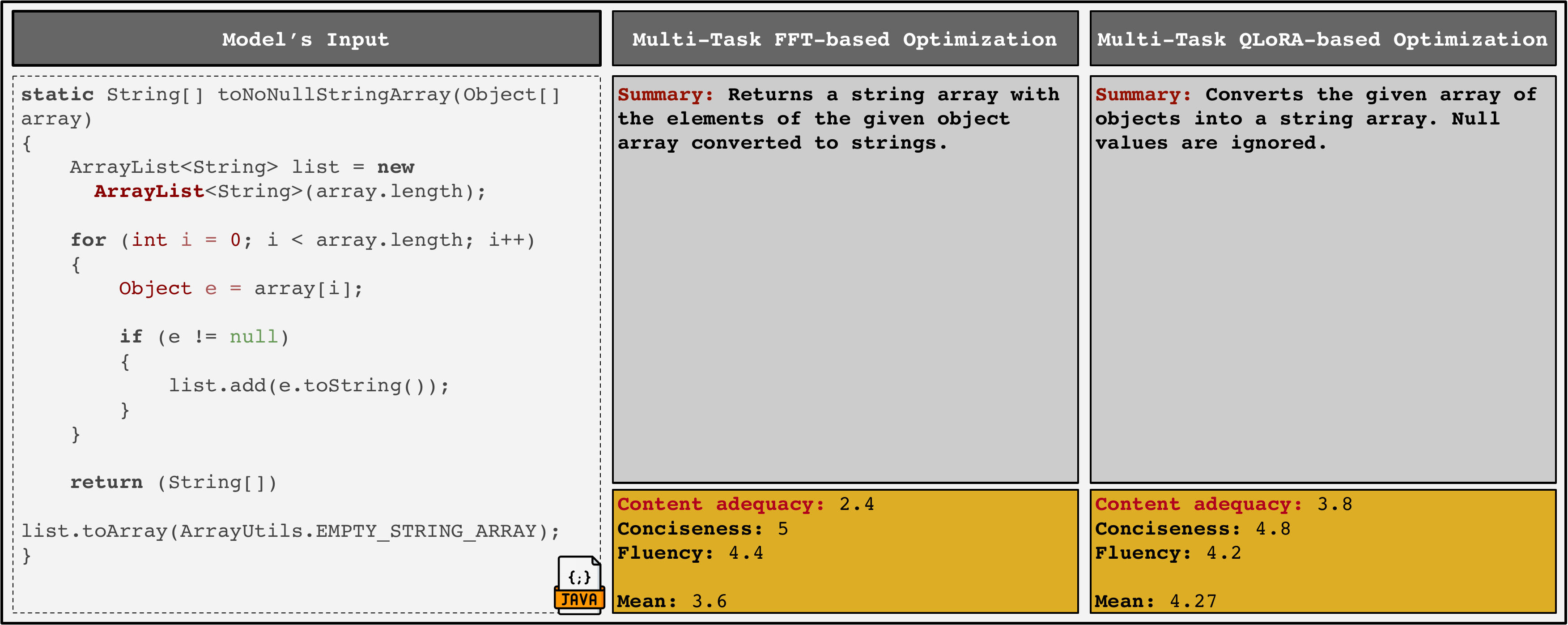}
\caption{Qualitative comparison of \java code summaries generated by \MTfft and \MTqlora. \MTqlora produces a higher quality summary with better content adequacy, while \MTfft generates a less accurate summary that misses the null-filtering behavior.}
\label{fig:summarization_example}
\end{figure*}

Figure~\ref{fig:summarization_example} illustrates a representative Java method where \MTqlora achieves higher CA than \MTfft. The method \newline \texttt{toNoNullStringArray} constructs a \texttt{String[]} by iterating over an \texttt{Object[]} input, converting elements via \texttt{toString()}, and \emph{skipping null entries} before returning the resulting array. MT-FFT produces a largely generic description (converting an object array to strings) but fails to state the null-filtering behavior, which is central to the method's contract. In contrast, \MTqlora explicitly captures this constraint (``Null values are ignored''), yielding a more semantically complete summary despite similar fluency and conciseness, and thereby improving content adequacy.



\section{Threats to Validity}
\label{sec:threats}

\textbf{Internal Validity.} Our experimental design may be influenced by several factors. First, hyperparameter selection follows configurations from prior work~\cite{afrin2025resource} rather than task-specific optimization, potentially underestimating the performance of certain configurations. While we maintain consistent hyperparameters across all experiments to ensure fair comparison, task-specific tuning might yield different relative performance patterns. Second, the early stopping criterion based on validation metrics may favor certain model configurations over others, particularly when comparing \MT versus \ST models with different convergence characteristics. 
\rev{Third, we focus on the simplest and most widely deployed \MT configuration --- proportional concatenation, with balanced sampling as a controlled alternative reported in \secref{sec:balanced-ablation} --- and whether more sophisticated strategies such as curriculum learning or task-aware parameter sharing yield additional gains remains an open question.} 
\revtwo{Fourth, the adapter configuration in \tabref{tab:qlora_config} is held fixed across all model scales, so the share of trainable parameters falls as capacity grows (0.60\% at 1.5B to 0.27\% at 7B; \tabref{tab:efficiency-comparison}). This is a design choice made to isolate the effect of scale, but it means adapter capacity per task is not matched across scales, and code translation --- the task with by far the smallest training share ($\sim$20K instances against $\sim$416K for each of the other two tasks) and the one our sampling ablation (\secref{sec:balanced-ablation}) identifies as most sensitive to data composition --- is the configuration most exposed to it. We regard this as a candidate factor behind the pronounced translation degradation observed at 7B in RQ$_1$ and RQ$_2$ rather than an established cause: isolating it would require a scale-matched adapter-capacity study or a per-task capacity analysis, neither of which the present experiments support, and we therefore leave characterizing this degradation to future work.}

\textbf{External Validity.} The generalizability of our findings faces several constraints. We evaluate using the Qwen2.5-Coder family (0.5B, 1.5B, 3B, and \rev{7B} parameters), and results may not directly transfer to other model architectures. \revtwo{To examine whether our findings extend beyond a single family, we replicated RQ$_1$ on CodeLlama-7b-Instruct under an identical protocol. The \STqlora advantage on translation and the \MTqlora advantage on \python summarization are both reproduced for CodeLlama, while the code-generation behavior differs between the two families. This replication covers RQ$_1$ at the 7B scale for one additional family; broader replication across more families is left to future work.} Our focus on three representative code-related tasks --- generation, summarization, and translation --- \rev{spans the three canonical input-output modalities for \lcms (NL$\rightarrow$Code, Code$\rightarrow$NL, Code$\rightarrow$Code), which is sufficient for characterizing the cross-task transfer dynamics studied in this work but does not exhaust the space of software engineering activities. Bug fixing, test generation, and vulnerability detection warrant separate study: the first two require paired (buggy, fixed) or (function, test) instances structurally distinct from the input-output pairs used here, and vulnerability detection is a classification task whose evaluation methodology differs substantively from generative tasks. Whether the QLoRA-specific patterns we report extend to these other task families is left for future work}. The employed datasets (CodeXGLUE and CoderEval) predominantly contain function-level code snippets, which, while standard in prior research, may not fully reflect project-level dependencies observed in large industrial systems. Finally, our experiments cover \java, \python, and C\#, offering diversity across paradigms, though future extensions to additional languages could further strengthen generalizability.

\textbf{Construct Validity.} The evaluation design relies on well-established metrics and tools to assess both functional and non-functional performance. For correctness, we combine execution-based (\passatone) and structural–semantic (CodeBLEU) measures; for summarization, we integrate lexical, semantic, and SIDE metrics, complemented by an LLM-as-a-judge protocol following a validated methodology~\cite{crupi2025effectiveness}. For code quality, we employ multiple static analyzers (\eg Pylint, PMD, SonarCloud, Roslyn) to triangulate results across diverse rule sets. Although static analysis tools may differ slightly in sensitivity across languages, this multi-tool strategy minimizes tool-specific biases and provides a comprehensive perspective on quality dimensions.

\textbf{Conclusion Validity.} Our statistical analyses employ robust non-parametric tests (Wilcoxon signed-rank and McNemar) with Holm-Bonferroni correction to control for multiple comparisons, complemented by effect size reporting to capture practical significance. These analyses are consistently applied across both RQ$_1$ (\MTqlora \vs \STqlora) and RQ$_2$ (\MTqlora \vs \MTfft), where model outputs allow for paired evaluation across configurations. Minor variations fall within expected stochastic bounds of LLM generation, and cross-task consistency supports the soundness and reproducibility of our findings.

\section{Conclusions and Future Work}
\label{sec:conclusion}
This study presents \rev{a systematic evaluation} of \MT QLoRA \revtwo{--- one representative method from the LoRA family ---} across code generation, translation, and summarization, examining both functional correctness and code quality with Qwen2.5-Coder models (0.5B, 1.5B, 3B\revtwo{, 7B}). \MT QLoRA achieves competitive performance versus \ST QLoRA while training a unified model. The 3B configuration demonstrates language-dependent effectiveness: 32.07\% \java \passatone versus 29.89\% \ST (+2.18 points) and 29.90\% \python summarization BLEU versus 23.31\% (+28.3\% relative). 
Against \MT Full Fine-Tuning, \MT QLoRA shows superiority for generation and \python summarization (21.05\% \vs\ 17.37\% \passatone; 29.90\% \vs\ 22.54\% BLEU at 3B), alongside quality gains of 32.0\% lower \python complexity at 3B and 6.3\% fewer \java PMD violations across the 0.5B--3B scales.

Our findings reveal that larger models consistently demonstrate superior ability to balance multiple objectives within QLoRA's parameter-efficient framework. Translation exhibits scale- and direction-dependent trade-offs, with \MT QLoRA underperforming for Java→C\# across all scales but showing competitive or improved results for C\#$\rightarrow$Java at \revtwo{smaller scales; at 7B, however, this limitation extends to both translation directions}. Code summarization demonstrates clear language-dependent effects: \python summarization benefits substantially from \MT training across all scales, while \java summarization favors \ST QLoRA, particularly at larger model capacities where task-specific training proves more effective.

Future work should investigate task-specific adapters to address translation instability, dynamic task weighting for language-specific transfer, \rev{curriculum learning strategies that schedule task exposure based on difficulty estimation, and task-aware parameter sharing that introduces learnable routing to selectively activate adapter subsets per task}. Extension to additional programming languages and software engineering tasks (\rev{such as} bug fixing, program repair\rev{, test generation, and vulnerability detection}) would further validate the generalizability of our findings. \revtwo{Because our study characterizes QLoRA as one representative method from the LoRA family, whether the observed patterns hold for PEFT approaches beyond that family (\eg adapters, prompt or prefix tuning) likewise remains open.}

\bibliographystyle{ACM-Reference-Format}
\bibliography{main}

@article{li2025preference,
  title={Preference leakage: A contamination problem in llm-as-a-judge},
  author={Li, Dawei and Sun, Renliang and Huang, Yue and Zhong, Ming and Jiang, Bohan and Han, Jiawei and Zhang, Xiangliang and Wang, Wei and Liu, Huan},
  journal={arXiv preprint arXiv:2502.01534},
  year={2025}
}

@article{weyssow2023exploring,
	title={Exploring parameter-efficient fine-tuning techniques for code generation with large language models},
	author={Weyssow, Martin and Zhou, Xin and Kim, Kisub and Lo, David and Sahraoui, Houari},
	journal={arXiv preprint arXiv:2308.10462},
	year={2023}
}

@article{wang2021codet5,
  title={Codet5: Identifier-aware unified pre-trained encoder-decoder models for code understanding and generation},
  author={Wang, Yue and Wang, Weishi and Joty, Shafiq and Hoi, Steven CH},
  journal={arXiv preprint arXiv:2109.00859},
  year={2021}
}

@misc{codellama2,
	title = {CodeLLama},
	howpublished = {\url{https://github.com/meta-llama/codellama/tree/main}}}

@misc{replication,
	title={Replication Package},
	url={https://github.com/alvi75/MultiTask-QLoRA-NFAnalysis}
}

@article{zhu2024deepseek,
	title={DeepSeek-Coder-V2: Breaking the Barrier of Closed-Source Models in Code Intelligence},
	author={Zhu, Qihao and Guo, Daya and Shao, Zhihong and Yang, Dejian and Wang, Peiyi and Xu, Runxin and Wu, Y and Li, Yukun and Gao, Huazuo and Ma, Shirong and others},
	journal={arXiv preprint arXiv:2406.11931},
	year={2024}
}

@article{nijkamp2022codegen,
  title={Codegen: An open large language model for code with multi-turn program synthesis},
  author={Nijkamp, Erik and Pang, Bo and Hayashi, Hiroaki and Tu, Lifu and Wang, Huan and Zhou, Yingbo and Savarese, Silvio and Xiong, Caiming},
  journal={arXiv preprint arXiv:2203.13474},
  year={2022}
}

@article{chen2021evaluating,
  title={Evaluating large language models trained on code},
  author={Chen, Mark and Tworek, Jerry and Jun, Heewoo and Yuan, Qiming and Pinto, Henrique Ponde De Oliveira and Kaplan, Jared and Edwards, Harri and Burda, Yuri and Joseph, Nicholas and Brockman, Greg and others},
  journal={arXiv preprint arXiv:2107.03374},
  year={2021}
}

@inproceedings{mastropaolo2021studying,
  title={Studying the usage of text-to-text transfer transformer to support code-related tasks},
  author={Mastropaolo, Antonio and Scalabrino, Simone and Cooper, Nathan and Palacio, David Nader and Poshyvanyk, Denys and Oliveto, Rocco and Bavota, Gabriele},
  booktitle={2021 IEEE/ACM 43rd International Conference on Software Engineering},
  pages={336--347},
  year={2021},
  organization={IEEE}
}

@inproceedings{xu2022systematic,
  title={A systematic evaluation of large language models of code},
  author={Xu, Frank F and Alon, Uri and Neubig, Graham and Hellendoorn, Vincent Josua},
  booktitle={Proceedings of the 6th ACM SIGPLAN International Symposium on Machine Programming},
  pages={1--10},
  year={2022}
}

@article{codellama,
  title={Code llama: Open foundation models for code},
  author={Roziere, Baptiste and Gehring, Jonas and Gloeckle, Fabian and Sootla, Sten and Gat, Itai and Tan, Xiaoqing Ellen and Adi, Yossi and Liu, Jingyu and Sauvestre, Romain and Remez, Tal and others},
  journal={arXiv preprint arXiv:2308.12950},
  year={2023}
}

@inproceedings{liu2024mftcoder,
	title={Mftcoder: Boosting code llms with multitask fine-tuning},
	author={Liu, Bingchang and Chen, Chaoyu and Gong, Zi and Liao, Cong and Wang, Huan and Lei, Zhichao and Liang, Ming and Chen, Dajun and Shen, Min and Zhou, Hailian and others},
	booktitle={Proceedings of the 30th ACM SIGKDD Conference on Knowledge Discovery and Data Mining},
	pages={5430--5441},
	year={2024}
}

@article{dettmers2024qlora,
  title={Qlora: Efficient finetuning of quantized llms},
  author={Dettmers, Tim and Pagnoni, Artidoro and Holtzman, Ari and Zettlemoyer, Luke},
  journal={Advances in Neural Information Processing Systems},
  volume={36},
  year={2024}
}

@article{yang2024multi,
  title={Multi-Objective Fine-Tuning for Enhanced Program Repair with LLMs},
  author={Yang, Boyang and Tian, Haoye and Ren, Jiadong and Zhang, Hongyu and Klein, Jacques and Bissyand{\'e}, Tegawend{\'e} F and Goues, Claire Le and Jin, Shunfu},
  journal={arXiv preprint arXiv:2404.12636},
  year={2024}
}

@article{Tunstall2023starchat-alpha,
  author = {Tunstall, Lewis and Lambert, Nathan and Rajani, Nazneen and Beeching, Edward and Le Scao, Teven and von Werra, Leandro and Han, Sheon and Schmid, Philipp and Rush, Alexander},
  title = {Creating a Coding Assistant with StarCoder},
  journal = {Hugging Face Blog},
  year = {2023},
  note = {https://huggingface.co/blog/starchat-alpha},
}

@article{mistral,
  title={Mistral 7B},
  author={Jiang, Albert Q and Sablayrolles, Alexandre and Mensch, Arthur and Bamford, Chris and Chaplot, Devendra Singh and Casas, Diego de las and Bressand, Florian and Lengyel, Gianna and Lample, Guillaume and Saulnier, Lucile and others},
  journal={arXiv preprint arXiv:2310.06825},
  year={2023}
}

@article{liu2022few,
  title={Few-shot parameter-efficient fine-tuning is better and cheaper than in-context learning},
  author={Liu, Haokun and Tam, Derek and Muqeeth, Mohammed and Mohta, Jay and Huang, Tenghao and Bansal, Mohit and Raffel, Colin A},
  journal={Advances in Neural Information Processing Systems},
  volume={35},
  pages={1950--1965},
  year={2022}
}

@article{zhang2022survey,
	title={A survey of automatic source code summarization},
	author={Zhang, Chunyan and Wang, Junchao and Zhou, Qinglei and Xu, Ting and Tang, Ke and Gui, Hairen and Liu, Fudong},
	journal={Symmetry},
	volume={14},
	number={3},
	pages={471},
	year={2022},
	publisher={MDPI}
}

@inproceedings{bleu,
	title={Bleu: a method for automatic evaluation of machine translation},
	author={Papineni, Kishore and Roukos, Salim and Ward, Todd and Zhu, Wei-Jing},
	booktitle={Proceedings of the 40th annual meeting of the Association for Computational Linguistics},
	pages={311--318},
	year={2002}
}

@inproceedings{meteor,
	title={METEOR: An automatic metric for MT evaluation with improved correlation with human judgments},
	author={Banerjee, Satanjeev and Lavie, Alon},
	booktitle={Proceedings of the acl workshop on intrinsic and extrinsic evaluation measures for machine translation and/or summarization},
	pages={65--72},
	year={2005}
}

@inproceedings{lin2004rouge,
	title={Rouge: A package for automatic evaluation of summaries},
	author={Lin, Chin-Yew},
	booktitle={Text summarization branches out},
	pages={74--81},
	year={2004}
}

@inproceedings{popovic2015chrf,
	title={chrF: character n-gram F-score for automatic MT evaluation},
	author={Popovi{\'c}, Maja},
	booktitle={Proceedings of the tenth workshop on statistical machine translation},
	pages={392--395},
	year={2015}
}

@article{zhang2019bertscore,
	title={Bertscore: Evaluating text generation with bert},
	author={Zhang, Tianyi and Kishore, Varsha and Wu, Felix and Weinberger, Kilian Q and Artzi, Yoav},
	journal={arXiv preprint arXiv:1904.09675},
	year={2019}
}

@inproceedings{afrin2025resource,
  title={Resource-Efficient \& Effective Code Summarization},
  author={Afrin, Saima and Call, Joseph and Nguyen, Khai-Nguyen and Chaparro, Oscar and Mastropaolo, Antonio},
  booktitle={2025 IEEE/ACM Second International Conference on AI Foundation Models and Software Engineering (Forge)},
  pages={224--235},
  year={2025},
  organization={IEEE}
}

@article{afrin2025systematic,
  title={A systematic literature review of parameter-efficient fine-tuning for large code models},
  author={Afrin, Saima and Haque, Md Zahidul and Mastropaolo, Antonio},
  journal={arXiv preprint arXiv:2504.21569},
  year={2025}
}

@article{weyssow2025exploring,
  title={Exploring parameter-efficient fine-tuning techniques for code generation with large language models},
  author={Weyssow, Martin and Zhou, Xin and Kim, Kisub and Lo, David and Sahraoui, Houari},
  journal={ACM Transactions on Software Engineering and Methodology},
  volume={34},
  number={7},
  pages={1--25},
  year={2025},
  publisher={ACM New York, NY}
}

@inproceedings{lu2023llama,
  title={Llama-reviewer: Advancing code review automation with large language models through parameter-efficient fine-tuning},
  author={Lu, Junyi and Yu, Lei and Li, Xiaojia and Yang, Li and Zuo, Chun},
  booktitle={2023 IEEE 34th International Symposium on Software Reliability Engineering (ISSRE)},
  pages={647--658},
  year={2023},
  organization={IEEE}
}

@article{su2024distilled,
  title={Distilled GPT for source code summarization},
  author={Su, Chia-Yi and McMillan, Collin},
  journal={Automated Software Engineering},
  volume={31},
  number={1},
  pages={22},
  year={2024},
  publisher={Springer}
}

@inproceedings{shi2023towards,
  title={Towards efficient fine-tuning of pre-trained code models: An experimental study and beyond},
  author={Shi, Ensheng and Wang, Yanlin and Zhang, Hongyu and Du, Lun and Han, Shi and Zhang, Dongmei and Sun, Hongbin},
  booktitle={Proceedings of the 32nd ACM SIGSOFT International Symposium on Software Testing and Analysis},
  pages={39--51},
  year={2023}
}

@article{mastropaolo2022using,
  title={Using transfer learning for code-related tasks},
  author={Mastropaolo, Antonio and Cooper, Nathan and Palacio, David Nader and Scalabrino, Simone and Poshyvanyk, Denys and Oliveto, Rocco and Bavota, Gabriele},
  journal={IEEE Transactions on Software Engineering},
  volume={49},
  number={4},
  pages={1580--1598},
  year={2022},
  publisher={IEEE}
}

@inproceedings{ahmed2024automatic,
  title={Automatic semantic augmentation of language model prompts (for code summarization)},
  author={Ahmed, Toufique and Pai, Kunal Suresh and Devanbu, Premkumar and Barr, Earl},
  booktitle={Proceedings of the IEEE/ACM 46th international conference on software engineering},
  pages={1--13},
  year={2024}
}

@article{vitale2025optimizing,
  title={Optimizing Datasets for Code Summarization: Is Code-Comment Coherence Enough?},
  author={Vitale, Antonio and Mastropaolo, Antonio and Oliveto, Rocco and Di Penta, Massimiliano and Scalabrino, Simone},
  journal={arXiv preprint arXiv:2502.07611},
  year={2025}
}

@inproceedings{ahmed2022few,
  title={Few-shot training llms for project-specific code-summarization},
  author={Ahmed, Toufique and Devanbu, Premkumar},
  booktitle={Proceedings of the 37th IEEE/ACM international conference on automated software engineering},
  pages={1--5},
  year={2022}
}

@inproceedings{wang2022no,
  title={No more fine-tuning? an experimental evaluation of prompt tuning in code intelligence},
  author={Wang, Chaozheng and Yang, Yuanhang and Gao, Cuiyun and Peng, Yun and Zhang, Hongyu and Lyu, Michael R},
  booktitle={Proceedings of the 30th ACM joint European software engineering conference and symposium on the foundations of software engineering},
  pages={382--394},
  year={2022}
}

@article{ayupov2022parameter,
  title={Parameter-efficient finetuning of transformers for source code},
  author={Ayupov, Shamil and Chirkova, Nadezhda},
  journal={arXiv preprint arXiv:2212.05901},
  year={2022}
}

@inproceedings{zlotchevski2022exploring,
  title={Exploring and evaluating personalized models for code generation},
  author={Zlotchevski, Andrei and Drain, Dawn and Svyatkovskiy, Alexey and Clement, Colin B and Sundaresan, Neel and Tufano, Michele},
  booktitle={Proceedings of the 30th ACM joint European software engineering conference and symposium on the foundations of software engineering},
  pages={1500--1508},
  year={2022}
}

@inproceedings{goel2022cross,
  title={On the cross-modal transfer from natural language to code through adapter modules},
  author={Goel, Divyam and Grover, Ramansh and Fard, Fatemeh H},
  booktitle={Proceedings of the 30th IEEE/ACM International Conference on Program Comprehension},
  pages={71--81},
  year={2022}
}

@article{saberi2024utilization,
  title={Utilization of pre-trained language models for adapter-based knowledge transfer in software engineering},
  author={Saberi, Iman and Fard, Fatemeh and Chen, Fuxiang},
  journal={Empirical Software Engineering},
  volume={29},
  number={4},
  pages={94},
  year={2024},
  publisher={Springer}
}

@inproceedings{choi2023codeprompt,
  title={CodePrompt: Task-agnostic prefix tuning for program and language generation},
  author={Choi, YunSeok and Lee, Jee-Hyong},
  booktitle={Findings of the Association for Computational Linguistics: ACL 2023},
  pages={5282--5297},
  year={2023}
}

@inproceedings{liu2023empirical,
  title={An empirical study of parameter-efficient fine-tuning methods for pre-trained code models},
  author={Liu, Jiaxing and Sha, Chaofeng and Peng, Xin},
  booktitle={2023 38th IEEE/ACM International Conference on Automated Software Engineering (ASE)},
  pages={397--408},
  year={2023},
  organization={IEEE}
}

@article{wang2023prompt,
  title={Prompt tuning in code intelligence: An experimental evaluation},
  author={Wang, Chaozheng and Yang, Yuanhang and Gao, Cuiyun and Peng, Yun and Zhang, Hongyu and Lyu, Michael R},
  journal={IEEE Transactions on Software Engineering},
  volume={49},
  number={11},
  pages={4869--4885},
  year={2023},
  publisher={IEEE}
}

@article{afrin2025quantization,
  title={Is Quantization a Deal-breaker? Empirical Insights from Large Code Models},
  author={Afrin, Saima and Xu, Bowen and Mastropaolo, Antonio},
  journal={arXiv preprint arXiv:2507.09665},
  year={2025}
}

@article{hui2024qwen2,
  title={Qwen2. 5-coder technical report},
  author={Hui, Binyuan and Yang, Jian and Cui, Zeyu and Yang, Jiaxi and Liu, Dayiheng and Zhang, Lei and Liu, Tianyu and Zhang, Jiajun and Yu, Bowen and Lu, Keming and others},
  journal={arXiv preprint arXiv:2409.12186},
  year={2024}
}

@article{roziere2023code,
  title={Code llama: Open foundation models for code},
  author={Roziere, Baptiste and Gehring, Jonas and Gloeckle, Fabian and Sootla, Sten and Gat, Itai and Tan, Xiaoqing Ellen and Adi, Yossi and Liu, Jingyu and Sauvestre, Romain and Remez, Tal and others},
  journal={arXiv preprint arXiv:2308.12950},
  year={2023}
}

@inproceedings{yu2024codereval,
  title={Codereval: A benchmark of pragmatic code generation with generative pre-trained models},
  author={Yu, Hao and Shen, Bo and Ran, Dezhi and Zhang, Jiaxin and Zhang, Qi and Ma, Yuchi and Liang, Guangtai and Li, Ying and Wang, Qianxiang and Xie, Tao},
  booktitle={Proceedings of the 46th IEEE/ACM International Conference on Software Engineering},
  pages={1--12},
  year={2024}
}

@article{lu2021codexglue,
  title={Codexglue: A machine learning benchmark dataset for code understanding and generation},
  author={Lu, Shuai and Guo, Daya and Ren, Shuo and Huang, Junjie and Svyatkovskiy, Alexey and Blanco, Ambrosio and Clement, Colin and Drain, Dawn and Jiang, Daxin and Tang, Duyu and others},
  journal={arXiv preprint arXiv:2102.04664},
  year={2021}
}

@article{crupi2025effectiveness,
  title={On the Effectiveness of LLM-as-a-judge for Code Generation and Summarization},
  author={Crupi, Giuseppe and Tufano, Rosalia and Velasco, Alejandro and Mastropaolo, Antonio and Poshyvanyk, Denys and Bavota, Gabriele},
  journal={IEEE Transactions on Software Engineering},
  year={2025},
  publisher={IEEE}
}

@article{ren2020codebleu,
  title={Codebleu: a method for automatic evaluation of code synthesis},
  author={Ren, Shuo and Guo, Daya and Lu, Shuai and Zhou, Long and Liu, Shujie and Tang, Duyu and Sundaresan, Neel and Zhou, Ming and Blanco, Ambrosio and Ma, Shuai},
  journal={arXiv preprint arXiv:2009.10297},
  year={2020}
}

@article{post2018call,
  title={A call for clarity in reporting BLEU scores},
  author={Post, Matt},
  journal={arXiv preprint arXiv:1804.08771},
  year={2018}
}

@inproceedings{papineni2002bleu,
  title={Bleu: a method for automatic evaluation of machine translation},
  author={Papineni, Kishore and Roukos, Salim and Ward, Todd and Zhu, Wei-Jing},
  booktitle={Proceedings of the 40th annual meeting of the Association for Computational Linguistics},
  pages={311--318},
  year={2002}
}

@inproceedings{mastropaolo2024evaluating,
  title={Evaluating code summarization techniques: A new metric and an empirical characterization},
  author={Mastropaolo, Antonio and Ciniselli, Matteo and Di Penta, Massimiliano and Bavota, Gabriele},
  booktitle={Proceedings of the IEEE/ACM 46th International Conference on Software Engineering},
  pages={1--13},
  year={2024}
}

@inproceedings{zhang2020retrieval,
  title={Retrieval-based neural source code summarization},
  author={Zhang, Jian and Wang, Xu and Zhang, Hongyu and Sun, Hailong and Liu, Xudong},
  booktitle={Proceedings of the ACM/IEEE 42nd International Conference on Software Engineering},
  pages={1385--1397},
  year={2020}
}

@inproceedings{leclair2020improved,
  title={Improved code summarization via a graph neural network},
  author={LeClair, Alexander and Haque, Sakib and Wu, Lingfei and McMillan, Collin},
  booktitle={Proceedings of the 28th international conference on program comprehension},
  pages={184--195},
  year={2020}
}

@article{yin2019lizard,
  title={Lizard: an extensible cyclomatic complexity analyzer},
  author={Yin, Terry},
  journal={Astrophysics Source Code Library},
  pages={ascl--1906},
  year={2019}
}

@misc{PylintTeam2025,
  author       = {{Pylint Team}},
  title        = {Pylint: Code Analysis for Python},
  howpublished = {\url{https://www.pylint.org/}},
  year         = {2025},
  note         = {Version 3.3.8, released August 9, 2025. Initial release in 2001. Accessed: 2025-10-12}
}

@inproceedings{siddiq2024quality,
  title={Quality assessment of chatgpt generated code and their use by developers},
  author={Siddiq, Mohammed Latif and Roney, Lindsay and Zhang, Jiahao and Santos, Joanna Cecilia Da Silva},
  booktitle={Proceedings of the 21st international conference on mining software repositories},
  pages={152--156},
  year={2024}
}

@article{liu2024refining,
  title={Refining chatgpt-generated code: Characterizing and mitigating code quality issues},
  author={Liu, Yue and Le-Cong, Thanh and Widyasari, Ratnadira and Tantithamthavorn, Chakkrit and Li, Li and Le, Xuan-Bach D and Lo, David},
  journal={ACM Transactions on Software Engineering and Methodology},
  volume={33},
  number={5},
  pages={1--26},
  year={2024},
  publisher={ACM New York, NY}
}

@misc{MicrosoftRoslyn2025,
  author       = {{Microsoft .NET Team}},
  title        = {The .NET Compiler Platform ({Roslyn})},
  howpublished = {\url{https://github.com/dotnet/roslyn}},
  year         = {2025},
  note         = {Open-source compiler and analysis platform for C\# and Visual Basic. Accessed: 2025-10-12}
}

@misc{OpenAI2025GPT5,
  author       = {{OpenAI}},
  title        = {Introducing GPT-5},
  howpublished = {\url{https://openai.com/index/introducing-gpt-5/}},
  year         = {2025},
  note         = {Accessed: 2025-10-12}
}

@article{wilcoxon1945individual,
  title={Individual comparisons by ranking methods},
  author={Wilcoxon, Frank},
  journal={Biometrics bulletin},
  volume={1},
  number={6},
  pages={80--83},
  year={1945},
  publisher={JSTOR}
}

@article{ciniselli2021empirical,
  title={An empirical study on the usage of transformer models for code completion},
  author={Ciniselli, Matteo and Cooper, Nathan and Pascarella, Luca and Mastropaolo, Antonio and Aghajani, Emad and Poshyvanyk, Denys and Di Penta, Massimiliano and Bavota, Gabriele},
  journal={IEEE Transactions on Software Engineering},
  volume={48},
  number={12},
  pages={4818--4837},
  year={2021},
  publisher={IEEE}
}

@inproceedings{du2024evaluating,
  title={Evaluating large language models in class-level code generation},
  author={Du, Xueying and Liu, Mingwei and Wang, Kaixin and Wang, Hanlin and Liu, Junwei and Chen, Yixuan and Feng, Jiayi and Sha, Chaofeng and Peng, Xin and Lou, Yiling},
  booktitle={Proceedings of the IEEE/ACM 46th International Conference on Software Engineering},
  pages={1--13},
  year={2024}
}

@article{wang2025software,
  title={Software Development Life Cycle Perspective: A Survey of Benchmarks for Code Large Language Models and Agents},
  author={Wang, Kaixin and Li, Tianlin and Zhang, Xiaoyu and Wang, Chong and Sun, Weisong and Liu, Yang and Shi, Bin},
  journal={arXiv preprint arXiv:2505.05283},
  year={2025}
}

@article{zhuo2024bigcodebench,
  title={Bigcodebench: Benchmarking code generation with diverse function calls and complex instructions},
  author={Zhuo, Terry Yue and Vu, Minh Chien and Chim, Jenny and Hu, Han and Yu, Wenhao and Widyasari, Ratnadira and Yusuf, Imam Nur Bani and Zhan, Haolan and He, Junda and Paul, Indraneil and others},
  journal={arXiv preprint arXiv:2406.15877},
  year={2024}
}

@article{quan2025codeelo,
  title={Codeelo: Benchmarking competition-level code generation of llms with human-comparable elo ratings},
  author={Quan, Shanghaoran and Yang, Jiaxi and Yu, Bowen and Zheng, Bo and Liu, Dayiheng and Yang, An and Ren, Xuancheng and Gao, Bofei and Miao, Yibo and Feng, Yunlong and others},
  journal={arXiv preprint arXiv:2501.01257},
  year={2025}
}

@article{tambon2025bugs,
  title={Bugs in large language models generated code: An empirical study},
  author={Tambon, Florian and Moradi-Dakhel, Arghavan and Nikanjam, Amin and Khomh, Foutse and Desmarais, Michel C and Antoniol, Giuliano},
  journal={Empirical Software Engineering},
  volume={30},
  number={3},
  pages={65},
  year={2025},
  publisher={Springer}
}

@article{wang2025beyond,
  title={Beyond functional correctness: Investigating coding style inconsistencies in large language models},
  author={Wang, Yanlin and Jiang, Tianyue and Liu, Mingwei and Chen, Jiachi and Mao, Mingzhi and Liu, Xilin and Ma, Yuchi and Zheng, Zibin},
  journal={Proceedings of the ACM on Software Engineering},
  volume={2},
  number={FSE},
  pages={690--712},
  year={2025},
  publisher={ACM New York, NY, USA}
}

@inproceedings{paul2024benchmarks,
  title={Benchmarks and metrics for evaluations of code generation: A critical review},
  author={Paul, Debalina Ghosh and Zhu, Hong and Bayley, Ian},
  booktitle={2024 IEEE International Conference on Artificial Intelligence Testing (AITest)},
  pages={87--94},
  year={2024},
  organization={IEEE}
}

@article{kharma2025security,
  title={Security and Quality in LLM-Generated Code: A Multi-Language, Multi-Model Analysis},
  author={Kharma, Mohammed and Choi, Soohyeon and AlKhanafseh, Mohammed and Mohaisen, David},
  journal={arXiv preprint arXiv:2502.01853},
  year={2025}
}

@article{liu2023your,
  title={Is your code generated by chatgpt really correct? rigorous evaluation of large language models for code generation},
  author={Liu, Jiawei and Xia, Chunqiu Steven and Wang, Yuyao and Zhang, Lingming},
  journal={Advances in Neural Information Processing Systems},
  volume={36},
  pages={21558--21572},
  year={2023}
}

@misc{sonarsource2025sonarcloud,
  author       = {SonarSource},
  title        = {SonarCloud},
  howpublished = {\url{https://docs.sonarsource.com/sonarqube-cloud/}},
  note         = {Accessed: 2025-10-18}
}

@article{hou2024large,
  title={Large language models for software engineering: A systematic literature review},
  author={Hou, Xinyi and Zhao, Yanjie and Liu, Yue and Yang, Zhou and Wang, Kailong and Li, Li and Luo, Xiapu and Lo, David and Grundy, John and Wang, Haoyu},
  journal={ACM Transactions on Software Engineering and Methodology},
  volume={33},
  number={8},
  pages={1--79},
  year={2024},
  publisher={ACM New York, NY}
}

@article{watson2022systematic,
  title={A systematic literature review on the use of deep learning in software engineering research},
  author={Watson, Cody and Cooper, Nathan and Palacio, David Nader and Moran, Kevin and Poshyvanyk, Denys},
  journal={ACM Transactions on Software Engineering and Methodology (TOSEM)},
  volume={31},
  number={2},
  pages={1--58},
  year={2022},
  publisher={ACM New York, NY}
}

@inproceedings{khoury2023secure,
  title={How secure is code generated by chatgpt?},
  author={Khoury, Rapha{\"e}l and Avila, Anderson R and Brunelle, Jacob and Camara, Baba Mamadou},
  booktitle={2023 IEEE international conference on systems, man, and cybernetics (SMC)},
  pages={2445--2451},
  year={2023},
  organization={IEEE}
}

@inproceedings{mastropaolo2023robustness,
  title={On the robustness of code generation techniques: An empirical study on github copilot},
  author={Mastropaolo, Antonio and Pascarella, Luca and Guglielmi, Emanuela and Ciniselli, Matteo and Scalabrino, Simone and Oliveto, Rocco and Bavota, Gabriele},
  booktitle={2023 IEEE/ACM 45th International Conference on Software Engineering (ICSE)},
  pages={2149--2160},
  year={2023},
  organization={IEEE}
}

@article{cassano2023multipl,
  title={Multipl-e: A scalable and polyglot approach to benchmarking neural code generation},
  author={Cassano, Federico and Gouwar, John and Nguyen, Daniel and Nguyen, Sydney and Phipps-Costin, Luna and Pinckney, Donald and Yee, Ming-Ho and Zi, Yangtian and Anderson, Carolyn Jane and Feldman, Molly Q and others},
  journal={IEEE Transactions on Software Engineering},
  volume={49},
  number={7},
  pages={3675--3691},
  year={2023},
  publisher={IEEE}
}

@article{li2022competition,
  title={Competition-level code generation with alphacode},
  author={Li, Yujia and Choi, David and Chung, Junyoung and Kushman, Nate and Schrittwieser, Julian and Leblond, R{\'e}mi and Eccles, Tom and Keeling, James and Gimeno, Felix and Dal Lago, Agustin and others},
  journal={Science},
  volume={378},
  number={6624},
  pages={1092--1097},
  year={2022},
  publisher={American Association for the Advancement of Science}
}

@article{anthropic2024claude,
  title={The claude 3 model family: Opus, sonnet, haiku},
  author={Anthropic, AI},
  journal={Claude-3 Model Card},
  volume={1},
  number={1},
  pages={4},
  year={2024}
}

@article{team2023gemini,
  title={Gemini: a family of highly capable multimodal models},
  author={Team, Gemini and Anil, Rohan and Borgeaud, Sebastian and Alayrac, Jean-Baptiste and Yu, Jiahui and Soricut, Radu and Schalkwyk, Johan and Dai, Andrew M and Hauth, Anja and Millican, Katie and others},
  journal={arXiv preprint arXiv:2312.11805},
  year={2023}
}

@article{shi2025efficient,
  title={Efficient and Green Large Language Models for Software Engineering: Literature Review, Vision, and the Road Ahead},
  author={Shi, Jieke and Yang, Zhou and Lo, David},
  journal={ACM Transactions on Software Engineering and Methodology},
  volume={34},
  number={5},
  pages={1--22},
  year={2025},
  publisher={ACM New York, NY}
}

@article{wang2022adamix,
  title={Adamix: Mixture-of-adaptations for parameter-efficient model tuning},
  author={Wang, Yaqing and Agarwal, Sahaj and Mukherjee, Subhabrata and Liu, Xiaodong and Gao, Jing and Awadallah, Ahmed Hassan and Gao, Jianfeng},
  journal={arXiv preprint arXiv:2205.12410},
  year={2022}
}

@article{sun2024source,
  title={Source code summarization in the era of large language models},
  author={Sun, Weisong and Miao, Yun and Li, Yuekang and Zhang, Hongyu and Fang, Chunrong and Liu, Yi and Deng, Gelei and Liu, Yang and Chen, Zhenyu},
  journal={arXiv preprint arXiv:2407.07959},
  year={2024}
}

@inproceedings{wang2023one,
  title={One adapter for all programming languages? adapter tuning for code search and summarization},
  author={Wang, Deze and Chen, Boxing and Li, Shanshan and Luo, Wei and Peng, Shaoliang and Dong, Wei and Liao, Xiangke},
  booktitle={2023 IEEE/ACM 45th International Conference on Software Engineering (ICSE)},
  pages={5--16},
  year={2023},
  organization={IEEE}
}

@inproceedings{chen2023pass,
  title={Pass-tuning: Towards structure-aware parameter-efficient tuning for code representation learning},
  author={Chen, Nuo and Sun, Qiushi and Wang, Jianing and Li, Xiang and Gao, Ming},
  booktitle={Findings of the Association for Computational Linguistics: EMNLP 2023},
  pages={577--591},
  year={2023}
}

@inproceedings{liu2024delving,
  title={Delving into parameter-efficient fine-tuning in code change learning: An empirical study},
  author={Liu, Shuo and Keung, Jacky and Yang, Zhen and Liu, Fang and Zhou, Qilin and Liao, Yihan},
  booktitle={2024 IEEE International Conference on Software Analysis, Evolution and Reengineering (SANER)},
  pages={465--476},
  year={2024},
  organization={IEEE}
}

@article{raffel2020exploring,
  title={Exploring the limits of transfer learning with a unified text-to-text transformer},
  author={Raffel, Colin and Shazeer, Noam and Roberts, Adam and Lee, Katherine and Narang, Sharan and Matena, Michael and Zhou, Yanqi and Li, Wei and Liu, Peter J},
  journal={Journal of machine learning research},
  volume={21},
  number={140},
  pages={1--67},
  year={2020}
}

@misc{pmd2025,
  author       = {{P. D. Team}},
  title        = {{PMD - Source Code Analyzer}},
  year         = {2025},
  note         = {Static code analysis tool for Java and other languages},
  howpublished = {\url{https://pmd.github.io}},
}

@article{wei2022emergent,
  title={Emergent abilities of large language models},
  author={Wei, Jason and Tay, Yi and Bommasani, Rishi and Raffel, Colin and Zoph, Barret and Borgeaud, Sebastian and Yogatama, Dani and Bosma, Maarten and Zhou, Denny and Metzler, Donald and others},
  journal={arXiv preprint arXiv:2206.07682},
  year={2022}
}

@article{qin2023chatgpt,
  title={Is ChatGPT a general-purpose natural language processing task solver?},
  author={Qin, Chengwei and Zhang, Aston and Zhang, Zhuosheng and Chen, Jiaao and Yasunaga, Michihiro and Yang, Diyi},
  journal={arXiv preprint arXiv:2302.06476},
  year={2023}
}

@inproceedings{shin2025prompt,
  title={Prompt Engineering or Fine-Tuning: An Empirical Assessment of LLMs for Code},
  author={Shin, Jiho and Tang, Clark and Mohati, Tahmineh and Nayebi, Maleknaz and Wang, Song and Hemmati, Hadi},
  booktitle={2025 IEEE/ACM 22nd International Conference on Mining Software Repositories (MSR)},
  pages={490--502},
  year={2025},
  organization={IEEE}
}

@article{frantar2022gptq,
  title={Gptq: Accurate post-training quantization for generative pre-trained transformers},
  author={Frantar, Elias and Ashkboos, Saleh and Hoefler, Torsten and Alistarh, Dan},
  journal={arXiv preprint arXiv:2210.17323},
  year={2022}
}

@article{dettmers2022gpt3,
  title={Gpt3. int8 (): 8-bit matrix multiplication for transformers at scale},
  author={Dettmers, Tim and Lewis, Mike and Belkada, Younes and Zettlemoyer, Luke},
  journal={Advances in neural information processing systems},
  volume={35},
  pages={30318--30332},
  year={2022}
}

@article{vitale2025toward,
  title={Toward Explaining Large Language Models in Software Engineering Tasks},
  author={Vitale, Antonio and Nguyen, Khai-Nguyen and Poshyvanyk, Denys and Oliveto, Rocco and Scalabrino, Simone and Mastropaolo, Antonio},
  journal={arXiv preprint arXiv:2512.20328},
  year={2025}
}

@inproceedings{burleigh2025beyond,
  title={Beyond the Hint: Using Self-Critique to Constrain LLM Feedback in Conversation-Based Assessment},
  author={Burleigh, Tyler and Han, Jenny and DiCerbo, Kristen},
  booktitle={Proceedings of the Artificial Intelligence in Measurement and Education Conference (AIME-Con): Coordinated Session Papers},
  pages={79--85},
  year={2025}
}

\appendix
\section{LLM-as-a-Judge Quality Assessment Visualizations}
\label{appendix:llm_judge}

\begin{figure*}[h]
    \centering
    \begin{subfigure}{0.9\textwidth}
        \includegraphics[width=\textwidth]{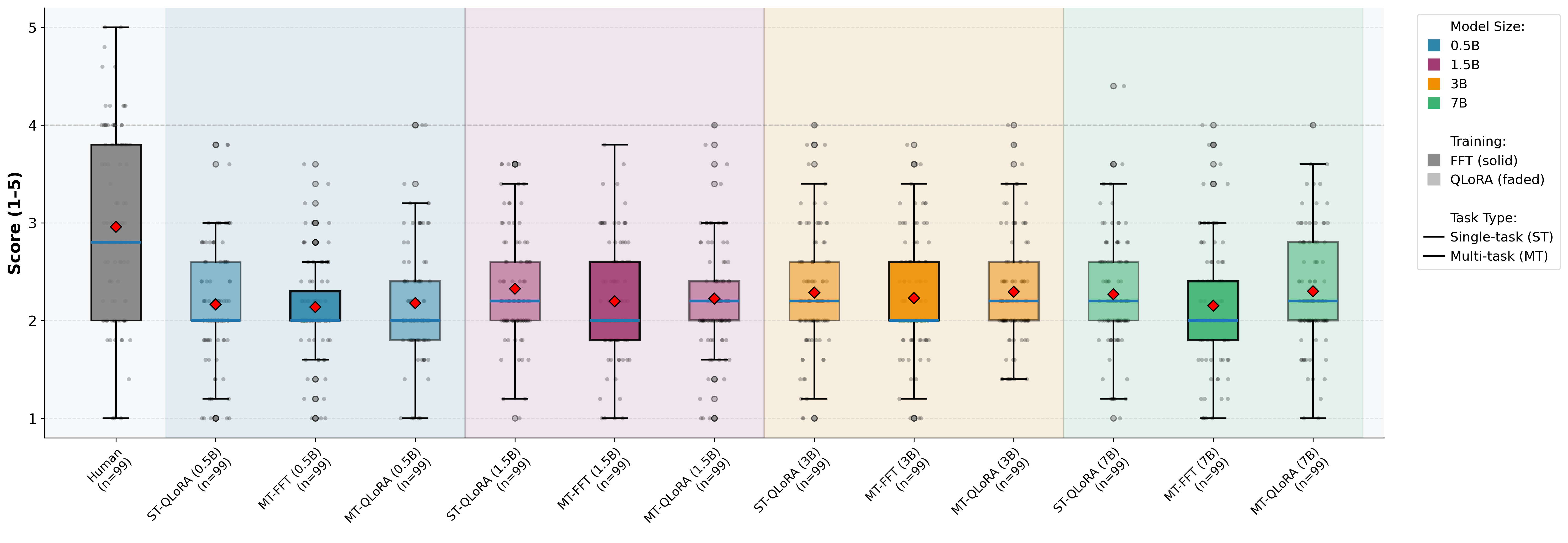}
        \caption{Content Adequacy}
        \label{fig:java_content}
    \end{subfigure}
    
    \begin{subfigure}{0.9\textwidth}
        \includegraphics[width=\textwidth]{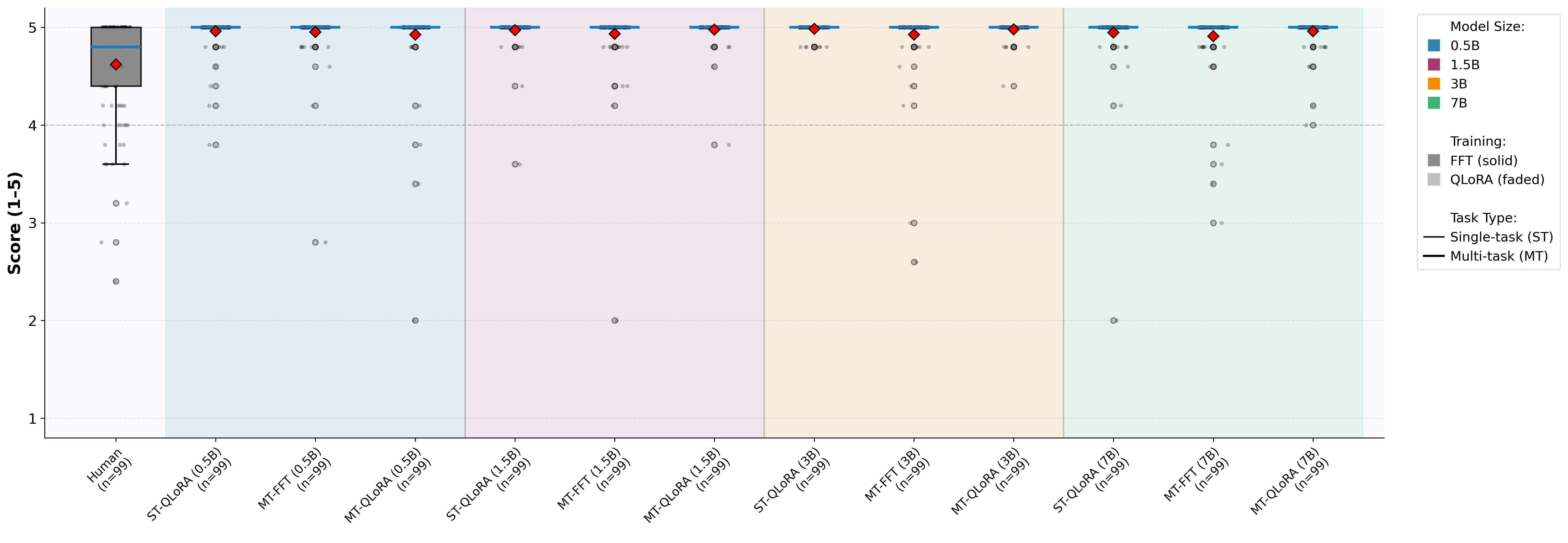}
        \caption{Conciseness}
        \label{fig:java_concise}
    \end{subfigure}
    
    \begin{subfigure}{0.9\textwidth}
        \includegraphics[width=\textwidth]{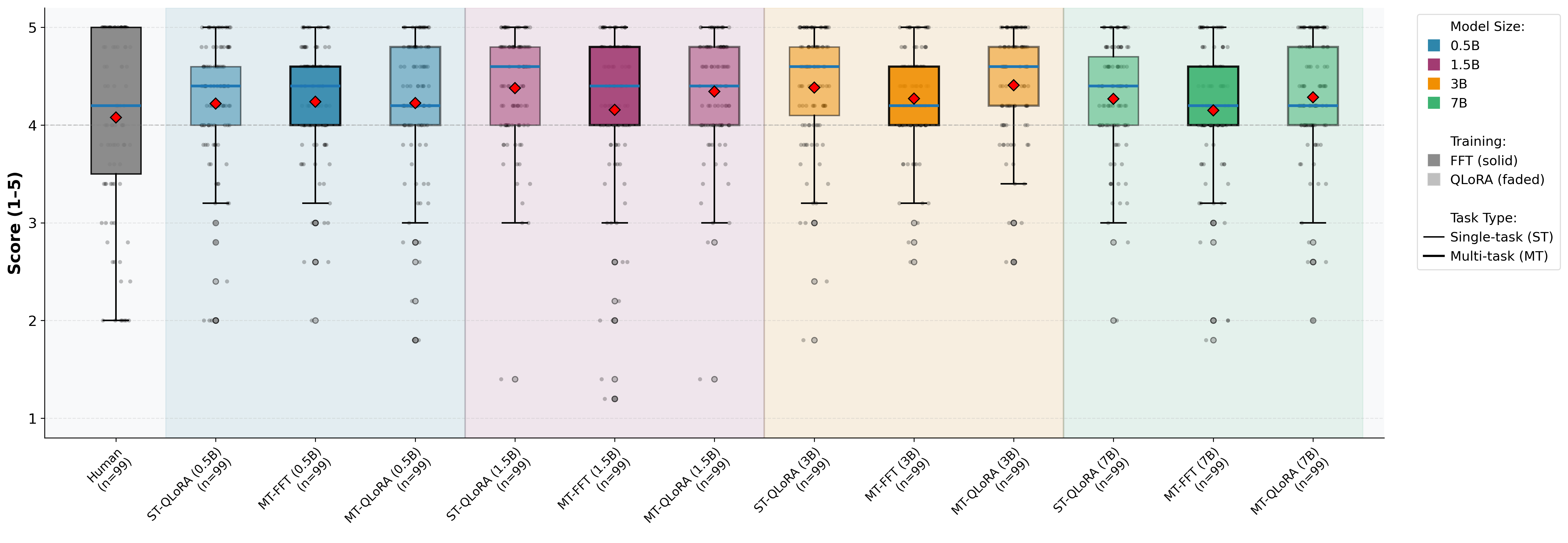}
        \caption{Fluency}
        \label{fig:java_fluency}
    \end{subfigure}
    \caption{Boxplots for Java code summarization across three quality dimensions. Distribution compares Human-written summaries against models at four scales (0.5B, 1.5B, 3B\rev{, 7B}) trained with Full Fine-Tuning (FFT, solid) or QLoRA (faded) in SingleTask and MultiTask configurations. Red diamonds indicate mean scores, blue lines indicate medians, background shading distinguishes model sizes.}
    \label{fig:llm_judge_java}
\end{figure*}

\begin{figure*}[h]
    \centering
    \begin{subfigure}{0.9\textwidth}
        \includegraphics[width=\textwidth]{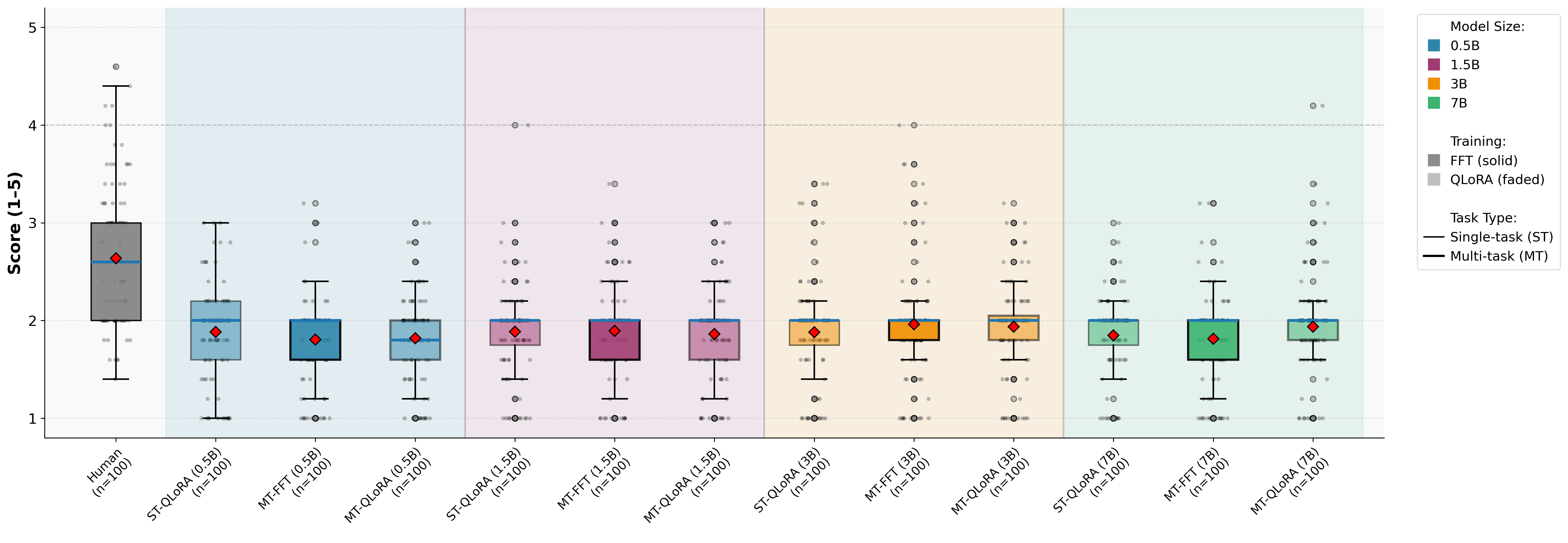}
        \caption{Content Adequacy}
        \label{fig:python_content}
    \end{subfigure}
    
    \begin{subfigure}{0.9\textwidth}
        \includegraphics[width=\textwidth]{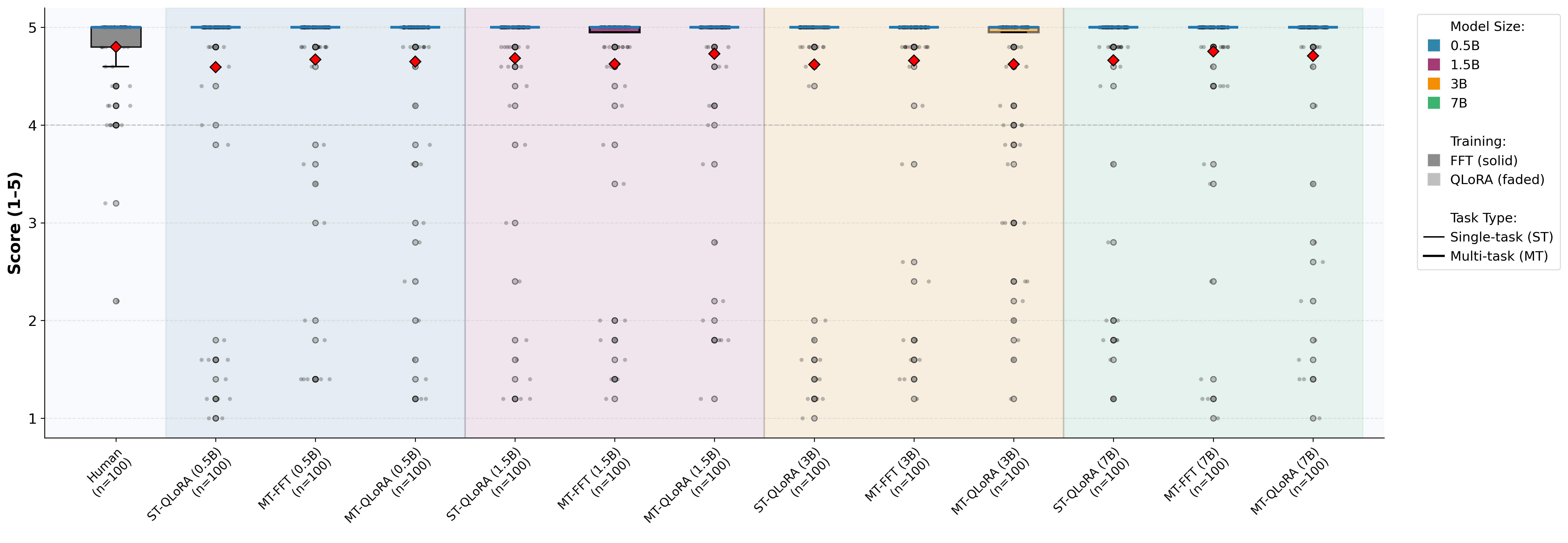}
        \caption{Conciseness}
        \label{fig:python_concise}
    \end{subfigure}
    
    \begin{subfigure}{0.9\textwidth}
        \includegraphics[width=\textwidth]{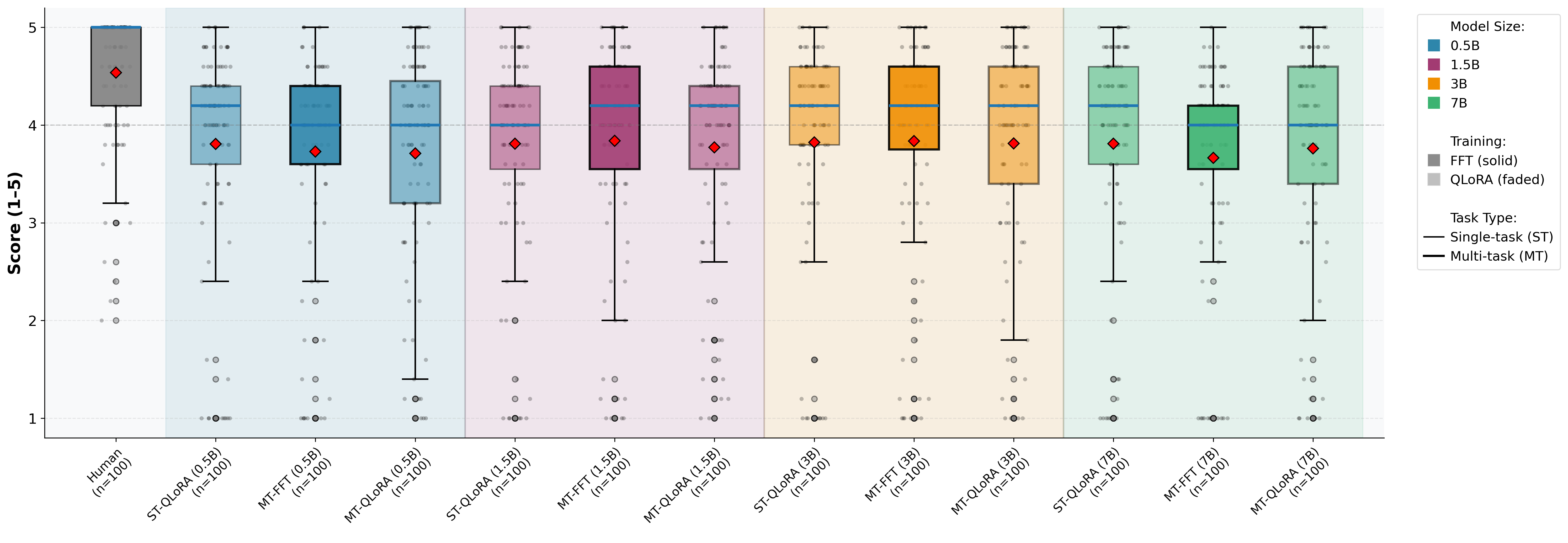}
        \caption{Fluency}
        \label{fig:python_fluency}
    \end{subfigure}
    \caption{Boxplots for \python code summarization across three quality dimensions. Distribution compares Human-written summaries against models at four scales (0.5B, 1.5B, 3B\rev{, 7B}) trained with Full Fine-Tuning (FFT, solid) or QLoRA (faded) in SingleTask and MultiTask configurations. Red diamonds indicate mean scores, blue lines indicate medians, background shading distinguishes model sizes.}
    \label{fig:llm_judge_python}
\end{figure*}

 This appendix presents the complete distribution of GPT-5 mini quality scores for code summarization across all experimental configurations. \figref{fig:llm_judge_java} and \figref{fig:llm_judge_python} display boxplots for three quality dimensions -- Content Adequacy, Conciseness, and Fluency -- comparing Human-written summaries against all training configurations.



\end{document}